**NATIONAL AND KAPODISTRIAN UNIVERSITY OF ATHENS**

**DEPARTMENT OF INFORMATICS AND TELECOMUNICATIONS**

**PROGRAM OF POSTGRADUATE STUDIES**

**DIPLOMA THESIS**

# Sharing of Semantically Enhanced Information for the Adaptive Execution of Business Processes

**Pigi D. Kouki**

**Supervisor: Aphrodite Tsalgatidou,** Assistant Professor NKUA
**Technical Support: Georgios Athanasopoulos,** PhD Candidate NKUA

**DIPLOMA THESIS**

Sharing of Semantically Enhanced Information for the Adaptive Execution of Business Processes

**Pigi D. Kouki**

Registration Number: M 984

**SUPERVISOR: Aphrodite Tsalgatidou,** Assistant Professor NKUA
**TECHNICAL SUPPORT: Georgios Athanasopoulos,** PhD Candidate NKUA

## ABSTRACT

Motivated from the Context Aware Computing, and more particularly from the Data-Driven Process Adaptation approach, we propose the Semantic Context Space (SCS) Engine which aims to facilitate the provision of adaptable business processes. The SCS Engine provides a space which stores semantically annotated data and it is open to other processes, systems, and external sources for information exchange. The specified implementation is inspired from the Semantic TupleSpace and uses the JavaSpace Service of the Jini Framework (changed to Apache River lately) as an underlying basis. The SCS Engine supplies an interface where a client can execute the following operations: (i) *write*: which inserts in the space available information along with its respective meta-information, (ii) *read*: which retrieves from the space information which meets specific meta-information constrains, and (iii) *take*: which retrieves and simultaneously deletes from the space information which meets specific meta-information constrains. In terms of this thesis the available types of meta-information are based on ontologies described in RDFS or WSML. The applicability of the SCS Engine implementation in the context of data-driven process adaptation has been ensured by an experimental evaluation of the provided operations. Ultimately, we refer to open issues which could be addressed to enrich the proposed Engine with additional features.

# TABLE OF CONTENTS

# LIST OF FIGURES

## LIST OF TABLES

# PREFACE

The work described in this thesis has been performed in the Service Oriented Computing, Software Engineering, and System Development Laboratory of the Department of Informatics and Telecommunications of the National and Kapodistrian University of Athens, during the time period May 2010 to May 2011. Mrs Aphrodite Tsalgatidou, Assistant Professor of the department of Informatics and Telecommunications, undertook the supervision of this thesis. Mr Georgios Athanasopoulos, PhD candidate of the department of Informatics and Telecommunications was responsible for the technical support. Mr Stathes Hadjiefthymiades, Assistant Professor of the department of Informatics and Telecommunications, was the examiner of this thesis.

# CHAPTER 1
# INTRODUCTION

## 1.1 Motivation and Objectives of the thesis

Within the context of the pervasive computing environment, the advent of *information sources* and *services* facilitate the emergence of adaptable systems [1]. Representative information sources are the emerging Sensor Web [2] and the materializations of the Web 2.0 and Web 3.0 paradigms, e.g. Social Networking applications and Agent based systems. Services include, apart from the conventional Web Services, additional or emerging types such as P2P, and Grid Services. Service processes can be considered as systems performing basic (e.g. invocation of a service) and structure (e.g. if, while) activities specified beforehand which produce and consume the desired data at execution time. The existing approaches such as WS-BPEL[3] allow only static processes, invoking conventional or semantic Web Services and ignoring available information sources or other types of services which may exist in the process environment; such information can be used to adapt the process execution accordingly.

Adaptable processes have been investigated from the discipline of Artificial Intelligence, where the focus is on the automated construction of orchestrations [4] (the so-called task plans) which include semantically annotated Web services. They can be considered as State Machine Models, where transitions are mapped to execution of service operations. As a result, services along with their respective chaining constitute the prime elements in a process. The main drawbacks of these approaches are two: (i) they do not take into consideration the emerging field of Context Aware Computing (CAC)[5] which exploits contextual information produced by external sources which may be relevant to the process and (ii) they do not deal with the interoperability concerns raised by the multiple instantiations of the service oriented computing paradigm [6].

Therefore, the proposal and implementation of an approach which can assure the development and execution of adaptable service processes is an imperative need. According to CAC, the collection and storing of context data from a wide range of context sources, such as sensors and Web Services, can be applied to promote mining of context data towards service adaptation in ubiquitous computing systems. Following this principle, we argue that a process should not take into consideration only the internal information produced by the execution of its operations (including the responses produced by service invocations). The service chain should also depend on the information produced by the external sources which exist in its environment during its development and execution. This information, referring to semantically annotated structured data, can be used for adapting with a high probability the execution of the process.

To provide processes able to adapt their execution accordingly, an approach that comprises an execution engine, along with a space representing the process's environment and special process adaptation algorithms has been first presented by Athanasopoulos and Tsalgatidou[1]. The space stores semantically annotated data and it is open to other processes, systems, and external sources for information exchange. The process adaptation algorithms specify appropriate adaptation plans to given processes and produce queries searching for relevant to the process information. These queries are executed in the space and when valid information is gathered, it is fed to the process execution engine. The provided process adaptation plans control the execution and adaptation of service chains based on the relevant information. This platform has been designed and it is on the implementation phase in terms of Envision, a European project which is analyzed later on in this section.

From an implementation point of view, the SCS Engine has been inspired from TupleSpace model [7] and uses JavaSpace [8] service of the Jini framework as its underlying basis. Extensions to the JavaSpace paradigm assure the support of semantically annotated information. This thesis elaborates on the design, implementation, and test of the SCS Engine.

### 1.1.1 Envision Project

ENVISION is a project funded by the European Union. It aims at providing an **ENVI**ronmental **S**ervices **I**nfrastructure with **ON**tologies that supports non ICT-skilled users (Information and Communication Technologies) in the process of semantic discovery, adaptive chaining, and process execution of environmental services. The innovative parts in ENVISION are: (i) on-the-Web enabling and packaging of technologies for their use by non ICT-skilled users, (ii) the support for migrating environmental models to be provided as models as a service (Maas), and (iii) the use of data streaming information for harvesting information for dynamic building of ontologies and adapting service execution [9]. The pilots which will be held in terms of ENVISION include landslide hazard assessment and oil spills decision support systems. The excepted goals of the project are (i) the promotion of using modeling tools available in the Web and (ii) easier accessibility to distributed sources of information.

The key components of the ENVISION are: (i) the Environmental Decision Portal, (ii) the ENVISION Ontology Infrastructure, and (iii) the Execution Infrastructure. The first component supports the creation of web-based applications enabled for dynamic discovery and visual service chaining. The second one supports the visual semantic annotation tools and multilingual ontology management. The last component consists of a semantic discovery catalogue, a semantic service mediator, and the adaptive execution infrastructure. The design and implementation of the Adaptive Execution Infrastructure is on the responsibility of S3Lab[1] (Service Oriented Computing – Software Engineering – Systems Development Laboratory) and it is inspired from the current transformation of environmental information systems to environmental services accessible over the web. As we already mentioned before, this component exploits available information within the process environment to succeed the adaptive and distributed execution of service chains.

## 1.2 High level goals and needs

In the context of this thesis our goals as regards the Semantic Context Space have been to facilitate:

(i) Mechanisms allowing information providers (sources) to write or remove semantically annotated information as well as consumers to retrieve this information located in a specific space.

(ii) The manipulation of data coming up from several types of information sources.

(iii) The manipulation of large data volumes.

(iv) The rapid information retrieval.

These requirements come from the following needs/observations:

(i) The solitary manipulation of data without any meta-information about what they conceptually represent has been proved inefficient especially when considering the quality of returned results in search operations. Additionally, in terms of process adaptation, the exchange of information among the process and its surrounding systems is critical.

[1] http://s3lab.di.uoa.gr/

(ii) The specified pilot scenarios defined in terms of Envision Project manipulate data that may come up from several types of information sources. Therefore, the system should be able to address the use of distinct information provided by external sources [10].

(iii) Environmental processes normally involve the exchange and manipulation of large data volumes, e.g. sensor information.

(iv) The sooner relevant information is retrieved from the SCS Engine, the more likely it is for the process optimizer to use this information in order to adapt its execution.

When valid and appropriate information is gathered byt the SCS Engine, it is fed to the process execution engine. The provided process adaptation plans control the execution and adaptation of service chains based on the relevant information.

## 1.3 Structure of the thesis

The rest of this thesis is organized as follows:

CHAPTER 2 presents the technological background used as a reference point for this thesis. More particularly, we first refer to Linda Coordination language, as this was the introducer of the notion of TupleSpace. Afterwards, we investigate the four most prominent up-to-now approaches to semantically enhanced TupleSpace: sTuples, Semantic Web Spaces, Triple Space Computing, and Conceptual Spaces. Following, we briefly describe WSMX, an execution environment for the dynamic discovery, selection, mediation, and invocation of Semantic Web Services. In the sequel, JavaSpaces of the Jini River technology are presented. We analyze the architecture, the supported interfaces, and the application model of the high-level JavaSpace service for building distributed and collaborative applications, as this is serving as the underlying basis of the SCS Engine. Finally, we make a comparative study of the aforementioned technologies based on specific categorization criteria and we explain why these approaches do not fit to our needs.

CHAPTER 3 elaborates on the SCS Engine. First, we present decisions on the implementation of the SCS Engine explaining (i) why we used the TupleSpace paradigm and the JavaSpaces as the underlying basis for our implementation, (ii) the extension points of the SCS Engine to the JavaSpace, and (iii) the structure of the data inserted in the engine. Subsequently, we refer to the proposed architecture and distinguish the three main subcomponents: Data Manager, Query Processor, and Meta-Information based Interface. Thereafter, we establish the basic supported operations. Initially, we explain how does the process of information retrieval-acquisition from ontologies work, by making special reference to the created structure which will be used later on to support the main operations of the engine. Ultimately, we determine the form and the functionality of the operations write, read, and take.

CHAPTER 4 evaluates the SCS Engine. First, we present the context in which the experiments were held. In the sequel, for each of the three basic operations offered by the SCS Engine (write, read, take) we refer to the criteria on which we execute the experiments. The write operation is evaluated in terms of (i) the size of the objects inserted into the space and (ii) the single-threaded or multi-threaded environment where a client executes these operations. The read operation is evaluated in terms of: (i) the size of the objects read from the space, (ii) the metric used for the retrieval of contextual information, and (iii) the single-threaded or multi-threaded environment where a client executes these operations. The take operation is evaluated in terms of: (i) the size of the objects taken from the space, (ii) the single-threaded or multi-threaded environment where a client executes these operations. After that we refer to characteristics of the results that we want to succeed and we quote the measurements obtained from the execution of dedicated experiments. A discussion on those results follows, explaining the behavior of the SCS Engine in each case. Finally, we compute the complexity that the SCS Engine adds to the JavaSpaces.

CHAPTER 5 concludes the thesis by providing a brief summary of what has been extensively presented. Afterwards, we explain the role of the SCS Engine in terms of the Adaptive Execution Infrastructure. We elaborate on the interfaces provided by the SCS Engine to facilitate the connection with the other two components of the Infrastructure, namely the Process Optimizer and the Orchestration Engine. Finally, we refer to aspects of our future work. These future extensions, when implemented, can be integrated with the SCS Engine providing a more powerful infrastructure for the maintenance and exchange of semantically annotated information.

# CHAPTER 2
# TECHNOLOGICAL BACKROUND

## 2.1 TupleSpace Computing

### 2.1.1 Linda Coordination Language

The notion of TupleSpace was first introduced in Linda language [11]. Linda is based on parallel computing and its ultimate goal was to add the characteristic of concurrency in sequential programming languages. It is defined as a model of coordination and communication among several parallel processes operating upon objects stored in and retrieved from shared, virtual, associative memory: the TupleSpace. Data contained in the TupleSpace are called tuples. A tuple consists of actual fields (typed value fields) and formal fields (non-valued typed fields). Actual fields include integer, Boolean, string values, etc. An example of a simple tuple is the following: <tuplename, ex:integer, FALSE>, where tuplename is the tuple's identifier, ex:integer a formal field and FALSE an actual field of type boolean. The supported coordination operations are four: (i) **in** atomically reads and removes a tuple from TupleSpace, (ii) **rd** non-destructively reads a TupleSpace, (iii) **out** writes a tuple into the TupleSpace, and (iv) **eval** evaluates tuples and writes the result into the TupleSpace. For the retrieval operations Linda introduces the concept of template. A template matches a tuple if they have equal number of fields and each template field matches the corresponding tuple field in type or value. For example, template <tuplename,5,ex:boolean>> matches tuple <tuplename, ex:integer, FALSE>. Furthermore, the retrieval operations are blocking, namely, if they wait until a matching tuple is written in the space. The main characteristics that made Linda famous are the following [12]:

1. Time and reference autonomy: there are no time dependencies between the data provider and the data reader, and also the producer and the consumer do not need to know one another's address.
2. Associative addressing: data are accessed in terms of what kind of data is requested.
3. The coordination implementation is independent of both the host platform and the programming language.
4. Location autonomy: the TupleSpace is independent of the providers' and readers' storage spaces.
5. Asynchrony and Concurrency are supported.

Linda implementations can be found for many programming languages, including Prolog, Python, C, Java, and Lisp. There are also several commercial implementations of the TupleSpace: (i) C-Linda or TCP-Linda from Scientific Computing Associates, founded by Martin Schultz, is targeted on supercomputers and clustered systems. (ii) TSpaces are a TupleSpace platform created by IBM. (iii) JavaSpaces is an implementation in Java by Sun, incorporated into the Jini project. Since JavaSpaces are the underlying basis for our implementation, we describe both Jini and JavaSpaces in Appendix 1 and Paragraph 2.2 respectively.

As we already mentioned, Linda was created to enable parallel processing in centralized and closed computer network environments. Therefore, its application to open distributed systems (with the Web on top) needs extensions in multiple layers. First, the optimization of coordination differs between open and closed systems. In close systems clients are known in advance, while in open systems clients can join and leave at will, without informing the system at start-up. Additionally, in a close system, shared models and types are predefined, while in an open system clients may not agree on these issues, so a middleware should resolve any heterogeneities. Other Linda weaknesses include (i) the set of provided operations which is too small and does not come up to the expectations of many applications and (ii) the type of tuples which is not acceptable for the Semantic Web principle which foresees that all information is encoded in triples of URIs: according to Linda, tuples with the same number of fields and the same field typing cannot be distinguished. In general, each extension to the Linda defines the additional issues that it takes care of.

### 2.1.2 Significant Approaches in Semantically enhanced TupleSpace Technology

The four most prominent semantic TupleSpace platforms are sTuples, Semantic Web spaces, Triple Space Computing, and Conceptual Spaces. Following we give a description of each platform. Afterwards, we summarize the main attributes of each platform, and finally we make a comparative study among them, based on predefined criteria.

#### 2.1.2.1 sTuples

sTuples constitute the first serious attempt towards a Semantic Web-enabled TupleSpace. The initial goal of the Nokia Research Center was to overcome the shortcomings of the Pervasive Computing. The TupleSpace model with the addition of some extensions considered to offer the solution; sTuples is the outcome of this effort. They use as a basis the JavaSpaces and extend them in three main aspects according to its authors [13]. The implementation is based on: i) the web ontology language DAML+OIL[14], which is the precursor of the OWL, ii) Racer[2], a description-logic reasoning engine, and iii) Vigil, a framework for communication and access of intelligent services in a pervasive environment. Vigil realizes the Smart Home scenario where mobile users can access devices (e.g. printers) over low bandwidth or short-range wireless networks. In sTuples it plays the role of the communication gateway by abstracting and translating communication protocols (e.g. Bluetooth). A description of the three main extension aspects follows.

1. **Tuple Representation and Template Matching**: The concept of Semantic Tuple is introduced and extended to represent data and service description. The tuple template matching is enriched by using on top of the object-based matching a new algorithm which also takes into account the semantic meta-information.

   Specifically, a *semantic tuple* is a JavaSpace object tuple which contains a field of type DAML+OIL Individual. This field contains either a set of statements about an instance of a service, or some data or a URL from which such a set of statements can be retrieved. Two are the possible types of a semantic tuple: *data tuple*, if it contains semantic information provided by a service/agent, and *service tuple*, if it contains advertisement of an available service.

   The exploitation of semantic tuples requires the creation of mechanisms which can manipulate them. The *Semantic Tuple Manager* is responsible for the manipulation of all interactions in space concerning semantic tuples (i.e. insertion, reading, and removal). When a semantic tuple is added (or removed) to the space, the DAML+OIL statements it contains are extracted and asserted in (or retracted from) the space's knowledge base, checking always for validity and consistency. The *Semantic Tuple Matcher* accomplishes the matching of templates to semantic tuples. Reasoning capabilities are provided by RACER, as stated before. A semantic tuple template is a semantic tuple, which DAML+OIL individual-typed field is based on a dedicated 'TupleTemplate' ontology. The matcher performs its matching through a series of steps. We underline that each tuple

[2] http://www.sts.tu-harburg.de/~r.f.moeller/racer/

that matches the template is given a relationship degree and the returned tuple is the one with the highest degree

2. **User Centric Reasoning**: The employment of agents from the field of Artificial Intelligence simplifies clients' interactions and incorporates user semantics in delivering data and services to the user. The prototype implementation defines three specialized agents: i) *Tuple Recommender Agent* supports customized recommendations in a plethora of data and service tuples. More particularly, it presents to the user services and data tuples that he is interested in and distracts his attention from unwanted data and services. The factors that are taken into account are temporal information, user preferences, and location. What the client should do is to register interests with this agent. ii) *Task-Execution Agent* acts as a proxy on behalf of the user by off-loading specific tasks which the user performs. Again, the client should register tasks with this agent. iii) *Publish-Subscribe Agent* dynamically delivers data to subscribed users. A service or an agent can publish data or events which are shared by multiple users. This can be achieved by writing a data tuple on the semantic space. The only thing user has to do is to send a subscription request to this agent.

3. **Semantic Infrastructure in Pervasive Computing**: The use of constructs, semantics, and Vigil[3] enable sTuples to serve as a semantic infrastructure in pervasive computing.

The future work of the sTuples included enhancements to the agents, especially in the recommender agent, migration from DAML+OIL to OWL, and automatic learning capabilities to the space. Nevertheless, from 2004 when the sTuples were introduced until now there is no evidence that the Nokia Research Center or anyone from the research community will continue with the specific implementation.

#### 2.1.2.2 Semantic Web Spaces

Semantic Web Spaces have been proposed by the Free University of Berlin. Based on the categorization proposed by Rossi & All [15], the initiators of the Semantic Web Spaces decided to extend the Linda Coordination Model in four aspects:

1. **New TupleSpace structure**: Since Semantic Web Spaces aim in reflecting the open and distributed nature of the web, the consideration of a single TupleSpace proved insufficient. The concept of scopes, introduced by Merrick & Wood[16], was reinterpreted to succeed a decentralized architecture of the Semantic Web Spaces. The use of scopes allows the administrator of the TupleSpace to define a structure of context in which a client operates on. Additionally, the new TupleSpace structure is formally represented with the use of an ontology which describes the typical components of the TupleSpace, e.g. sub-spaces and supported tuple types. To conclude, scopes and ontologies provide an explicit means to model access rights, trust policies, and active contexts of clients of the space, which lead to a flexible and efficient management of the TupleSpace.

2. **New types of tuples**: To represent Semantic Web Information through tuples, Semantic Web Spaces adopted the expressivity of common Semantic Web languages, Resource Description Framework (RDF Schema or RDF(S))[17] and OWL[18]. RDF triples are of the form <subject, predicate, object>, where each of the three fields contains a URI, which in turn identifies a resource. An extra ID field which globally identifies a tuple is added.

[3] http://ebiquity.umbc.edu/project/html/id/32/

This also allows addressing separately tuples sharing the same content. Special constrains of the RDF which are handled are: i) *blank nodes* (nodes in the RDF graph which are not identified by global identifiers), ii) *containers and collections* (special RDF objects which represent a set of resources), and iii) *reified statements* (a means to reference RDF statements) [19].

3. **New coordination primitives**: To transit from the classical data-centered to the new semantics-aware TupleSpace, the notion of *information and data view* of the TupleSpace was introduced. In the *data view*, all tuples are seen as plain data, without semantics, like in traditional Linda systems. In the *information view*, RDF tuples have specific semantics: the set of the RDF tuples is seen as an RDF graph. The aforementioned consideration affects the Linda operations. In the data view, the traditional our, rd, and in operations should be reformed in order to conform to consistency and satisfiability constrains. The new operations are called *outr, rdr,* and *inr* respectively. For the information view, new operations are needed to take into account the truth Semantic Web knowledge. Operation *claim* asserts tuples in the information view, operation *endorse* reads tuples from the information view, operation *excerpt* supports a solution for multiple reads, and finally operation *retract* removes a tuple from the set without denying its truthfulness.

4. **New matchings**: To take advantage of the new ontological knowledge, a revision of the standard Linda matching approach, as it is simply based on checking the equality of number of fields, equality of field constrains and binding of field variables. These extensions were also imposed to manage the newly defined tuple types. Two different types of matching are supported. Both of them takes into consideration the types associated with the subject, predicate, and object fields of each RDF tuple *Matching RDF abstract syntax* refers to the data view and considers only tuples identified in the space as RDF tuples. *Matching RDF Semantics* refers to the information view and considers the tuples as resources which have a special predefined meaning. The use of reasoning services can assure the improvement of the precision of the returned results.

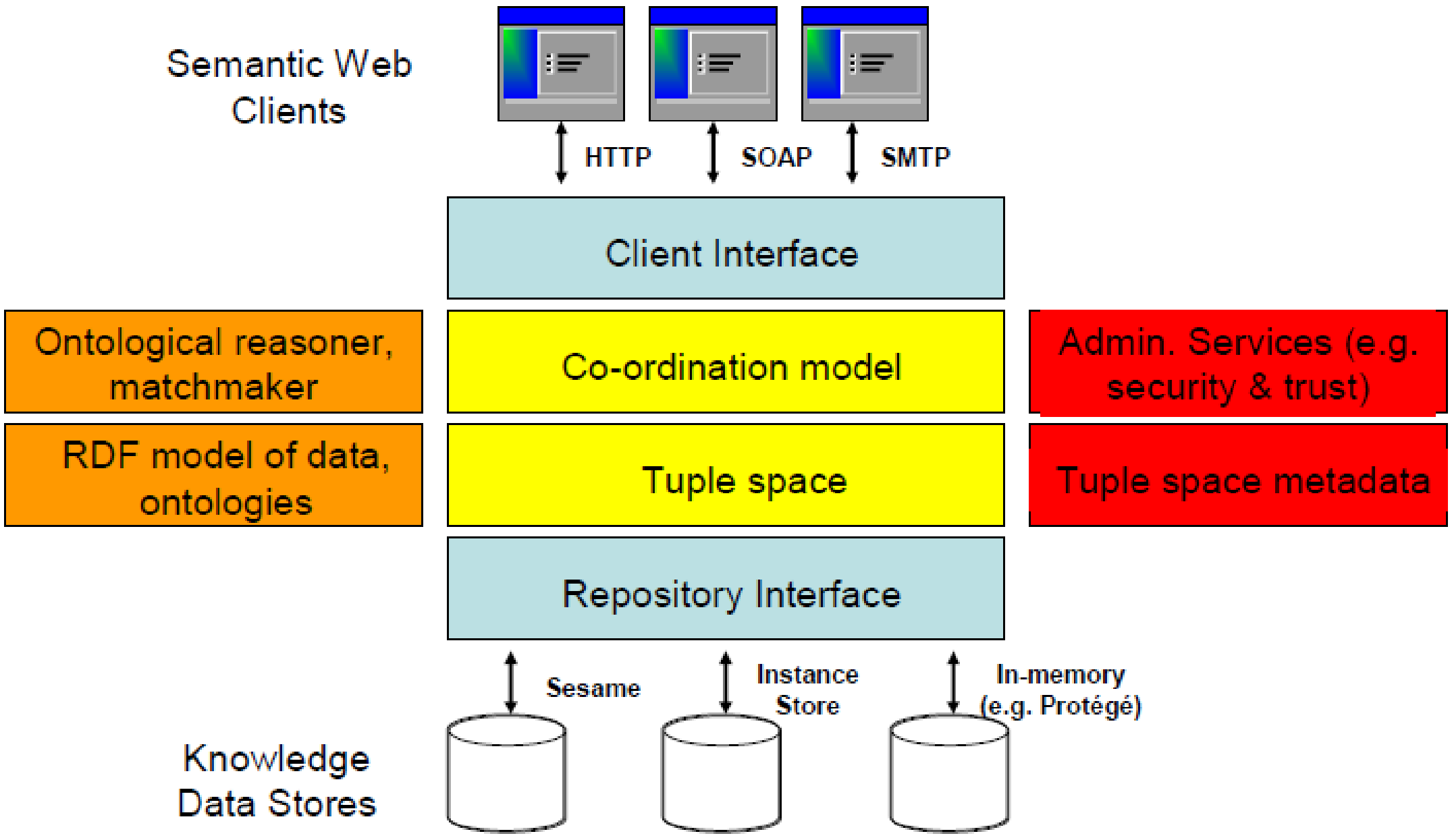

Figure 1 : High level architecture of Semantic Web Spaces [19]

Conforming to the aforementioned extensions, Figure 1 depicts the Semantic Web Spaces architecture. Based on Linda coordination model, it includes as main components the Coordination model and the TupleSpace. Features such as security and trust require a set of metadata which is build additionally on top of TupleSpace. TupleSpace is also extended with the construction of an RDF data model over the tuples, to semantically describe them, and with the construction of corresponding ontologies. Furthermore, there is a mechanism to associate RDF statements with their referenced ontologies. Finally, Semantic Web Services, like any middleware platform, must be independent of the underlying computing system. To succeed this goal, it separates the system's kernel from: a) the clients, using a Client Interface and b) the Knowledge Data Stores, which accumulates the information represented in the logical memory of the TupleSpace, using a Repository Interface.

Semantic Web Spaces are built around five building blocks. A short description of each one follows:

**Semantic data and organizational model** has two responsibilities. First, it represents RDF information in dedicated tuples known as RDFTuple. Second, it considers that an agent has two views upon a TupleSpace consisting of RDFTuples: a data view and an information view. Its organization model is inspired from scopes. According to Merrick & Wood [16], a scope is a viewpoint through which certain tuples can be seen. Any given tuple may be visible from several different scopes, while the latter can be created from names or by operations on existing scopes. The organization model introduces an application of scopes, called contexts. Contexts improve the scalability of open distributed Linda systems and enrich interaction patterns without expanding the number of co-ordination primitives.

**Coordination model** of Semantic Web Spaces extends Linda's TupleSpace-based co-ordination model with dedicated primitives for RDFTuple interaction at the syntactic and semantic level. It is also enriched with an Ontological Reasoner, which interprets ontologies according to their formal semantics or drawing inferences and checks for their satisfiability. The functionality of Coordination Model is also enhanced through Administration Services which fulfill issues related to security and trust.

**Collaborative and consensus-making model** is necessary since Semantic Web Spaces are modeled on the principles of Semantic Web, which implies the existence of heterogeneity of content and semantics shared in the space. As a result, this model offers mediation on two aspects to assure communication: a) Content and b) semantic mediation. These mediation operations take place through the provision of content and the semantic mapping of information by a semantic matching algorithm which seeks and applies this information when matching templates to RDFTuples[20].

**Security and trust model** extends the Linda coordination model; it deals with issues related to the security of the coordination model. There is a top level access policy controlled by the persons with the permission to create and control access to context. Access policies include policy rules for agents and policy rules for spaces. A list with access rights on contexts and spaces exists for each agent. Similarly, a list of agents with access rights on each space exists. Notice also that space access policies override agent access policies.

**Architecture model** is depicted in Figure 2. From left to write it consists of three components. The first component publishes Semantic Web information. The second component retrieves Semantic Web information using tuple matching heuristics, while the third component provides secure execution of the aforementioned activities. From top to bottom the architecture comprises three layers. The first two layers correspond to the information and data view while the third layer deals with the persistent storage of the TupleSpace information. From bottom to top, the TupleSpace system manages raw data, syntactic virtual data (Linda tuples), and semantic virtual data (RDF tuples) [20]. Finally the architecture provides an I/O interface through which the platform communicates with other systems.

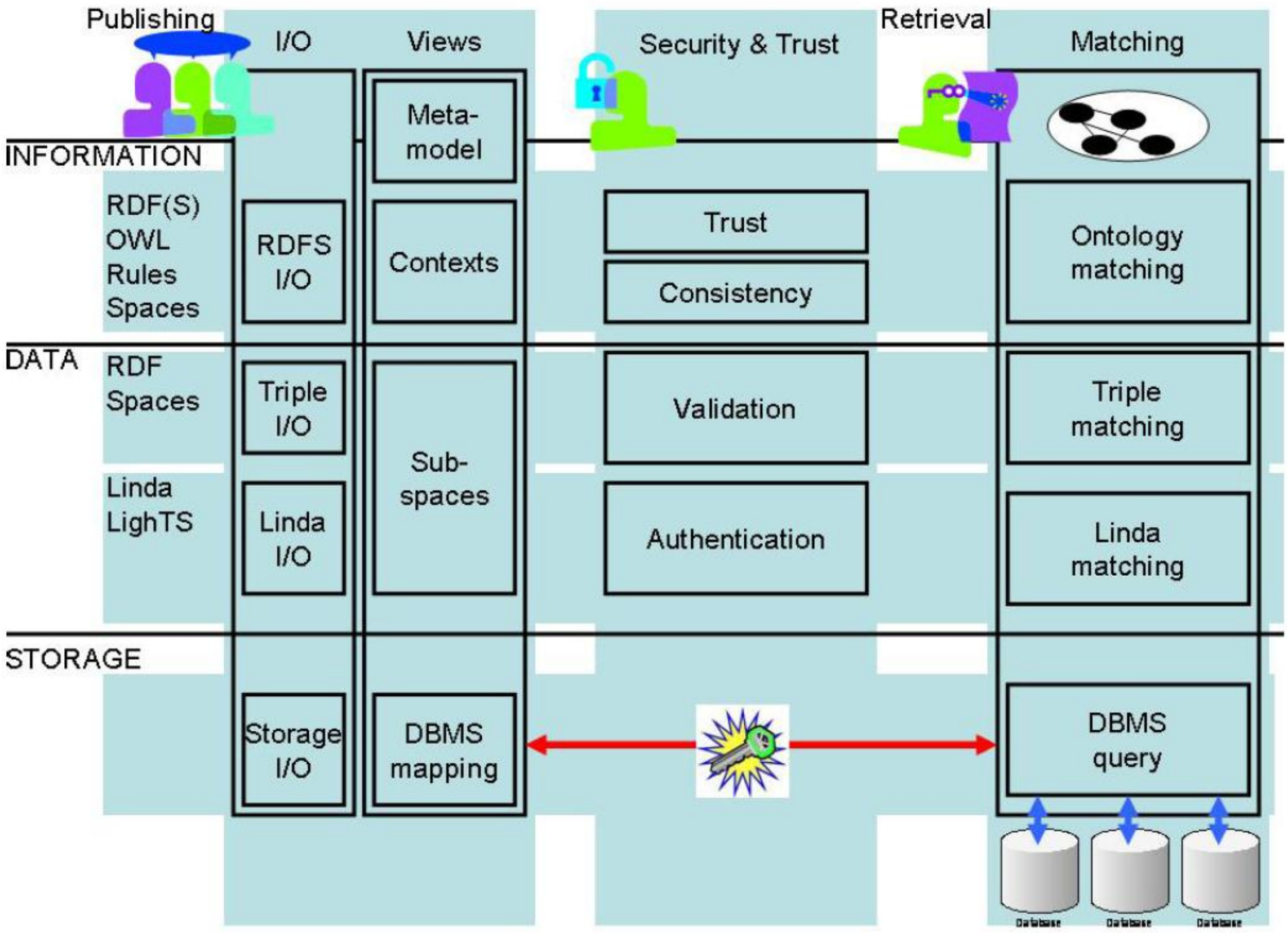

Figure 2: Implementation architecture of Semantic Web Spaces [21]

### 2.1.2.3 Triple Space Computing (TSC)

The vision of the Triple Space Computing triggered from the following observation: although the term "Web Services" refers to something compliant with the rules of Web, in practice this is not true, as Web Services do not follow the Web paradigm of 'persistently publish and read' [22]. Persistently publish and read apply to web pages: a web developer can publish a web site at any time, using any Internet Service Provider. Publication means that a web page, belonging to the published web site, is accessible to anyone in the world, suffice she/he knows the respective URL. Persistent publication means that power or network failure do not end to content loss. Reading means that using the appropriate URL, anyone can access the content of the web site, while the only requirement is that publication time must precede time of reading. This, simple to understand, paradigm has a lot of advantages and should be adopted by any type of data communication used by software systems, including Web Services technology. Until now, Web Services act as black boxes; they publish only conversations, based on the hidden content of messages, and not information. This technique except for violating the Web paradigm, it also requires a strong coupling between the sender and the receiver. Communication via message exchange requires a stable connection during the conversation, a common data representation, and the establishment of a contract between the two partners.

Based on the aforementioned, D. Fensel introduced the concept of Triple Space Computing (TSC) and argued that "Triple Space may become the web for machines as the web based on HTML became the Web for humans" [22]. TSC establishes the mechanism to publish communication data as the Web paradigm of 'persistently publish and read'. It follows the same goals for the Semantic Web services as the Web for humans: re-define and expand current communication paradigm. The Semantic Web Technology uses RDF for the description of triples, which create a natural link from the space-based computing paradigm into the Semantic Web. After the introduction of TSC, several approaches have been suggested ([23], [24], [25], [26], [27]). Following, we present the three most significant approaches, which extend the original proposal.

- ➔ First extension in D. Fensel's work was proposed by Bussler [23]. He defines a Minimal Triple Space Computer Architecture, based on the following six basic elements: (i) a **Data Model** which defines that objects written and read by Web Services are RDF triples uniquely identified by URIs. (ii) **Triple Space Clients** which write and read triples in parallel or sequentially, according to their security rights. (iii) **Triple Space Server** which hosts many triple spaces. The architecture consists of four components: (a) a *storage Component* responsible for storing the triples using databases, RDF databases, file systems etc., (b) an *HTTP communication component* for the receipt of HTTP calls that implement the Triple Space Transfer Protocol TSTP (TSTP is analyzed in the follow), (c) a *Triple Space Transfer Protocol (TSTP) operation component* which implements the functionality of writing and reading triples: it receives the write and read commands from the HTTP communication components, invokes the appropriate TSC server operation, and returns the result, (d) a *TSC server* for the implementation of write and read operations, as well as the appropriate error handling. Finally, we note that Triple Space Clients do not know about Triple Space Servers but only about virtual Triple Spaces. As a result, operations responsible for creating, deleting, and emptying triple spaces have to be defined. (iv) **Triple Space**, a virtual concept implemented by triple space servers. A Triple Space has to be part of one implementation; however, one implementation can host many triple spaces. In any case, each Triple Space should be distinguished so as write and read operations will be executed in the right space. To succeed that, we assign to each Triple Space a unique URI as an identifier. (v) **Triple Space Transfer Protocol (TSTP)** is used to initiate a connection between clients and Triple Space servers for the execution of write and read operations. Usually Triple Space Transfer Protocol is mapped to HTTP protocol for simplicity reasons. In a different implementation all web servers should implement this new protocol. (vi) **Minimal Triple Space Application Programming Interface (API)** provides the means through which clients can write and read single or multiple triples in a concrete Triple Space and servers can execute basic administrative operations, i.e. create a new Triple Space.

  As these mentioned elements cannot assure an actual application, Bussler extends his approach by introducing additional elements: Rich semantics for read and write operations, Destruction operations, Constraints definition, Transaction support, Ontology definition, Access Security, Transmission Security, Non-Repudiation, History and Archive, Location Directory, and Versioning.

- ➔ Another approach was proposed by Martin-Recuerda and Sapkota [24], which extends the proposal of Bussler [23] by introducing a richer coordination mechanism, based on the combination of tuple space computing and the publish-subscribe paradigm that also decouples process flows, thus allowing the flow decoupling. Flow decoupling, decouples main flows of space users from the generation or reception of notifications. So, users are not blocked while producing or receiving data and consumers can receive notifications while performing some concurrent activity. The approach also provides a transactional model implementing the two-phase protocol at the coordination model. Specifically, transactions should involve distributed resources, which are the spaces. The use of managers is also essential. A Resource Manager allows transactional access to some resources. A Transaction Manager coordinates one or multiple Resource Managers. Applications are those which communicate with the Transaction and Resource Managers.

TSC can be applied in Web Service Execution Environment (WSMX) [28], an execution environment for the dynamic discovery, selection, mediation, and invocation of Semantic Web Services described using Web Service Modeling Ontology (WSMO) [29]. A detailed description of the WSMX is given in Paragraph 2.1.3 due to its high importance.

➔ The third approach is coming from the TSC project [27]. For a decentralized and distributed infrastructure the approach implements a hybrid infrastructure, called super-peer, which combines the advantages of pure P2P and client/server systems. According to super-peer systems, TSC distinguishes two levels: the upper level which consists of forceful servers and the lower level, which comprises clients with limited computational resources. The above distinctions lead to three kinds of nodes. *Servers* are persistent peers that can partially host many triple spaces. Consequently, management operations have to be available on a triple space server to create, delete and empty triple spaces: store primary and secondary replicas of the data published; support versioning services; provide an access point for light clients to the peer network; maintain and execute searching services for evaluating complex queries; implement subscription mechanisms related with the contents stored; provide security and trust services; balance workload and monitor requests from other nodes and subscriptions and advertisements from publishers and consumers [30]. *Heavy Clients* are equipped with storage and searching capabilities. By this way, users are capable of working off-line with their in their own copy of the Triple Space. *Light-Clients* are lightweight devices–peers, which can write triples, read triples, at the same time or sequentially, as well as visualize data stored on Triple Spaces.

For the specification of Triple Space architecture the authors of the TSC [31] define the Triple Space Kernel, a software component which can be used to implement both Triple Space servers and heavy clients. More particularly, Triple Space implementation is based on the extension of the Coordinated Shared Objects (CORSO) middleware [27], which provides transactions and replication of Java object structures. To share triple spaces and named graphs among participating nodes we map them onto CORSO objects using a distinct Object Identifier. A large number of participating nodes runs the kernel implementation and forms the virtual shared memory space of CORSO. A CORSO kernel provides access to the shared objects and executes locally supplemented activities, such as read/write/take. Figure 3 illustrates the main components comprised in the architecture. A brief explanation of each one follows:

*Operation layer* is on top of this architecture, providing an access point to users. The *Security Framework*, located in Operation layer, incorporates simple security measures that are stored in a triple form and provided by the space infrastructure. The *security management API* defines and changes security configurations (e.g. access control for spaces or named graphs). The *mediation engine*, also located in Operation layer, resolves heterogeneity issues by providing mappings in order to resolve possible mismatches among different RDF triples. The *mediation management API* provides methods to turn on/off the usage of mediation engine, as well as to add, remove, and replace mediation rules. The *coordination layer* implements transaction management, assuring that concurrent operations are processed consistently. *Data access layer* acts as a gateway to the underlying data storage and provides resolution of semantic templates, limited reasoning support, and abstraction from the actual storage framework [30].

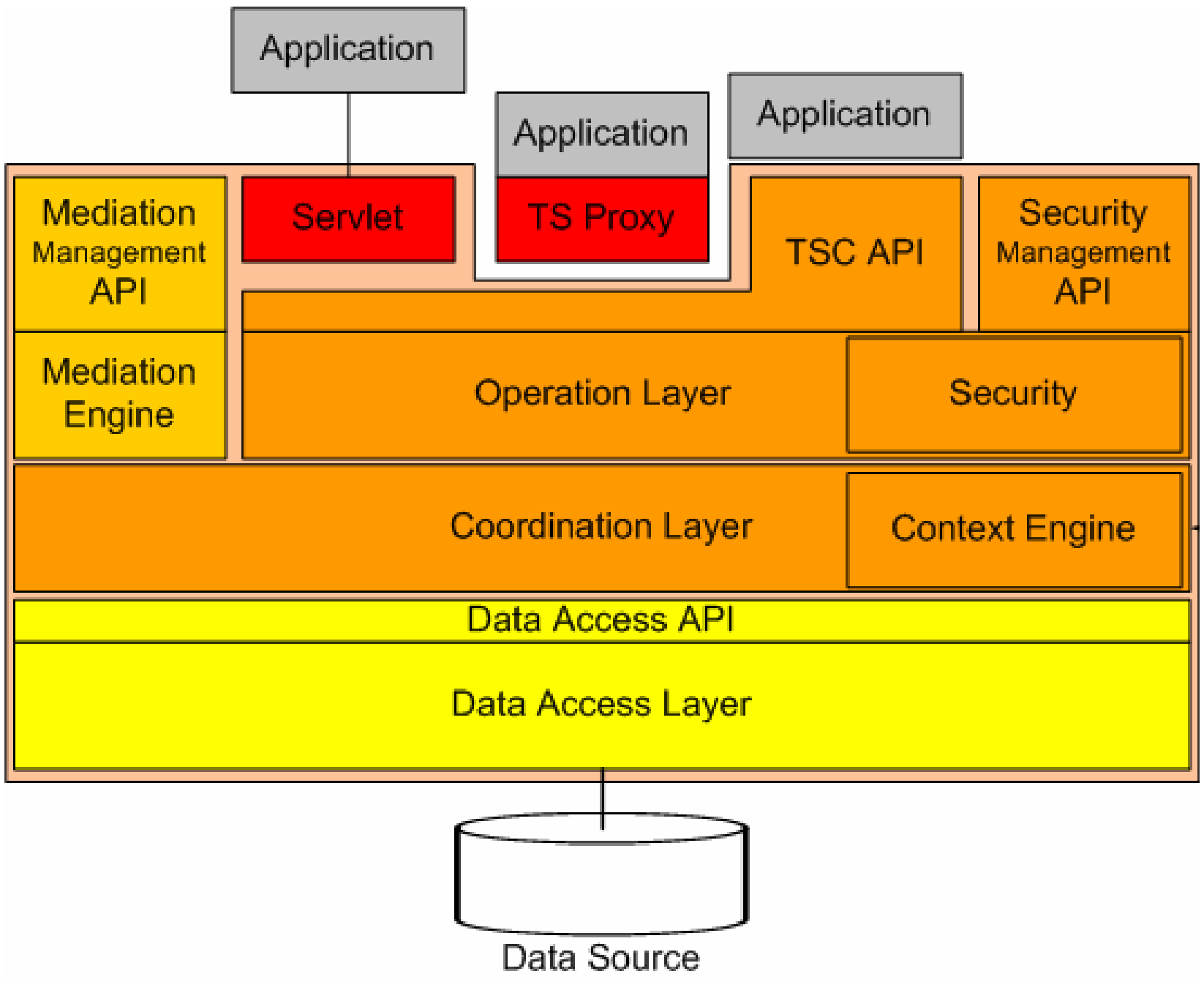

Figure 3: High level Architecture of TS Kernel [31]

### 2.1.2.4 Conceptual Spaces (CSpaces)

CSpaces aims to re-elaborate the Semantic Web proposal by minimizing syntactic data representation. In fact, the original goals of CSpaces are two: a) the extension of Triple Space Computing and b) study their applicability in different scenarios apart from Web Services. The basic objective was the constitution of a conceptual and architectural model that can appropriately characterize most of the requirements and functionality that the Semantic Web demands. CSpaces try to resolve well known problems in the Semantic Web area: dichotomy problem, scalability problem, publishing problem, heterogeneity problem, inconsistency problem, security problem, and trust problem.

CSpaces characterizes the Semantic Web around seven building blocks: Semantic Data and Schema Model, Organizational model, Coordination model, Semantic interoperability and consensus-making model, Security and trust model, Knowledge access model, and Architecture model (or Blue Storm). Figure 4 depicts how these building blocks are combined.

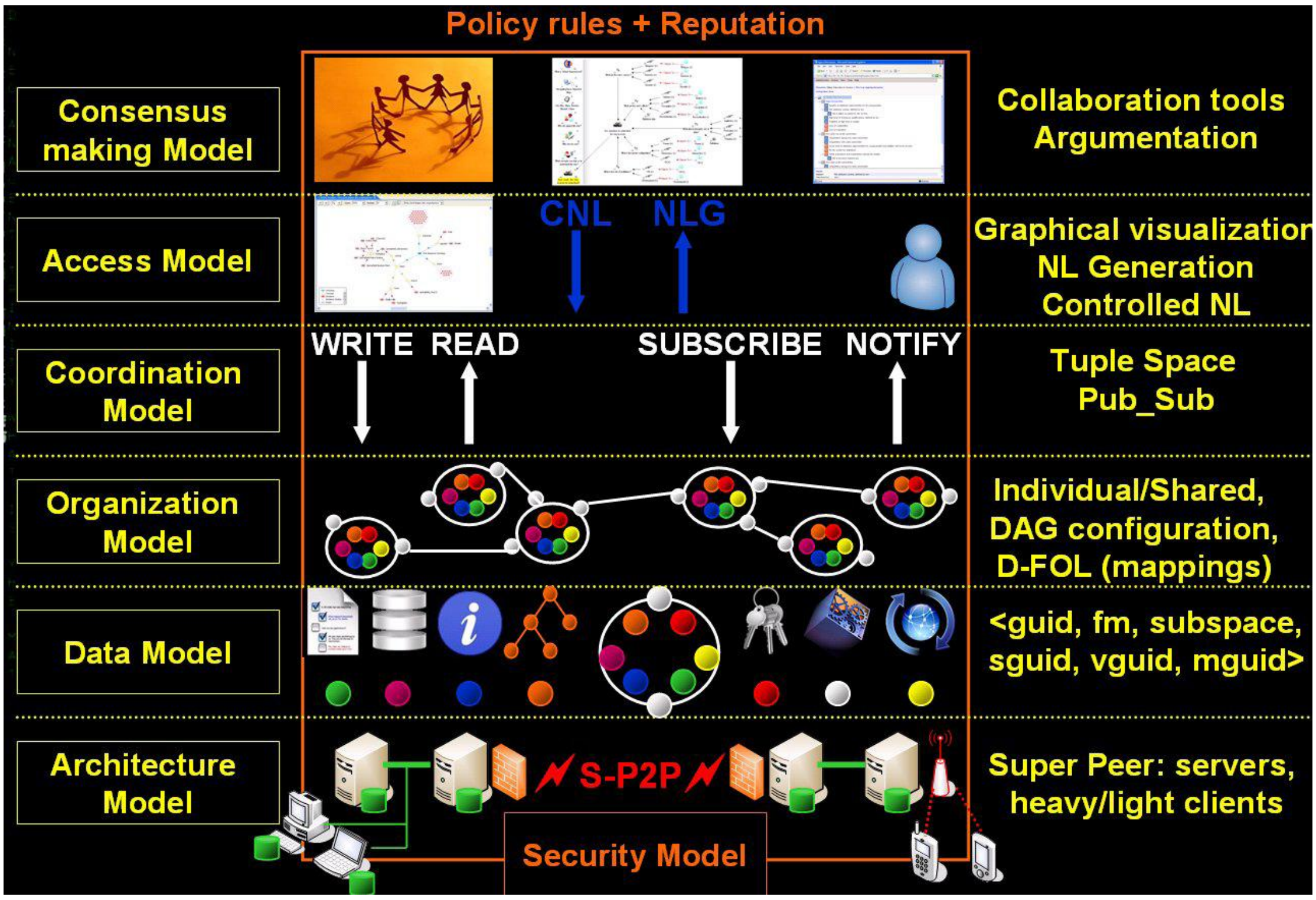

Figure 4: CSpaces building blocks [30]

Following we give a brief description of the models of CSpaces.

**Semantic Data and Schema Model** defines the notion of a CSpace. A CSpace is a knowledge container which stores the description of data elements and the relations between them. Each tuple in a CSpace is of the following form [32]:

<guid, fm, type, subspace, sguid, vguid, mguid>

Each field is defined below:

***guid***: a global unique id for the logical formula

***fm***: a fist order logical formula

***type***: an identifier of the formal language used in fm field

**subspace**: a description of the subset of the CSpace where each tuple belongs to. There are seven different subspaces defined for each CSpace: (i) *Domain Theory subspace* stores a logical theory which gives an explicit, partial account of a conceptualization [30]. The combination of a concept, a relation, a function, an axiom, and a rule build a Domain Theory. (ii) *Instances subspace* stores the concepts individuals as defined in the Domain Theory and the values of their attributes *sguid*: a global unique identifier of the CSpace. (iii) *Metadata subspace* constitutes an ontological description of the CSpace itself. (iv) *Annotations subspace* defines links between a concept (or an instance) and an information resource. These links can be described via the use of RDF or Topic Maps representations. (v) *Security and Trust Information subspace* is required as CSpaces can have associated access rights. It is described in terms of policy rules, reputation information, and social-network-based trust management [22]. (vi) *Mapping and transformation rules* subspace ensures interoperability between related individual and shared CSpaces by identifying common ontological terms, relations, and instances. (vii) *Subscriptions and Advertisements subspace* is responsible for storing queries that identify the information that is requested by consumers and will be published by producers.

***vguid***: a version global unique identifier responsible for distinguishing each version of the logical formula

***mguid***: the identifier of the member of the CSpace that stores the tuple. We note here that CSpaces permit the reputation of the stored data

**Organizational model** of CSpaces avoids the adoption of a centralized system. It does not presume that all users have to agree on a common set of rules, schemas, and data. Instead of this, CSpaces offer two types: the *Individual CSpaces* and the *Shared CSpaces*. *Individual CSpaces* offer a knowledge container, where the user can store data based on hers/his own conceptualization. The semantics stored there can be private, namely, only the owner of the space has access on them, restricted, namely, a limited number of individuals have access on them, or public, namely, all can access them. *Shared CSpaces* are conceptual spaces where several users have agreed on a common formal representation and a common conceptualization (e.g. common domain theories, instances, annotations) to share common point of views, interests and to achieve interoperability among individual CSpaces. The combination of restricted and public data previously reported is used to create Shared CSpaces. More particularly, Individual CSpaces when combined create a new space shared by all the users, the Shared CSpace. Accordingly, different Shared CSpace can be combined to create new Shared CSpaces which are bigger knowledge containers. We note here that related CSpaces can be connected with mapping and transformation rules that allow the execution of reasoning processes in a distributed fashion. From an implementation point of view, the whole organization can be considered as a tree where Individual CSpaces are the leaves and Shared CSpaces are the intermediate nodes. For the realization of the aforementioned the creators of CSpaces [33] adopted the proposal for organizing Knowledge Bases in a Directed Acyclic Graph (DAG) configuration [34]. This choice is appropriate for distributed and related knowledge containers and especially in distributed queries, because cyclic references are avoided.

**Coordination model** is defined on top of mediated, semantic and persistent communication channels (Shared CSpaces) that simultaneously represent places for knowledge storage. The concept “persistent publish and read” has been applied as a simple coordination model for TupleSpace computing, as well as to Semantic Web Services. A notification and subscription mechanism enhances the TupleSpace by allowing an asynchronous interaction from the consumers/reader side. Coordination model in CSpaces integrates TupleSpace and publish-subscribe operations, namely, it combines the concept “persistent publish and read” with the concept “publish and subscribe”. The API of the Coordination model in CSpaces is similar to Triple Space, with the following differences: a) CSpaces deal with triples instead of tuples and b) the API of CSpace has to consider that agents write information based on the logical theories stored in their own individual CSpaces and not on the destination CSpaces.

**Semantic interoperability and consensus-making model** has as an objective to reach an agreement on the specification of a knowledge base and a set of mapping and transformation rules. Ontologies where first created to offer people a common language in order to share knowledge. Nevertheless, until now, ontologies where seen as formal specifications of conceptualizations and they were rarely built to be shared and reused. This model tries to convert the ontologies to shared specifications of conceptualizations. To reach this goal some principles of Human Centered Computer approaches must be followed, while users and applications interact with each other to create new Shared CSpaces.

**Security and trust model** is important in CSpaces due to its openness and decentralization: a node of the CSpace does not mean that it is also trustworthy. So, this model supports the hiding of private and the projection of public information in a distributed information infrastructure, so as to guarantee a valid and trusted system. The challenges which should be addressed according to Gutiérrez & al. [35] comprise authentication, confidentiality, integrity, non-repudiation, availability, and end-to-end security. Security and trust model has six main characteristics. First, all users that read and publish information in CSpaces need a digital identity which facilitates the verification of access rights and the association of reputation values. According to the second characteristic, i.e. comparable trust, published trust information should be syntactically and semantically comparable. Third, explicit trust requires trust relationships to be visible by the agents that are registered in a CSpace and stored in the CSpace. Next, the existence of opinions indicates a subjective agent's measurement about another's trustworthiness, which is defined by a set of trust levels. Furthermore, for each agent that is member of a CSpace, a local reputation score is calculated based on the opinions of the rest members of the CSpace. Lastly, for each agent, a global reputation score is calculated from local scores and stored in the Shared CSpaces that also maintain a general information catalog of all CSpaces.

**Knowledge access model** is responsible for the creation, transfer, and comprehension of the information stored in knowledge bases. Mechanisms facilitating the visualization, exploration, and editing of this information are included in the model. For the visualization, techniques catering for the graphical representation of ontologies are considered. For the exploration, Natural Language Generation (NLG) [36] is used for knowledge access in which semantic data descriptions are presented in a user friendly way. Finally, for the editing, Controlled Natural Languages (CNL)[4] help users who are not experts in logician to edit information. CNL are subsets of natural languages whose grammars and dictionaries have been restricted to reduce complexity.

**Architecture model (Blue Storm)** follows the main principles of the Semantic Web architecture, assuring scalability, distribution, and decentralization. It also requires mechanisms for asynchronous communication and for organization of metadata around Individual and Shared CSpaces. The architecture model is hybrid one, based on P2P and client-server infrastructure in which agents store, read, and share information. A client-server P2P configuration separates the system in two levels. The upper level is composed by well-connected and powerful servers, and the lower level is constituted of clients with limited computational resources which are temporarily available. The first approach of the architecture defines three kinds of nodes. First, CSpace-servers mainly store primary and secondary replicas of the data published in individual and shared CSpaces, provide an access point for CSpace clients to the peer network, and maintain and execute reasoning services for evaluating complex queries. Second, CSpace-heavy-clients provide a storage infrastructure and reasoning support to let users to work off-line with their own individual and shared spaces. Finally, CSpace-light-clients include the presentation infrastructure to write query-edit operations and visualize knowledge contents stored on CSpace-servers.

### 2.1.3 WSMX

Web Services Execution Environment (WSMX) [28] is an execution environment for dynamic discovery, selection, mediation, invocation, and interoperation of Semantic Web Services (SWSs). The research goal in WSMX was to provide an architecture for Semantic Web Service based systems. WSMX is based on Service Oriented Architecture (SOA) and on Web Services

[4] http://sites.google.com/site/controllednaturallanguage/

Modeling Ontology (WSMO) [29] and its underlying formalized representational languages family WSML - in fact it is a reference implementation for WSMO. The development process of WSMX includes the establishment of a conceptual model, the definition of its execution semantics, the development of system architecture, the design of the software, and the building of a working system implementation.

#### 2.1.3.1 Main Components

Following, we shortly describe the main components designed and implemented in WSMX.

**Core Component** is considered as the kernel of WSMX, as it manages all the other components. It includes the business logic of the system, the events engine, the internal workflow engine, and the distributed components loading. Until now, Core Element is offered as a central module of WSMX.

**Resource Manager** is responsible for managing the repositories of WSMX. WSMX provides five repositories (four for storing definitions of any WSMO entity and one for storing non-WSMO related objects): Web Services Repository handles semantic description of Web services, Goals Repository handles semantic description of general goals, Ontology Repository handles ontologies which are stored in the registry and describe the semantics of particular domains, Mediator Repository handles mediators that are stored in the registry, and Data Repository deals with a variety of system-specific data produced by the system during run-time. Assume that a component needs to write, read, or modify data from any of the five repositories. Then, it has to invoke the Resource Manager, because no component can access any repository directly.

**Service Discovery** is the element which finds services that meet the needs of users' requests. Service Discovery can be one of the following three types: Goal Discovery, Web Service Discovery, and Service Discovery. There are three available approaches for any type of Service Discovery. First, Keyword-based Discovery matches keywords from service descriptions with a goal description. Second, Lightweight Semantic Discovery uses specific vocabularies that have an explicit semantic for the matching process. Third, Heavyweight Semantic Discovery uses relations between inputs and outputs of services, as well as inputs and outputs of goals. Thus far, WSMX focuses on Web Service Discovery in combination with Keyword-based Discovery.

**Service Selection** provides techniques to choose the best or the optimal web service from a list of available services, which potentially satisfies the requested goal. These techniques vary from simple selection criteria to complex selection of variants involving in interactions with a service requester.

**Data and Process Mediators** offer data and process mediation accordingly. When two entities cannot communicate because they use different syntax or semantics, then data mediation offers a solution. Data Mediators transform data from different sources, based on their semantic similarities as expressed by their reconciled conceptualizations [28]. Respectively, when two entities cannot communicate because they have different communication patterns, then mediation offers a solution. Process Mediator adjusts the different patterns to make them match. This can be achieved through several methods including grouping several messages into a single one or changing their order.

**Communication Manager** comprises invoker and receiver. These two subcomponents provide communication from service requester to service provider and communication from service provider to service requester. SOAP or proprietary protocols with an adapter framework can be used for the communication.

**Choreography Engine** caters for a match between a requestor's communication pattern and a provider's communication pattern. Choreography of a Web service defines the way a requester can interact with it, while in most cases a requestor of a service has its own communication pattern. As a result, if requester's communication pattern is different from provider's communication pattern, communication is disabled. Choreography Engine in combination with the Process Mediator resolve such kind of mismatches.

**Web Services Modeling Toolkit** is a framework for rapid creation and deployment of homogeneous tools for Semantic Web service. After the achievement of the above goals the following step is the centralization of these tools in a common application. Consequently, users will have to install only one application which will provide all the available tools. An initial set of tools includes a WSML Editor for editing WSML and publishing it to WSMO repositories, a WSMX Monitor for monitoring the state of the WSMX environment, a WSMX Mediation tool for creating mappings between ontologies and a WSMX Management tool for managing the WSMX environment.[28].

**Reasoner** have to be WSML – Compliant, providing the following reasoning services: mapping process of mediation, validation of a possible composition of services, indication that a service in a process is executable in a given context, finding capabilities that exactly match the requester's goal and finding capabilities that exactly subsume the requester's goal.

### 2.1.3.2 Architecture

As we have already mentioned, WSMX is based on Service Oriented Architecture (SOA). Consequently, it is a software system consisting of a set of collaborating software components with well-defined interfaces that are used to perform a task [28]. These components can execute in different locations communicating over a network connection. This fact creates requirements such as management of latency, memory access, concurrency and failure of subsystems.

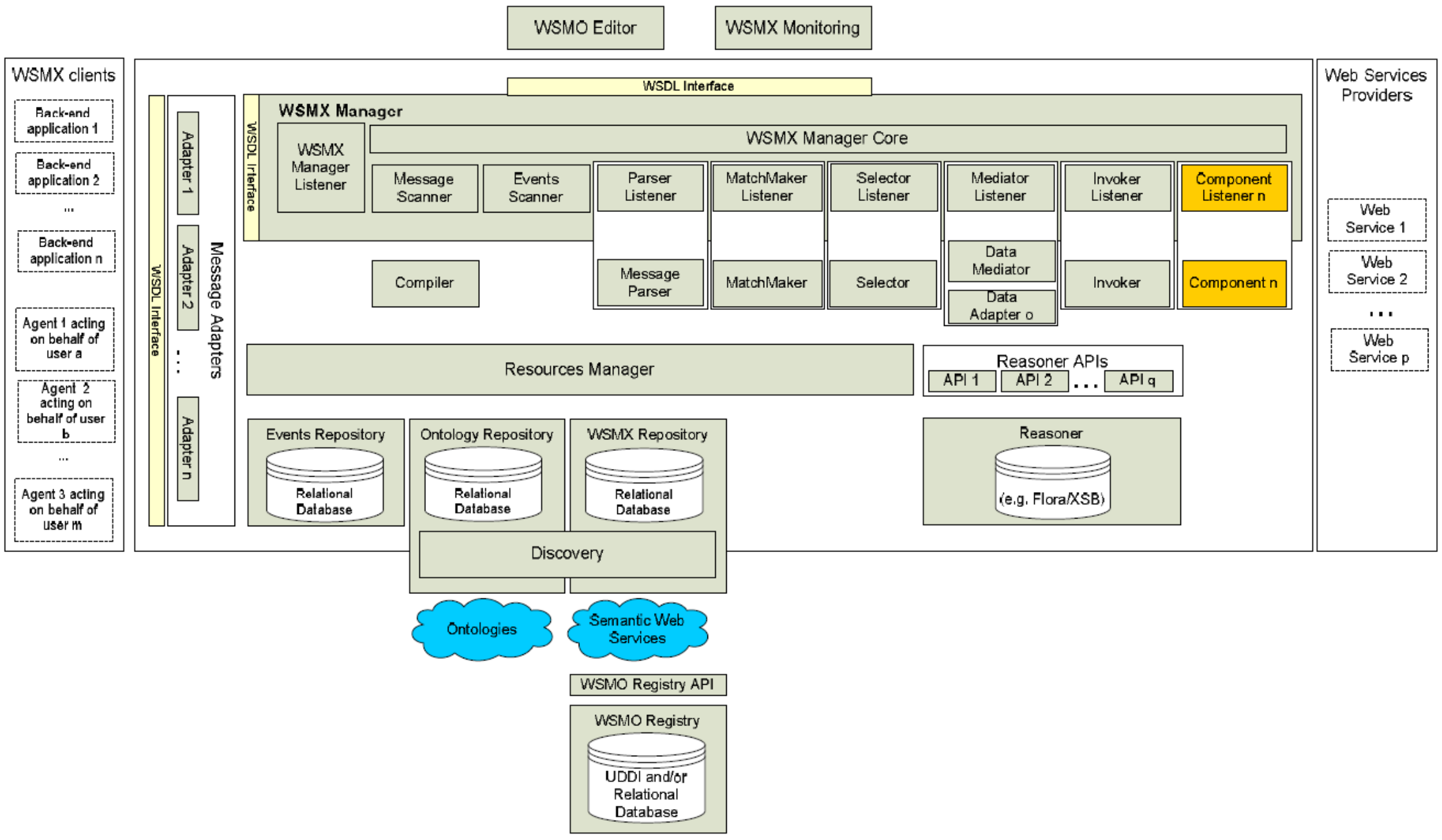

Figure 5: WSMX architecture [28]

Figure 5 illustrates the WSMX architecture which consists of the main concepts described above. Each component provides services – each of which is a logical unit of system code, application code, and persistency layers. The characteristic of these components is loose coupling, so they can be plugged-in or plugged-out from the system. We note that these WSMX components can be replaced by third party components which support the same functionality. To offer this functional characteristic, WSMX gives a detail description for the interface of each component.

## 2.2 JavaSpaces

The JavaSpaces technology [37] is a simple and powerful high-level tool for building distributed applications. It can also be used as a coordination tool. JavaSpaces should be thought of as a place in which it is possible to store and share information in the form of objects: processes can cooperate through the flow of objects into and out of one or more spaces. Based on that, JavaSpaces provide a simple API that is easy to learn and yet expressive for building sophisticated distributed applications. As we already mentioned in Paragraph 2.1.1 the aforementioned programming model has its roots in Linda.

### 2.2.1 Architecture

JavaSpaces is part of the Java Jini technology. JavaSpaces export proxies to clients dynamically, via the Jini lookup service (that is why they are naturally distributed). They consist of a combination of space proxies and space servers. A proxy constitutes the part of the space that is dynamically loaded to clients, while the space server is the part that stores, matches, and distributes objects. Figure 6 illustrates the JavaSpaces overview.

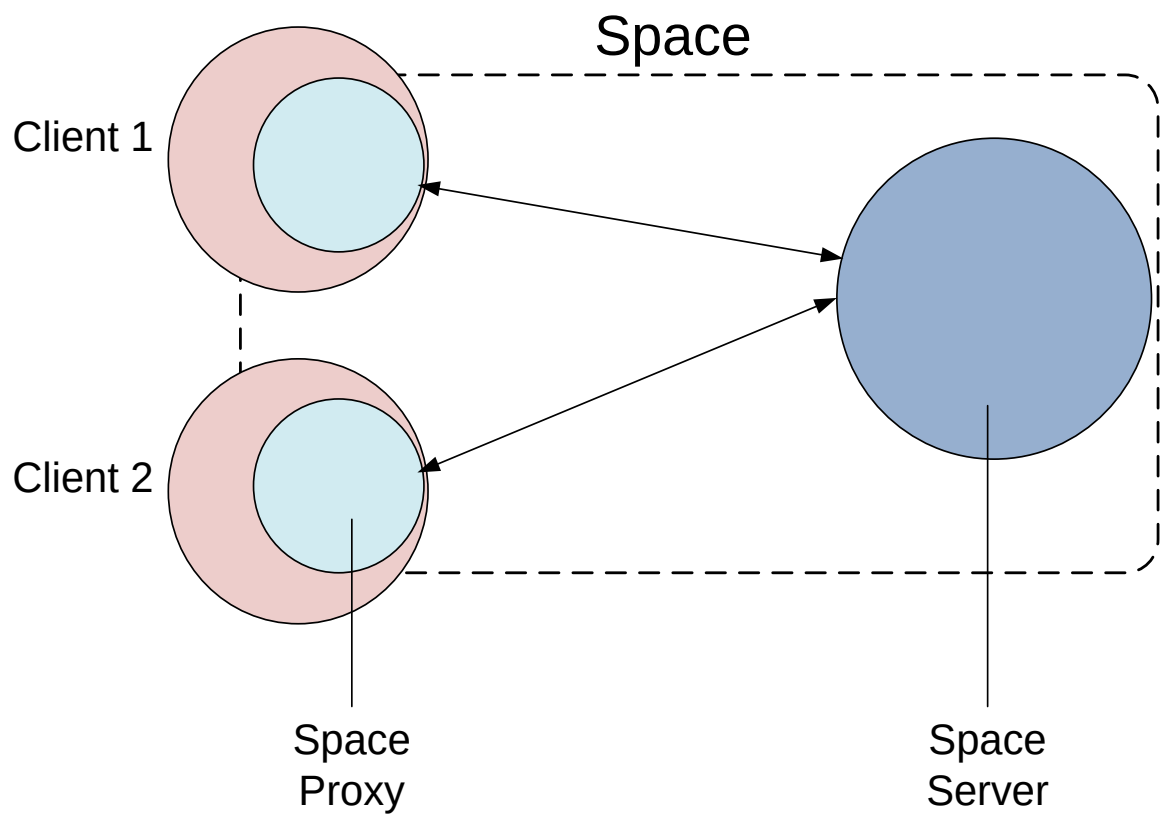

Figure 6 : JavaSpaces overview

Figure 7 depicts the internal structure of the space server, the space client, and their communication.

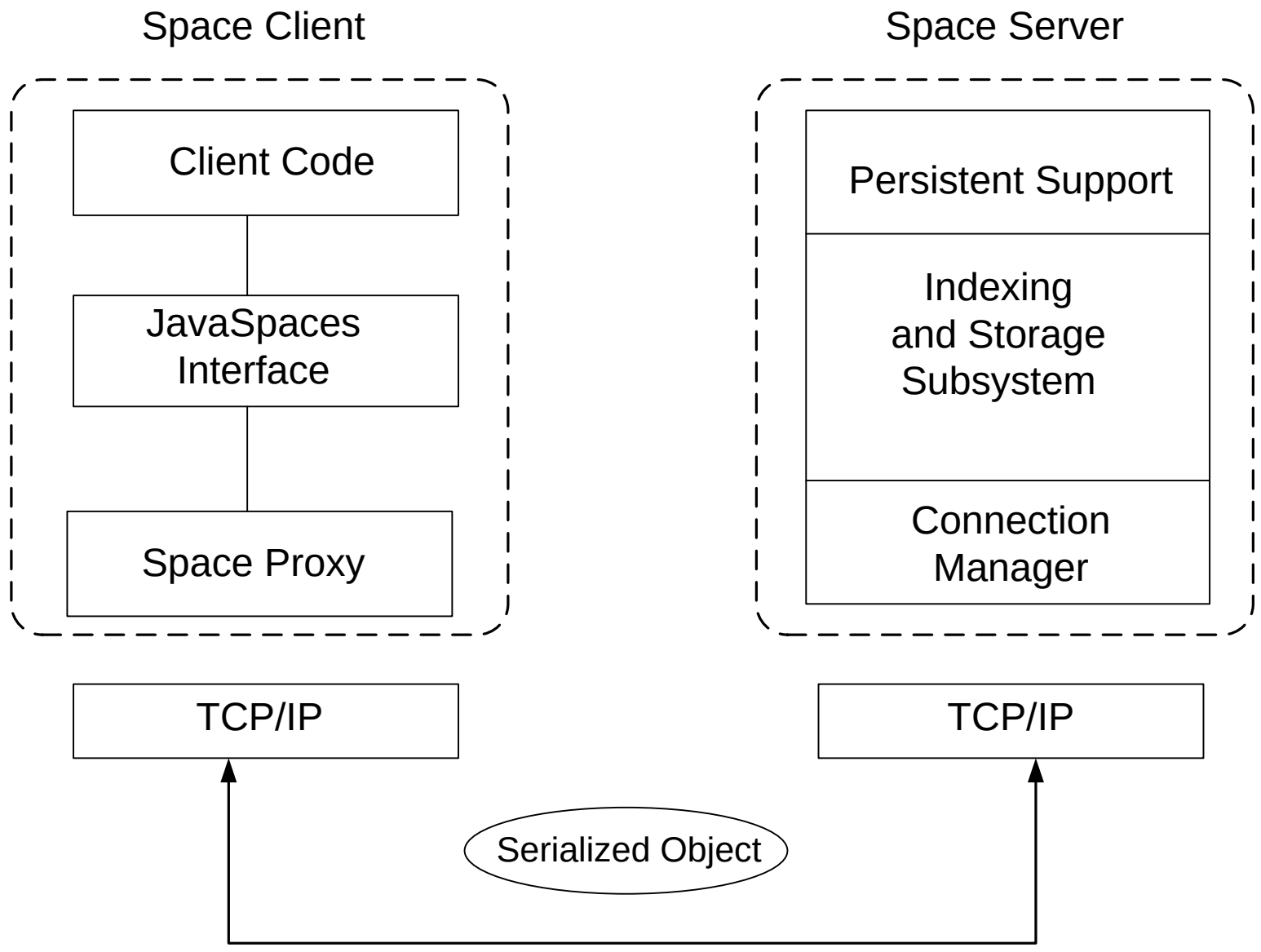

Figure 7: JavaSpaces internal architecture

From left to right we give a description of each component:

**Client Code** “manipulates” the space by writing, reading, and taking objects into and from the space. To do so, it uses the JavaSpaces Interface.

**JavaSpaces interface** defines the relationship between JavaSpaces client services and the JavaSpaces implementation by offering to the Client Code the templates of all the supported operations. The supported Javaspaces Interface is described in detail in Paragraph 2.2.2.

**Space Proxy** constitutes the implementation of the client side of the space. We note here that there may exist various implementations of JavaSpaces implement both the proxy and the spaces server in different ways and. The only prerequisite is that any implementation complies with the JavaSpaces specification. The proxy is also responsible for preparing objects to be transported across the network. Almost always, JavaSpaces implementations use the built-in RMI and serialization mechanisms to prepare objects for the network. This is where objects are serialized before they are streamed across the network.

**TCP/IP** is the network protocol which is responsible for sending packets of information across the network, to and from the space server. This protocol can be replaced with anyone else offering the same functionality, but for simplicity reasons all the implementations use this one.

**Connection Manager** is responsible for managing the client connections to the space. The space server should in theory support any number of concurrent connections and handle requests fairly for all of the connected clients.

**Indexing and Storage subsystem** can be considered as “the heart” of the system, as it is responsible for storing and matching objects.

**Persistent Spaces** ensures that if the Java Virtual Machine (JVM)[38] that they are running in fails for any reason, then the objects contained within the server will not be lost, along with other information. This is succeeded with the use of non-volatile memory, such as local disk drives.

### 2.2.2 JavaSpaces Interface – Supported Operations

JavaSpaces give the opportunity to clients to "manipulate" the objects stored in the space. In a Java program, all objects are derived from the base class Object. In JavaSpaces, all objects must also implement the Entry interface. So every object in a space is of type Entry. Entries may represent a very simple unique identity, or may contain a very detailed complex computerized model. Clients can *write* entries into the space, *read* entries stored in the space, and *take* entries from the space, namely read entries and then remove them from the space. Clients can also register by calling the *notify* method to receive notifications. These operations are of great importance and they are analyzed subsequently:

**write**

The form of the write method as it is defined in the JavaSpaces interface is:

```
Lease write (Entry entry, Transaction txn, long lease)
   throws TransactionException, RemoteException;
```

As one can easily assume, this method is a store operation and it is used to write an object into the space. More particularly, parameter entry defines the object that will be inserted in the space. Parameter txn indicates that the entry will be written under an optional transaction. If the txn parameter is null, then the written entry becomes visible to all space clients immediately, otherwise entries become visible outside that transaction only when it commits. The use of non-null price in the txn parameter is essential, otherwise, in case the client crashes between taking the entry and writing it back, the entry is lost forever. Finally, parameter lease specifies the requested time an entry lives in the space and it is measured in milliseconds. The method returns a parameter of type Lease which encapsulates the amount of time granted to the entry to live within the space. To become clearer, leases granted by the write method are determined as part of the negotiation between the client and the space, and are finally set by the space implementation. [39] According to the general rule, the implementation is free to return a lease that is shorter than the requested lease, but not longer. The exceptions that can be thrown are TransactionException and RemoteException. The former is thrown if an error occurs with the transaction subsystem, while the latter is thrown if there is a problem with communication between the proxy and the JavaSpace, or an exception propagates out of the space.

**read**

The form of the read method as it is defined in the JavaSpaces interface is:

```
Entry read(Entry tmpl, Transaction txn, long timeout)
   throws UnusableEntryException, InterruptedException,
          TransactionException, RemoteException;
```

As one can easily assume, this method is a search operation and it is used to get hold of a copy of an object from the space. More particularly, parameter tmpl defines the template that the space will use to perform the matching. Matching is done at the entry field level by comparing the serialized form of each of the non-null fields in all templates (a byte array) with entries existing in the space of the template's type or subtypes[39]. A null field in a template is considered as a wildcard, and as a result it will match any value. Parameter txn indicates the scope of the transaction where the read operation can search. So, if a non-null transaction is passed to the read method, then the space will match an entry that is visible to that transaction, including entries written to the space under a null transaction. Furthermore, if a read operation invoked with a non-null txn parameter has returned an entry, then no other transaction can take that entry until the former either commits or aborts. Finally, parameter timeout specifies how long the client is willing to wait until he/she receives a result entry. The exceptions that can be thrown are four. Exceptions TransactionException and RemoteException have the same semantics as before. UnusableEntryException is thrown if any serialized field of the entry being read cannot be deserialized, while an InterruptedException is thrown if the thread which executs the read method is interrupted.

**take**

The form of the take method as it is defined in the JavaSpaces interface is:

```
Entry take(Entry tmpl, Transaction txn, long timeout)
  throws UnusableEntryException, TransactionException,
         InterruptedException, RemoteException;
```

This method is a search-and-delete operation and it is used to retrieve and then delete a matching object from the space. The take method has the same set of parameters and possible exceptions as the read method. The only difference is on the semantics: if a matching entry is returned, then it has to be removed from the space.

**notify**

The form of the notify method as it is defined in the JavaSpaces interface is:

```
EventRegistration notify(Entry tmpl, Transaction txn,
      RemoteEventListener listener,
      long lease, MarshalledObject handback)
        throws TransactionException, RemoteException;
```

In a nutshell, this method notifies a specified object when entries that match the given template are written into the space, so it sets up repeated search operations as entries are written to the space. For any further details concerning this method, the reader may refer to the book “JavaSpaces in Practice” [39].

### 2.2.3 Application Model

In JavaSpaces technology distributed applications are modeled as a flow of objects between participants. Figure 8 depicts how a JavaSpaces technology-based application does look like.

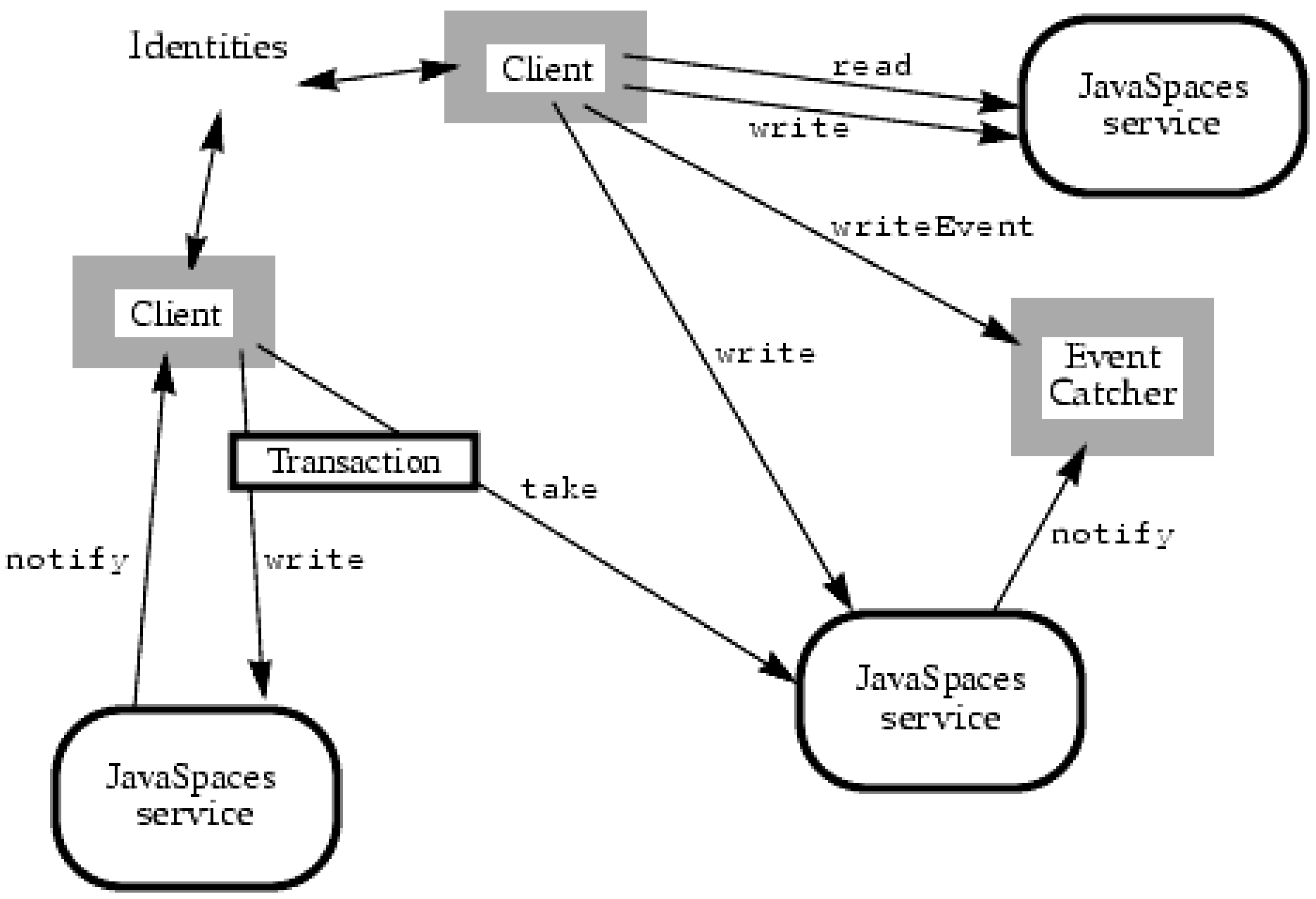

Figure 8: A typical JavaSpaces Technology Application [40]

A client can interact with as many JavaSpaces services as needed and has the ability to perform operations (write, read, take) which map entries to templates onto JavaSpaces services. Such operations may be contained in a transaction so that all or none of the operations take place.

## 2.3 Comparative Study

In this chapter we investigate the behavior of sTuples, SWS, TSC, CSpaces, WSMX, and JavaSpaces in terms of four basic categorization criteria. These criteria list comprises:

Criterion 1: the addition of semantic annotations to the information stored and retrieved from the TupleSpace. sTuples, SWS, TSC, CSpaces, and WSMX support new tuple types encapsulating the data formalized using Semantic Web languages. The language used in sTuples is DAML+OIL, in SWS, TSC, and CSpaces the language is RDF(S), while in WSMX the language is WSML. On the other hand, JavaSpaces do not semantically annotate the shared information.

Criterion 2: the scope of each approach. sTuples active in the field of pervasive computing and more particularly they are applied in mobile systems. SWS envisioned becoming a middleware for coordinating knowledge processes on the whole Semantic Web. TSC focuses on Web Services technology and establishes the mechanism to publish communication data as the Web paradigm of 'persistently publish and read'.  CSpaces are an extension of Triple Space Computing and study their applicability in different scenarios in the Semantic Web, apart from Web Services. Their upper goal is the distributed knowledge management. WSMX apply the principles of TSC to provide an execution environment for dynamic discovery, selection, mediation, invocation, and interoperation of Semantic Web Services. JavaSpaces technology is a simple and powerful high-level coordinator tool, able to be applied in the building of distributed applications. So, JavaSpaces show that the TupleSpace paradigm can be applied in numerous applications, while sTuples shows that semantic TupleSpace is not only to be applied to the Semantic Web or Semantic Web services but also to many other areas where distributed, heterogeneous and autonomous agents interact in open large-scale systems [41].

Criterion 3: the possibly reasoning engines used. sTuples use Racer reasoner to simplify clients' interactions and to incorporate user semantics in delivering data and services to the user. SWS use an Ontological Reasoner to interpret ontologies according to their formal semantics, to draw inferences, and to check for their satisfiability. Also, reasoning services are used to improve the precision of the returned results. Generally, TSC do not use any reasoners, apart from the TSC project [27] which provides limited reasoning support. CSpaces can be connected with mapping and transformation rules that allow the execution of reasoning processes in order to check for consistency. Moreover, reasoning services are executing to evaluate complex queries. WSMX use a reasoner compliant with WSML (e.g. Flora[42], XSB [42]) able to provide mapping process of mediation, validation of a possible composition of services, indication that a service in a process is executable in a given context, capabilities that exactly match the requester's goal, and capabilities that exactly subsume the requester's goal. Finally, since JavaSpaces do not semantically enhance the exchanged information, there was no reason to use a reasoned.

Criterion 4: the main blocks that each approach is build around. sTuples, TSC, and WSMX are built around four building blocks: Semantic Data Model, Organizational Model, Coordination Model, and Architecture Model. SWS and CSpaces are built around six building blocks: Semantic Data Model, Organizational Model, Coordination Model, Collaborative and Consensual Model, Security and Trust Model, and Architecture Model. JavaSpaces are built around three building blocks: Organizational Model, Coordination Model, and Architecture Model.

Table 1 summurises the collected results.

| | sTuples | Semantic Web Spaces | Triple Space Computing | Conseptual Spaces | WSMX | JavaSpaces |
|---|---|---|---|---|---|---|
| Semantic Annotations | DAML+OIL | RDF(S) | RDF(S) | RDF(S) | WSML | - |
| Scope of the Approach | Mobile Systems | Semantic Web | Semantic Web Services | Distributed Knowledge Management | Semantic Web Services | Distributed Applications |
| Reasoners | Racer | Ontological Reasoner | - | Reasoning Processes | Flora, XSB | - |
| Main building blocks | Semantic Data Model, Organizational Model, Coordination Model, Architecture Model | Semantic Data Model, Organizatio-nal Model, Coordination Model, Collaborative and Consensual Model, Security and Trust Model, Architecture Model | Semantic Data Model, Organizatio-nal Model, Coordination Model, Architecture Model | Semantic Data Model, Organizatio-nal Model, Coordination Model, Collaborative and Consensual Model, Security and Trust Model, Architecture Model | Semantic Data Model, Organizatio-nal Model, Coordina-tion Model, Architecture Model | Organizatio-nal Model, Coordina-tion Model, Architectu-re Model |

Table 1: Basic characteristics of the six technologies in terms of four main criteria

Based on these criteria we can analyze the reason why we did not use an existing semantically enhanced approach for the adaptation of bpel processes.

Criterion 1: it is obvious that none of sTuples, TSC, SWS, CSpaces, and WSMX deals with both RDFS and WSML meta-information. Worse, JavaSpaces do not deal at all with the semantically enhancement of information.

Criterion 2: the functionality of these approaches extends to a wide range. To this end their design and implementation is very complex. Additionally, their scope is not related to the field we are interested in. The exception here is the JavaSpaces technology.

Criterion 3: apart from JavaSpaces, all the other approaches use reasoners to succeed the desired functionality. It is widely known that the use of reasoners increases the system's response time.

Criterion 4: two of the approaches, SWS and CSpaces, are also built around Collaborative and Consensual Model and Security and Trust Model, thus becoming even more complex.

Table 2, Table 3, Table 4, Table 5, Table 6, and Table 7 explain why each of these approaches does not suit our case.

| sTuples | Does it fit to our needs? | Why? |
|---|---|---|
| Semantic Annotations | No | Support only for DAML+OIL annotations |
| Scope of the Approach | No | Oriented in mobile systems |
| Reasoners | No | Time consuming |
| Main building blocks | Yes | - |

Table 2: Evaluation of sTuples in terms of our needs

| Semantic Web Spaces | Does it fit to our needs? | Why? |
|---|---|---|
| Semantic Annotations | No | Support only for RDF(S) and not for WSML |
| Scope of the Approach | No | Oriented in Semantic Web |
| Reasoners | No | Time Consuming |
| Main building blocks | No | High Complexity |

Table 3: Evaluation of Semantic Web Spaces in terms of our needs

| Triple Space Computing | Does it fit to our needs? | Why? |
|---|---|---|
| Semantic Annotations | No | Support only for RDF(S) and not for WSML |
| Scope of the Approach | No | Oriented in Semantic Web Services |
| Reasoners | Yes | Non-Time Consuming |
| Main building blocks | Yes | - |

Table 4: Evaluation of Triple Space Computing in terms of our needs

| Conseptual Spaces | Does it fit to our needs? | Why? |
|---|---|---|
| Semantic Annotations | No | Support only for RDF(S) and not for WSML |
| Scope of the Approach | No | Oriented in Distributed Knowledge Management |
| Reasoners | No | Time Consuming |
| Main building blocks | No | High Complexity |

Table 5:  Evaluation of Conseptual Spaces in terms of our needs

| WSMX | Does it fit to our needs? | Why? |
|---|---|---|
| Semantic Annotations | No | Support only for WSML and not for RDF(S) |
| Scope of the Approach | No | Oriented in Semantic Web Services |
| Reasoners | No | Time Consuming |
| Main building blocks | Yes | - |

Table 6: Evaluation of WSMX in terms of our needs

| JavaSpaces | Does it fit to our needs? | Why? |
|---|---|---|
| Semantic Annotations | No | No support for semantic annotations |
| Scope of the Approach | Yes | Oriented in distributed applications |
| Reasoners | Yes | - |
| Main building blocks | Yes | - |

Table 7:  Evaluation of JavaSpaces in terms of our needs

Last but not least we should refer to the fact that all the aforementioned technologies store persistently the manipulated information in structures such as databases or files. Subsequently, the execution time for write and read operation is bigger than the respective execution time for write and read operation of information kept in-memory. JavaSpaces provide two possible solutions: (a) an implementation where information is stored persistently in a dedicated structure and (b) an implementation where information is kept in-memory.

All in all, the most simple model able to be used as an implementation basis for our approach is the JavaSpaces with the proper extensions to assure the semantically annotation of information.

# CHAPTER 3
# SEMANTIC CONTEXT SPACE ENGINE

The main goal of the SCS Engine is to facilitate the provision of adaptable processes by providing an open mechanism for data acquisition, which supports the collection and sharing of information elements. Information elements refer to structured data that are semantically annotated and contained within a specific space. A space is considered to be the process's environment, which is open to other processes and systems, e.g. agent based systems and sensor networks, for information exchange. The acquisition mechanism is independent of the metadata primitives used for the annotation of information elements and supports their logical organization into groups, i.e. similarly called scopes.

The SCS Engine provides to its clients a basic set of operations which include writing, grouping, and retrieving information elements. Specifically the core features of the SCS Engine are:

- the acquisition of semantically enhanced Java based information elements,
- the support for the logical grouping of information, the so-called "information scopes" (e.g. information pertaining to weather conditions),
- the specification of associations among information scopes, and
- the support for multiple types of meta-information models (e.g. Simple Text, WordNet[43], OWL [44], Web Service Modeling Language (WSML) [45], Resource Description Framework (RDFS)[17])) and associated meta-information search engines.

The objective of this thesis focuses on the acquisition of semantically enhanced information and on the support for RDFS and WSML meta-information models along with the associated meta-information search engines. The mechanisms for the information scopes along with the operations among them have not been implemented in terms of this thesis. Nevertheless, in the sequel we make reference to them because they are of high importance.

## 3.1 Design Decisions on the SCS Engine

### 3.1.1 Implementation Basis

SCS Engine incorporates appropriate semantic annotations to the TupleSpace information model and has been designed to have the features described above. As we already mentioned in CHAPTER 2, a TupleSpace is an implementation of the associative memory paradigm for parallel/distributed computing. It provides a repository of tuples that can be accessed concurrently. As an illustrative example, consider that there is a group of processors that produce pieces of data and a group of processors that use the data. Producers post their data as tuples in the space, and then the consumers retrieve data from the space that match a certain pattern. TupleSpace may be thought of as a form of distributed shared memory. The TupleSpace model has been extensively applied in the coordination of distributed and parallel systems. Yet, its use by the Service-Oriented Computing paradigm is an emerging trend that has been accompanied by proprietary extensions ([41], [46]).

Although services provide a layer of abstraction that facilitates the interoperation of systems over the web, the merits of the TupleSpace paradigm will further enhance the Service Oriented model. These merits include: i) decoupling of process components, ii) associative based addressing (i.e. data is referenced by its content and not by its address), and iii) support for the provision of synchronous and asynchronous communication patterns ([47], [11]). More to that, such properties will foster the formation of new types of collaboration schemes among Service-Oriented systems and other existing or emerging systems such as Agent-based systems, Sensor and Grid applications.

Nonetheless, one may argue that the SCS Engine could be implemented on top of a database management system, e.g. an RDBMS such as Oracle[5], or MySQL[6]. However, the rigidness of the employed information models and their complex implementation does not render them a first class choice. This has also been the reason why existing implementations of semantically enhanced spaces have been mostly based on TupleSpace implementations [41].

Since there are plenty open-source implementations of the TupleSpace paradigm, we decided not to implement it from scratch. Instead, we use the JavaSpace service [37] of the Jini framework [48] as a pure implementation of the TupleSpace paradigm and we make the required extensions to semantically annotate the information stored and retrieved from the TupleSpace. As we saw in CHAPTER 2, JavaSpaces provide a set of basic operations for the manipulation of Java Objects which additionally implement the Entry Interface. A client can write an Entry in the space (operation write), make a copy of an Entry in the space (operation read), and retrieve an Entry from the space (operation take). Since SCS Engine's primary goal is to supply an open mechanism for data acquisition, which supports the collection and sharing of information elements, operations write, read, and take should be supported. Considering also that the real information is of the form of Java Objects, the typed-based Index Tree of the Data Manager can be totally based on JavaSpaces. Furthermore, the scalability of this framework allows us to make all the extensions required for (i) the semantically annotation of information and (ii) the support for logical grouping of information in information scopes and for association among those scopes.

The selection of the JavaSpace service instead of other open source TupleSpace, e.g. TSpace [49] or an XML based TupleSpace such as XMLSpace [50], was primarily driven by our need for flexibility over the supported meta-information models and the related query engines, as well as the need for supporting Java based information elements. Adhering to the "separation of concerns" design principle, the provided open-source implementation of the JavaSpace service accommodates a simple yet extensible information model that is independent of the implemented functionality [10].

### 3.1.2 Extensions

The accommodated API is an extended version of the Linda model [11] based on JavaSpaces, providing the writing, reading (discovery), and retrieval of semantically annotated information elements, either from the whole space or a specific scope.

To support the semantically annotation of information the form of the basic operations should change: write, read, and take should also take into consideration the meta-information which is linked with the real information. Afterwards, a new mechanism, the Meta-Info based Index Tree, should be introduced to store in a dedicated structure in the space those meta-information elements. This mechanism is also responsible for connecting the meta-information with the real information that is associated with.

To support scope management operations our model should add new operations able to create, discover, and remove scope(s), as well as create and remove affiliations among those scopes. These additions will lead to new forms of the basic operations which will also take into account the scope where we want to write, read, or take the semantically annotated information. Furthermore, a new mechanism, the Scope Manager located in the Data Manager, should also be introduced to store the created scopes. This mechanism is also in charge of associating the meta-information stored in the Meta-info based Index Tree with the respective scope.

[5] http://www.oracle.com/index.html
[6] http://www.mysql.com/

Linda model should also meet the need for persistent querying (continuous querying) for information elements. The provided operations accommodate for a basic set of functionality, which facilitates the extensibility of the SCS engine with additional higher level operations and mechanism. For example, inference engines may use the semantics and the information available at the SCS Engine to facilitate the extraction of higher level knowledge, e.g. the combination and/or refinement of information elements such as the current temperature and wind speed at specific region may provide indications of heat waves. In general, a user may access these operations through a Java based interface [1].

Since the role of the SCS Engine is to exploit semantically annotated information stemming from external sources we need a special mechanism for the selection of this information. External sources could be data providing services, databases, humans playing the role of users, or another external data source which has to be specified. The process of accumulating information from those sources is as follows: if the data source is a data providing service, the way to obtain information is through service calls; if the data source is a database, the way to obtain information is through queries; in case the data source is a human there is no need for such a mechanism, as he/she will fill in the data space with relevant semantically annotated information. In the presented implementation we consider that external sources are humans or programs that are able to execute the provided write operation.

### 3.1.3 Information Model

Each information entity stored in SCS Engine should be in a specific form, as illustrated in Figure 9. The main attributes are: the unique identifier of each information element; a class named *Lease* which adds temporal properties, as it represents a fixed period of time in which the information element is considered to be valid; and a class called *MetaInformation* which is responsible for keeping all the semantic information related to the information element. The *MetaInformation* is annotated with semantic attributes which can be OWL, RDFS, WORDNET, TEXT, XML, OBJECT, or WSML. In our implementation, we support WSML and RDFS-based attributes. Additionally, *MetaInformation* comprises *Features* which provide a generic container for holding implementation related properties which facilitate the processing of meta-information elements [51].

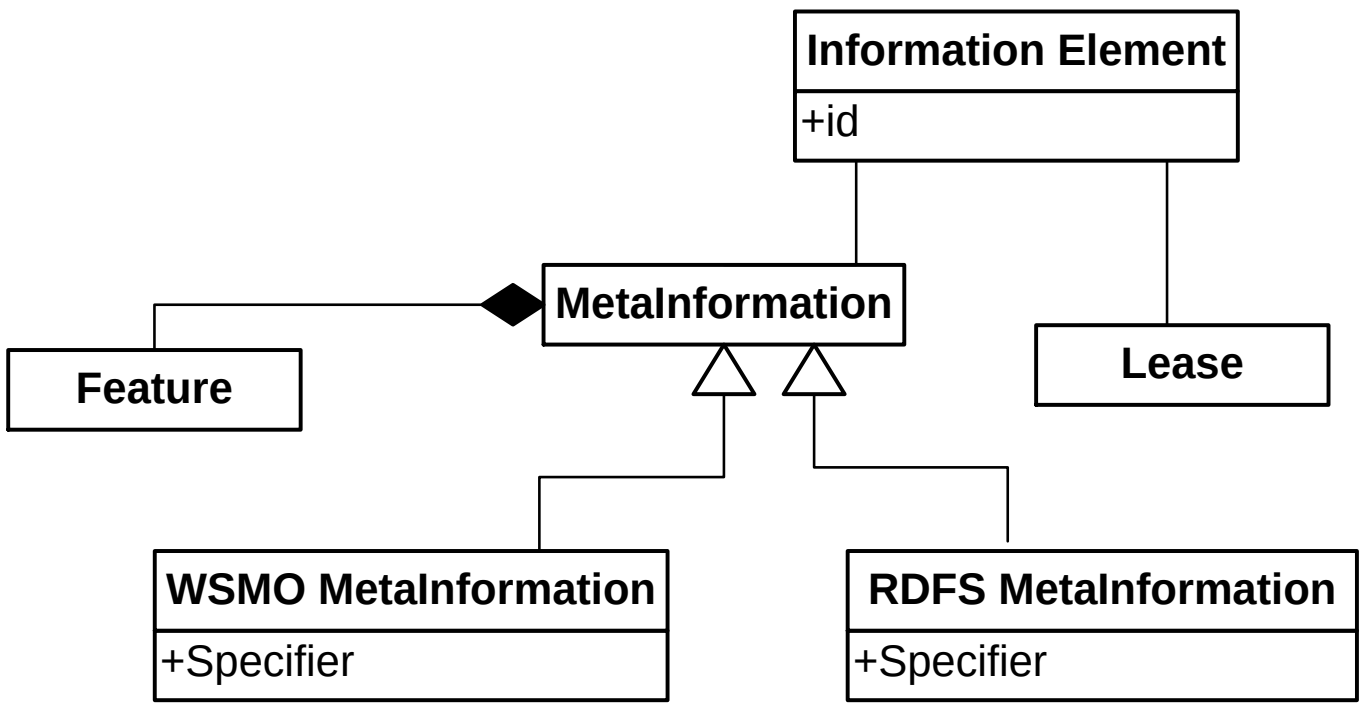

Figure 9: Core Elements of the Information Model of the SCS Engine

## 3.2 Semantic Context Space Architecture

Figure 10 graphically illustrates the architecture of SCS Engine. The architecture is plug-in, namely we propose a framework that allows future additions of modules without breaking the existing code base. As one may easily infer three components comprise the whole Engine and provide the required levels of efficiency, flexibility and extensibility. More specifically these components are the *Data Manager*, the *Query Processor* and the *Mata-Info based Interface*. The description of each component follows.

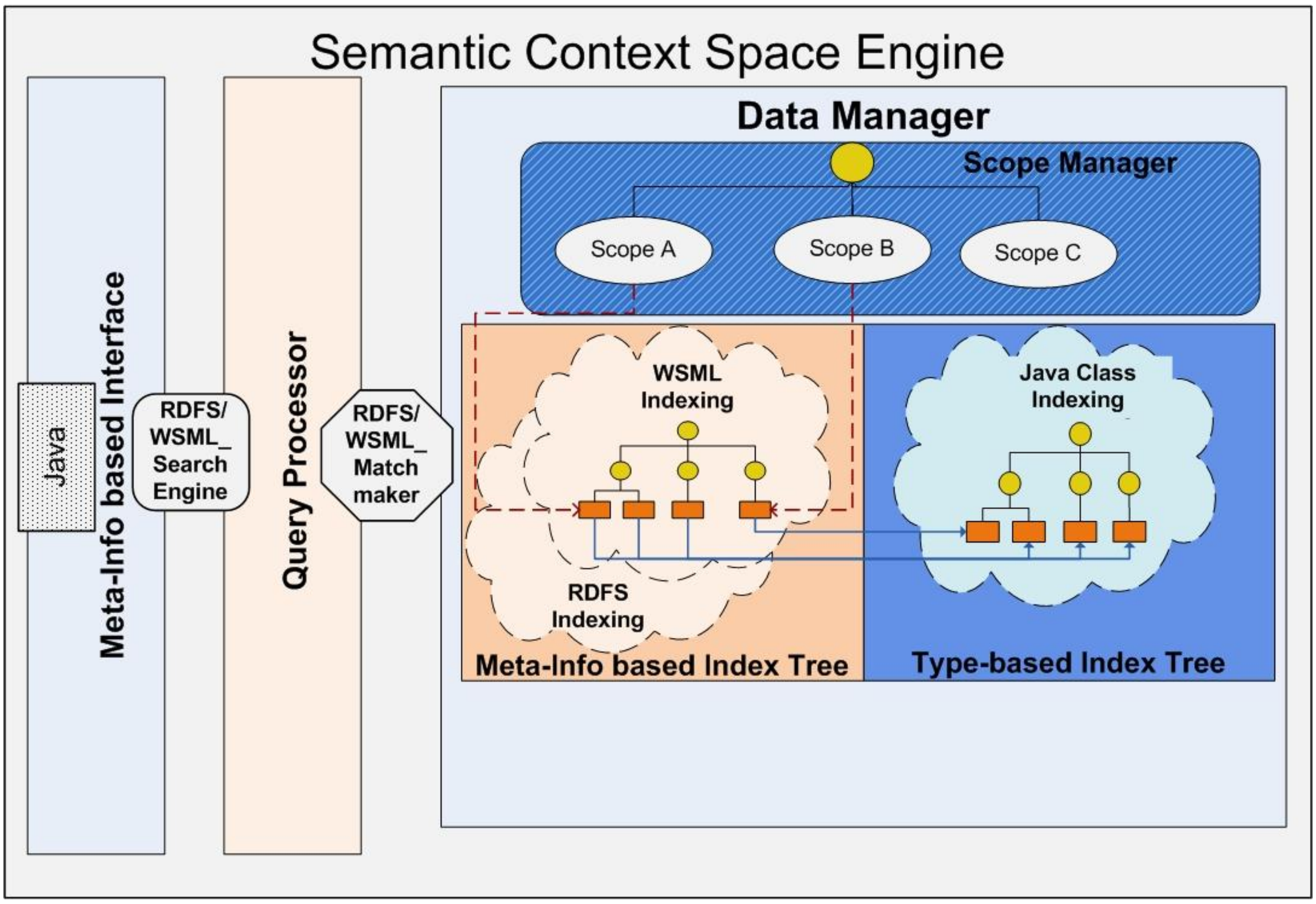

Figure 10: SCS Engine internal architecture

### 3.2.1 Data Manager

Data Manager consists of the Scope Manager, the Meta-info Based Index Tree, and the Type-based Index Tree. This is the core-element of the architecture, as it is responsible for the storage, maintenance, and indexing of provided information. Indexing arises from the need to execute queries as quickly as possible. The Meta-info Based Index Tree comprises the whole MetaInformation stored in the SCS Engine in a specific structure. The creation of the Meta-info Based Index Tree requires the knowledge about the semantics of the elements for insertion. As we mentioned before, these semantics are fed to the SCS Engine through ontologies described in RDFS or WSML. In order to extract information from these ontologies, we need a framework for RDFS files and a framework for WSML files, which will store (if necessary), parse, and query RDFS and respectively WSML data. The Type-based Index Tree comprises the real Java Objects stored in the SCS Engine. The creation of the Java Class Indexing is totally based on the JavaSpaces. The Scope Manager is responsible for the sustenance of scopes, however, it has not been implemented in this thesis.

### 3.2.2 Query Processor

Query Processor uses an extendable set of pluggable query engines to discover appropriate information entities. In general, the discovery of information within a space is performed via the use of appropriate Search Engines and Matchmaker components. Particularly, a query engine exploits a matchmaker for searching within specific types of Meta-Information elements (in our case WSML-based Meta-Information elements and RDFS-based Meta-Information elements) and InformationEntity elements (in our case Java based types). We are interested in the search of MetaInformation elements. Upon the receipt of a specific type of request, the Query Processor dispatches the related query to the Search Engine (in our case it is a RDFS/WSML_SearchEngine). The Search Engine uses a Matchmaker (in our case it is a RDFS/WSML_Matchmaker) to perform the discovery of the requested information. The matchmaker calculates a similarity degree for each element and returns back to the requestor the total similarity degree. In the general case, Query Processor may use an extendable set of pluggable query engines (e.g. WordNet search engine, OWL search engine), along with the appropriate matchmakers (e.g. WordNet matchmaker, OWL matchmaker), to search within specific types of Meta-Information elements (e.g. WordNet based Meta-Information). Figure 11 graphically illustrates the process of discovering MetaInformation Elements.

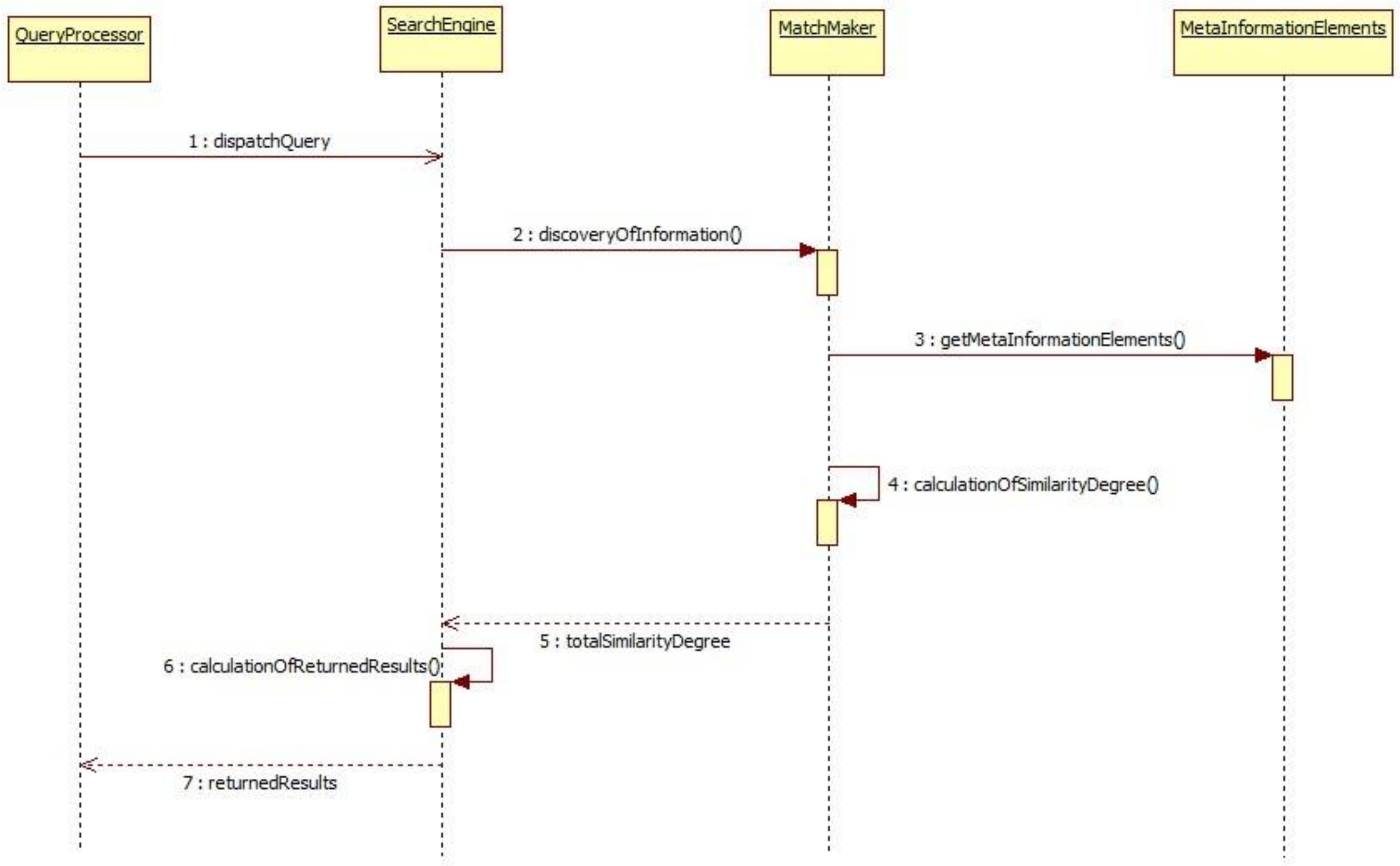

Figure 11: Sequence Diagram for the discovery of MetaInformation Elements

### 3.2.3 Meta-Information based Interface

Meta-Information based Interface implements all interaction points of the SCS Engine with its environment. A dedicated API will give the capability of executing two types of operations: *content related* operations and *scope management* operations. Content related operations consist of 3 individual operations: (i) "add operation" takes as input an entity along with its meta-information and adds this contextual information in space, (ii) search operation, called "read operation" takes as input a semantic query, then it executes it in space, and finally returns a list with all matching entities, (iii) remove operation, called "take operation" takes as input a semantic query, executes it, and removes from space all matching entities. The implementation details of these operations are described in Paragraph 3.3. Scope management operations include six operations: (i) createScope operation accepts two arguments, name and meta-information, and creates a new scope with these arguments, (ii) removeScope accepts the argument name and deletes from space the scope instance which corresponds to the name, (iii) findScope takes as input a query and returns a list with all scopes in space with the specific name, (iv) getScopes returns a list of all scopes in space, (v) addAffiliation accepts 3 arguments, fromScope, toScope, affiliationType, and creates specific type of bonds with other scopes that may reside at dispersed context spaces, (vi) removeAffiliation is the same as addAffiliation but instead of creating an affiliation, it removes it. In this thesis we focus on the implementation of content related operations, while the support for scope related operations is part of our future work.

## 3.3 SCS Engine Interface

The supported actions of the current implementation include the write, read, and take of semantically annotated information.

An important step for the proper functioning of the SCS Engine is the initialization, which includes the loading of ontologies which will be used by the supported actions. The loading of ontologies includes the extraction of information such as concept-superconcept relations; this is analyzed in the next section. Generally, this step is useful for the creation of the Meta-info based Index Tree and it is kept in memory. Its main advantage is that it accelerates the speed of the write, read, and take operations.

### 3.3.1 Engine Initialization

The SCS Engine retrieves information from ontologies described in RDFS or WSML which refers to the concept-superconcept relations. More particularly, for each concept of the ontology we need to know all its parents with the correct order. This knowledge is stored in a dedicated structure (separate from RDFS and WSML) and is further exploited by the Meta-Info based index tree of the Data Manager, during the write, read, and take operations. Figure 12 depicts the interaction between the Meta-info based index tree and the ontologies provided through the created structures in RDFS and WSML.

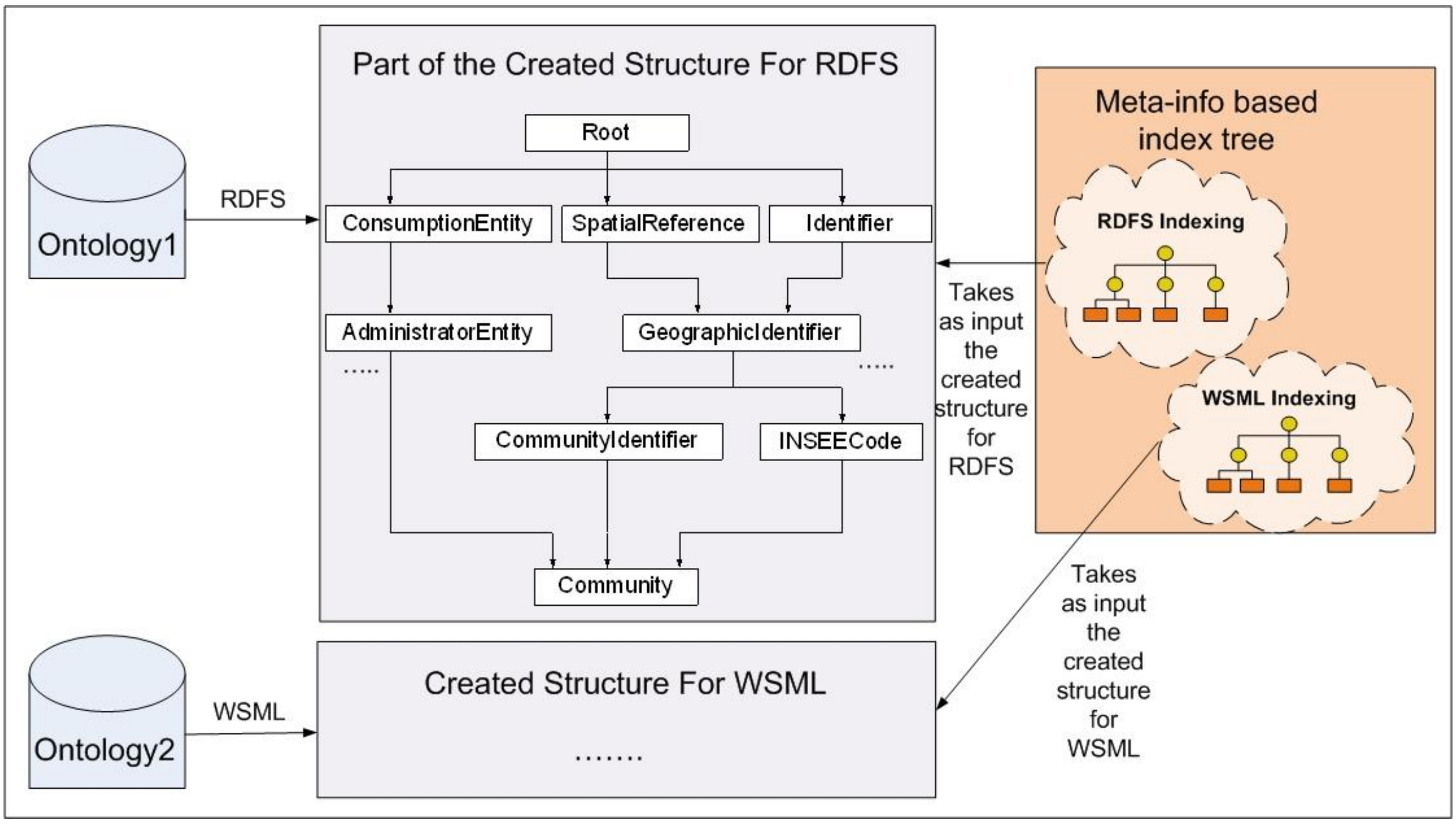

Figure 12: Interaction between the Meta-info based index tree and the ontologies through the created structures in RDFS and WSML

The first step for the information retrieval comprises the parsing of the RDFS and WSML files. The parsing tools differ according to the type of file. For the parsing of RDFS data we use *Sesame Repository*[7]. Sesame is an open source Java framework for storage and querying of RDF data, consisting of a set of libraries. The framework is fully extensible and configurable with respect to storage mechanisms, inference engines, RDF file formats, query result formats and query languages. Sesame supports SPARQL[52] and SeRQL [53] querying, a memory-based and a disk-based RDF store and RDF Schema inference engines. It also supports most popular RDF file formats and query result formats. Sesame is currently developed as a community project, with Aduna[8] as the project leader.

[7] http://www.openrdf.org/

For the parsing of WSML data we use the Ontology Representation and Data Integration *(ORDI) Framework*[9]. ORDI is an open-source ontology middleware based on Java. Among other things, ORDI provides support for integration of different structured ontologies, backward compatibility with the existing RDF specifications and the SPARQL query language, efficient processing and storage of meta-data or context information, and easy management of data from several sources within the same repository.

We need to state here that we choose to use Sesame Repository for the parsing of RDFS ontologies instead of ORDI because Sesame is easier in use and much more efficient.

#### 3.3.1.1 Initialization Process Description

To better illustrate the initialization process we use an example which represents what information we want to extract and how this is done.

Given an RDFS ontology, e.g. SwingDomainOntology[10] [11]which describes 343 concepts, for each concept of this ontology, we extract all the possible concept-paths (from now and on, we will refer to the concept-path as ConceptPath) leading to the concept from the root of the ontology. Considering for example the ConceptPath for the concept *Frog* illustrated in Figure 13, and the set of all ConceptPaths for the concept *Community* illustrated in Figure 14. Note that for the concept *Frog*, there is only one ConceptPath, while for the concept *Community*, there are five possible ConceptPaths. Analytically:

*ConceptPath for the concept Frog*

Root-Organism-Animal-Vertebrate-Amphibian-Frog,

*Set of all ConceptPaths for the concept Community*

Root-ConsumptionEntity-AdministrativeEntity-Community

Root-SpatialReference-GeographicIdentifier-CommunityIdentifier-Community

Root- SpatialReference-GeographicIdentifier-INSEECode-Community

Root-Identifier-GeographicIdentifier-CommunityIdentifier-Community

Root-Identifier- GeographicIdentifier-INSEECode-Community

Each ConceptPath is a list of concepts, while the set of all ConceptPaths is named after *ConceptPathList.*

[8] http://www.aduna-software.com/

[9] http://ordi.sourceforge.net/quickstart.html

[10] http://kenai.com/projects/envision/sources/runtime/show/SimpleUseCase/LinkedData%20Examples?rev=164

[11] Look at the Appendix 2 for more information

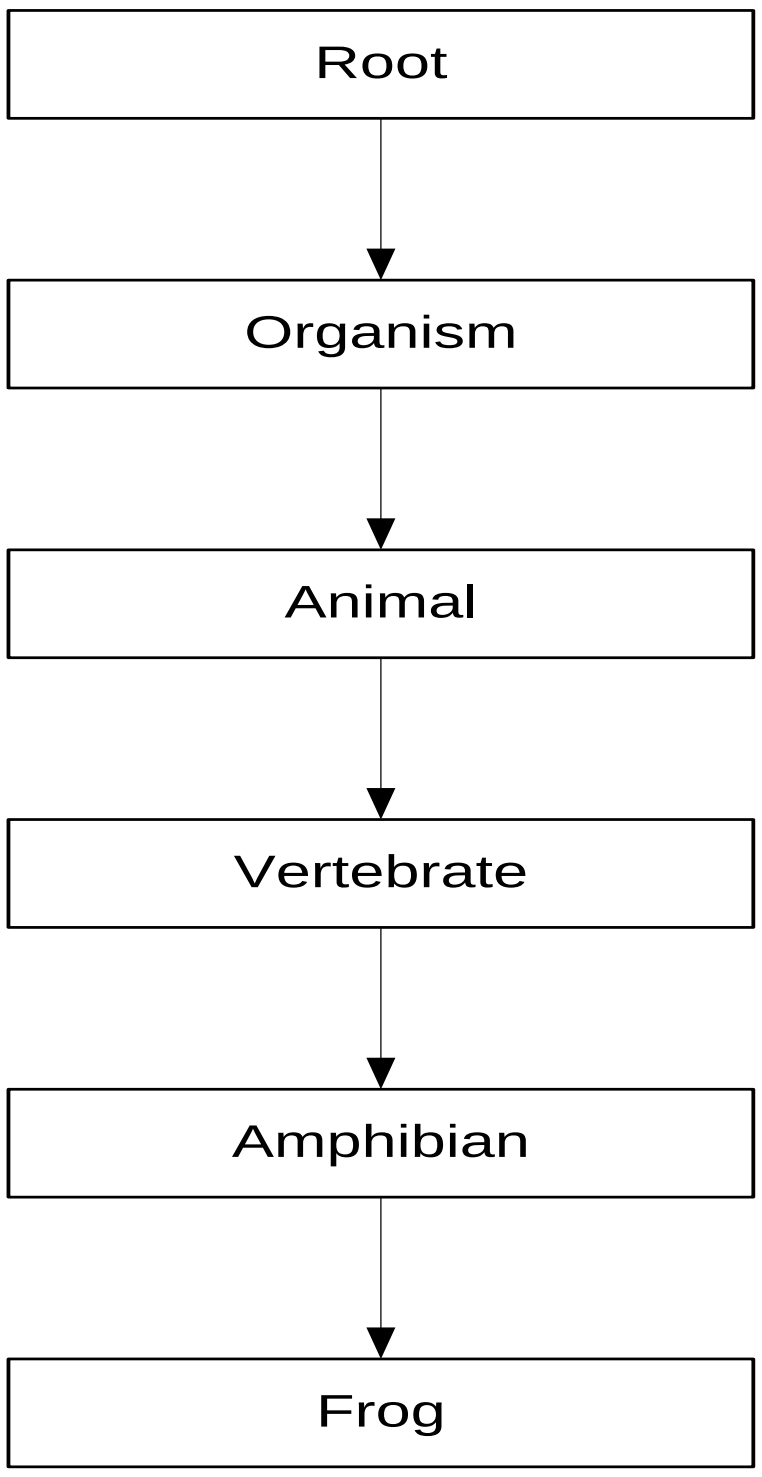

Figure 13: The ConceptPath for the concept Frog

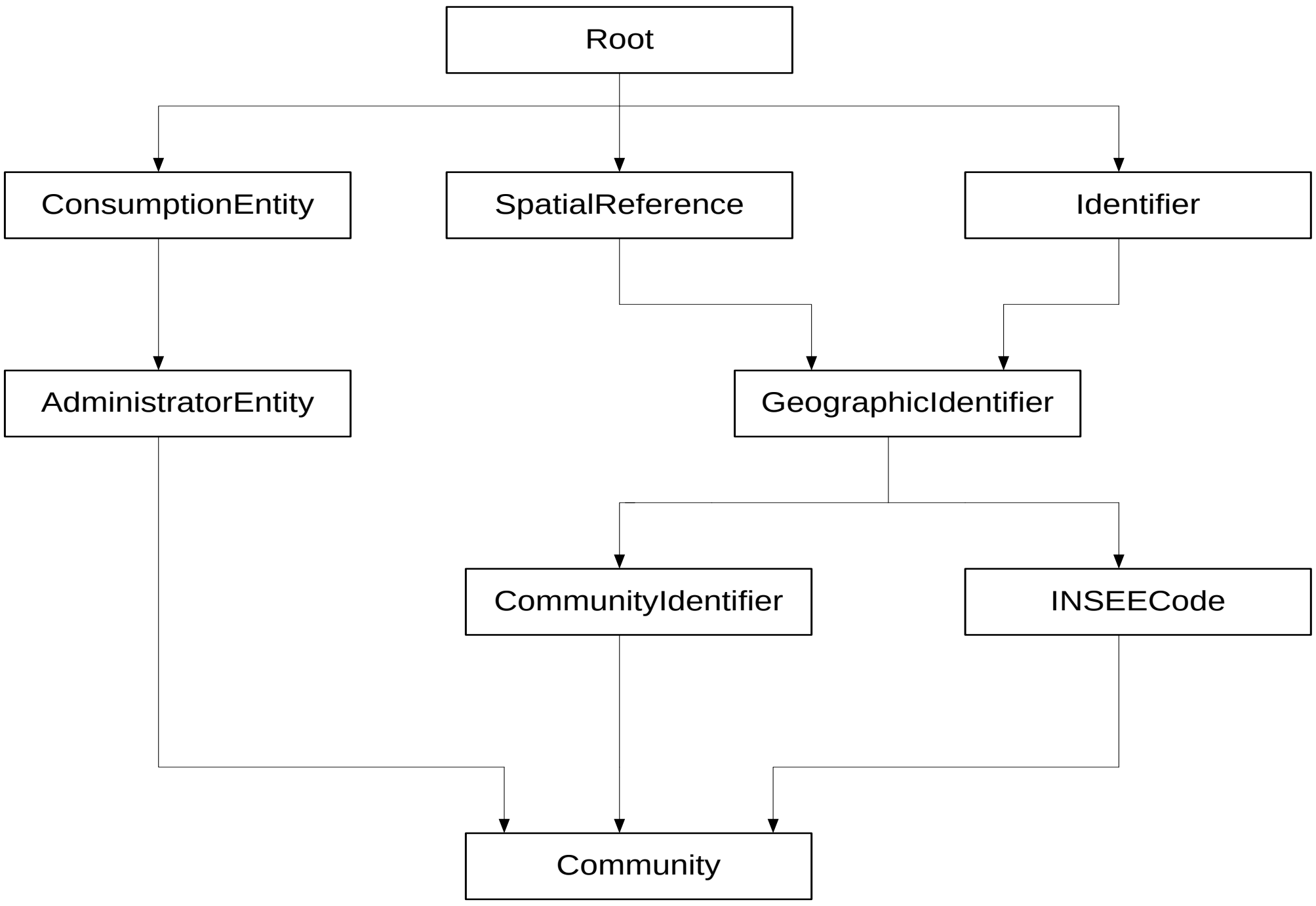

Figure 14: The set of all ConceptPaths for the concept Community

Following, we present the methodology which computes the *ConceptPathList* of each concept in the ontology. The methodology can be divided in three parts:

***First part – Algorithm***: The input parameter of this algorithm is a list of all pairs <concept, superconcept>. It is named after *ConceptSuperConcept* and it is produced from the parsing of the ontology described in RDFS or WSML. For example, according to the examples given for the *Frog* and the *Community* a part of that list is the following:

*part of the ConceptSuperConcept list*

…

<Frog- Amphibian>

<Amphibian - Vertebrate>

<Vertebrate - Animal>

<Animal - Organism>

…

<Community - AdministrativeEntity>

<AdministrativeEntity - ConsumptionEntity>

<Community - CommunityIdentifier>

<CommunityIdentifier - GeographicIdentifier>

<GeographicIdentifier - SpatialReference>

<Community - INSEECode>

<INSEECode - GeographicIdentifier>

<GeographicIdentifier - Identifier>

…

This algorithm is responsible for the creation of a structure, named after *ChildParentsList*. Each element of the *ChildParentsList* consists of a pair of the form: *<concept, ListOfSuperConcepts>*, where *ListOfSuperConcepts* is a list of concepts.

For example, for the concepts *Frog* and *Community*, we respectively have:

*ChildParentsList for the concept Frog in the form <concept, ListOfSuperConcepts>*

<Frog, <Amphibian>>

<Amphibian, <Vertebrate>>

<Vertebrate, <Animal>>

<Animal, <Organism>>

<Organism, Null>

*ChildParentsList for the concept Community in the form <concept, ListOfSuperConcepts>*

<Community, <AdministrativeEntity, CommunityIdentifier, INSEECode>>

<AdministrativeEntity, <ConsumptionEntity>>

<CommunityIdentifier, <GeographicIdentifier>>

<INSEECode, < GeographicIdentifier>>

<ConsumptionEntity, <Null>>

<GeographicIdentifier, <SpatialReference, Identifier>>

<SpatialReference, <Null>>

<Identifier, <Null>>

The described algorithm, which executes for each of the concepts of the given ontology, is:

**Algorithm: Create the ChildParentsList of a concept**

```
concept
ConceptSuperConcept
ConceptPathList function createChildParentsList
    create a list of <concept, ListOfSuperConcepts>, named after ChildParentsList
    find the direct super-concepts of the concept through the use of the
    ConceptSuperConcept
    insert the pair <concept, direct-super-concepts> to the ChildParentsList
    for each element <concept, direct-super-concepts> of the ChildParentsList do
         for each current-direct-super-concept do
             find its direct super-concepts through the use of the
             ConceptSuperConcept
             if ChildParentsList does not include the pair <current-super-concept,
             direct-super- concepts> then
                 insert the pair <current-direct-super-concept, direct-super-
                 concepts> to the ChildParentsList
             end if
         end for
    end for
end function
```

***Second part – Algorithm:*** this part deals with the building of the intermediate *ConceptPathList*. The process starts by taking the first element of the *ChildParentsList* and creating as many ConceptPaths as the direct superconcepts of this element. For example, for the concepts *Frog* and *Community*, we respectively have:

*initial ConceptPathList for the concept Frog*

<Amphibian-Frog>

*initial ConceptPathList for the concept Community*

<AdministrativeEntity - Community>

<CommunityIdentifier - Community>

<INSEECode - Community>

Each of the second elements of the *ChildParentsList* is compared against the first element of the initial *ConceptPathList*. If they are equal, then the *ConceptPathList is updated*. We distinguish two cases:

- In the first case, the *ListOfSuperConcepts* contains just one concept, so we only have to add this concept to the start of each element of the current *ConceptPathList*. We exemplify by giving two examples:

  First Example: Assuming that we want to create the intermediate *ConceptPathList* for the concept *Frog*, having the initial *ConceptPathList* and the *ChildParentsList*. In this case, the initial *ConceptPathList* contains only one element. We compare the second concept of the *ChildParentsList*, which is *Amphibian* with the first concept of the initial *ConceptPathList* which is *Amphibian*. These two elements are the same; as the size of the *ListOfSuperConcepts is one* the concept contained it, namely *Vertebrate*, is placed to the start of the first element of the initial *ConceptPathList*. As a result, the new intermediate *ConceptPathList* is:

<u>*intermediate ConceptPathList for the concept Frog*</u>

<Vertebrate-Amphibian-Frog>

Continuing similarly and after checking all the elements of the *ChildParentsList*, we end up with the following *ConceptPathList*:

<u>*final ConceptPathList for the concept Frog*</u>

<Organism, Animal, Vertebrate, Amphibian, Frog>

As one can easily infer, in the simple case of the concept *Frog*, the process always falls into the first case. Thus, after checking all the elements of the *ChildParentsList*, we end up with the creation of the final *ConceptPathList.*

<u>Second Example</u>: Assuming that we want to create the intermediate *ConceptPathList* for the concept *Community*, having the initial *ConceptPathList* and the *ChildParentsList*. In this case, the initial *ConceptPathList* contains three elements. We compare the second concept of the *ChildParentsList*, which is *AdministrativeEntity* with the first concept of the initial *ConceptPathList* which is *AdministrativeEntity* as well. As the size of the *ListOfSuperConcepts* is one, the concept contained in there, namely *AdministrativeEntity*, should be placed to the start of the first element of the initial *ConceptPathList*. As a result, the first element of the *ConceptPathList* is *<ConsumptionEntity - AdministrativeEntity - Community>*. Continuing similarly and after checking the fifth element of the *ChildParentsList*, we conclude with the following intermediate *ConceptPathList:*

<u>*intermediate ConceptPathList for the concept Community*</u>

<ConsumptionEntity - AdministrativeEntity - Community>

<GeographicIdentifier - CommunityIdentifier - Community>

<GeographicIdentifier - INSEECode - Community>

- In the second case, the *ListOfSuperConcepts* contains more than one concept, so we have to create additional ConceptPaths and add them to the structure of the *ConceptPathList*.

  For example, assume that we have the intermediate *ConceptPathList* for the concept *Community* analyzed in the first case, and we compare the sixth concept of the *ChildParentsList*, which is *GeographicIdentifier,* with the first concept of the intermediate *ConceptPathList* which is *ConsumptionEntity*. The elements are not the same, so we compare again the *GeographicIdentifier* with the second concept of the intermediate *ConceptPathList* which is *ConsumptionEntity*. These elements are the same, so we check the size of the *ListOfSuperConcepts* of the sixth element. The size is two (it contains the concepts *<SpatialReference, Identifier>*), so we have to duplicate the list *<GeographicIdentifier - CommunityIdentifier - Community>*, and at the start of the first list we add the concept *SpatialReference*, while at the start of the second list, we add the concept *Identifier*. The resulting *ConceptPathList* is the following:

*final ConceptPathList for the concept Community*

<ConsumptionEntity-AdministrativeEntity-Community>

<SpatialReference-GeographicIdentifier-CommunityIdentifier-Community>

< SpatialReference-GeographicIdentifier-INSEECode-Community>

<Identifier-GeographicIdentifier-CommunityIdentifier-Community>

<Identifier- GeographicIdentifier-INSEECode-Community>

The described algorithm, which runs for all the concepts of the given ontology, is:

**Algorithm: Find the ConceptPathList of a concept**

```
concept
ChildParentsList
function createConceptPathList
    take the first element of the ChildParentsList
    create the initial ConceptPathList adding in there as many elements as the size
    of the ListOfSuperConcepts. (Initially, each element is of the form
    <superconcept,concept>)
    for each other element of the ChildParentsList, beginning from the second, do
         take the concept
         for each ConceptPath of the current ConceptPathList do
              take the first concept of the current ConceptPath
              if the two concepts are equal then
                   if ListOfSuperConcepts.size ==1 then
                      add the superconcept corresponding to the equal concept to
                      the start of the ChildParentsList
                   else
                      create as many copies of the current ConceptPath as the size
                      of the ListOfSuperConcepts minus one
                      take the first element of the ListOfSuperConcepts
                      for each copy – ConceptPath do
                         add at the start of the ConceptPath the current element of
                         the ListOfSuperConcepts
                         take the next element of the ListOfSuperConcepts
                      end for
                      add the created ConceptPaths to the ConceptPathList
                   end if
              end if
         end for
    end for
end function
```

***Third part – Algorithm:*** The ultimate goal is to gather all the pairs <concept, ConceptPathList> in a single in-memory structure, which will be accessed by the Meta-Info based index tree. Up to here, the created structure is illustrated in Figure 15.

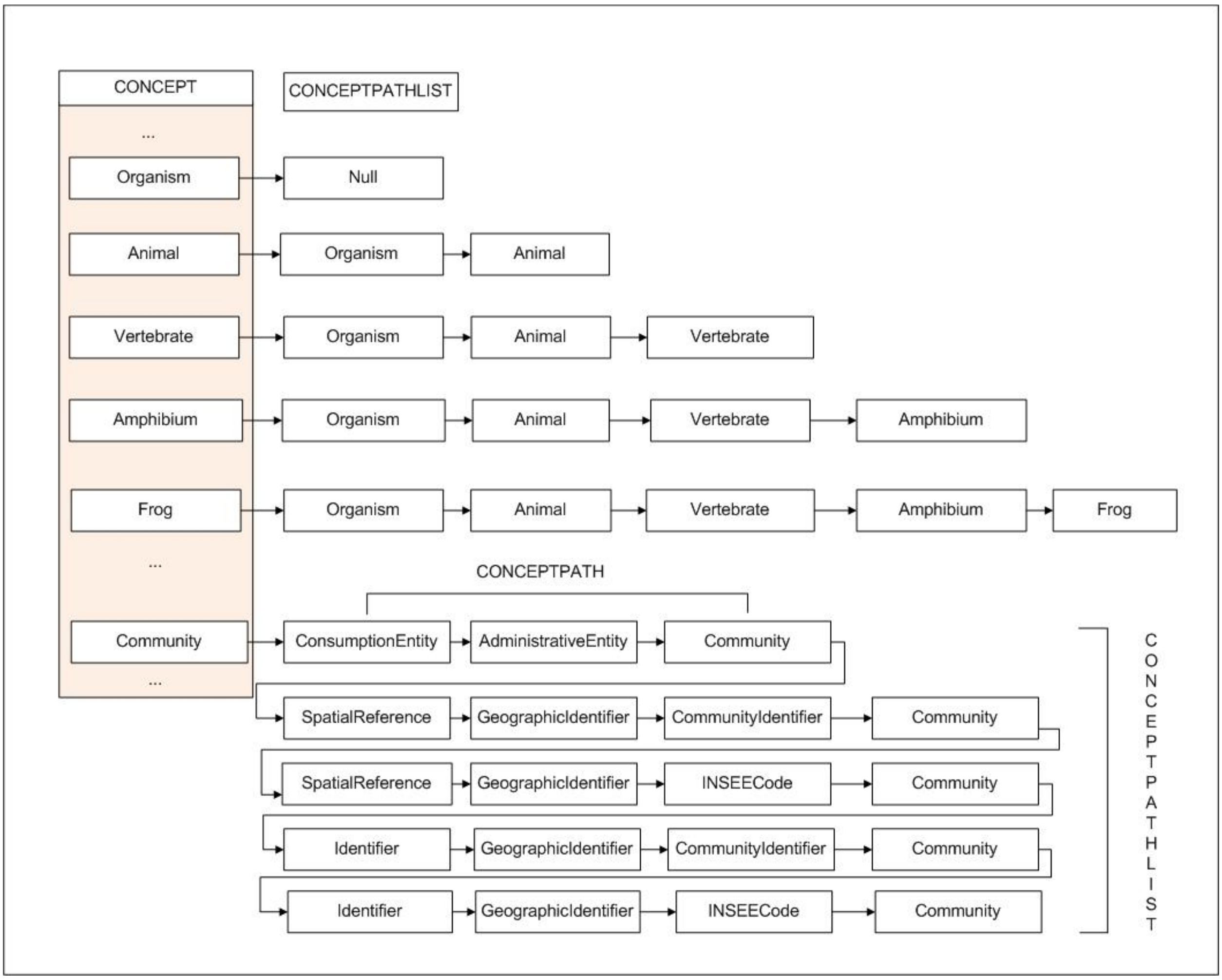

Figure 15: Part of the Map<concept, ConceptPathList>

As we want this structure to be small in size and efficient for search operations, we decided not to store each concept, which is represented as a URI, along with the ConceptPathList, which is represented as a list of lists of URIs. Instead, we perform to the URI a very simple hash function, the hashCode[12]. This function takes as input a string and computes an integer according to the following type:

s[0]*31^(n-1) + s[1]*31^(n-2) + ... + s[n-1]

As with any general hash function, collisions are possible, but for the needs of this thesis which does not include ontologies describing more than 8500 concepts, this function operates without causing any collision. Other kinds of hash functions, e.g. SHA (Secure Hash Algorithm)[13], can be considered in the case of ontologies with higher numbers of concepts.

So, we replace the *ConceptPath* with the *HashPath* and the *ConceptPathList* with the *HashPathList*.

For example, for the concept *Frog* we have respectively the *ConceptPath* and the *HashPath*:

ConceptPath: <Organism – Animal – Vertebrate – Amphibian – Frog>

HashPath: <2136954006 – 1622454272 – 55446222 – 323295868 – 270054104>

[12] http://download.oracle.com/javase/1.4.2/docs/api/java/lang/String.html#hashCode()
[13] http://en.wikipedia.org/wiki/Secure_Hash_Algorithm

For the concept Community we have respectively the *ConceptPathList* and the *HashPathList*:

ConceptPathList: <ConsumptionEntity-AdministrativeEntity-Community>
<SpatialReference-GeographicIdentifier-CommunityIdentifier-Community>
< SpatialReference-GeographicIdentifier-INSEECode-Community>
<Identifier-GeographicIdentifier-CommunityIdentifier-Community>
<Identifier- GeographicIdentifier-INSEECode-Community>

HashPathList: <650811966 – 284601877 – 1630185>
<1278058827 – 229587324 – 1347250258 – 1630185>
<1278058827 – 229587324 – 2086797093 – 1630185>
<952694669 – 229587324 – 2086797093 – 1630185>
<952694669 – 229587324 – 1347250258 – 1630185>

As a result, Figure 15 is transformed to the structure illustrated in Figure 16.

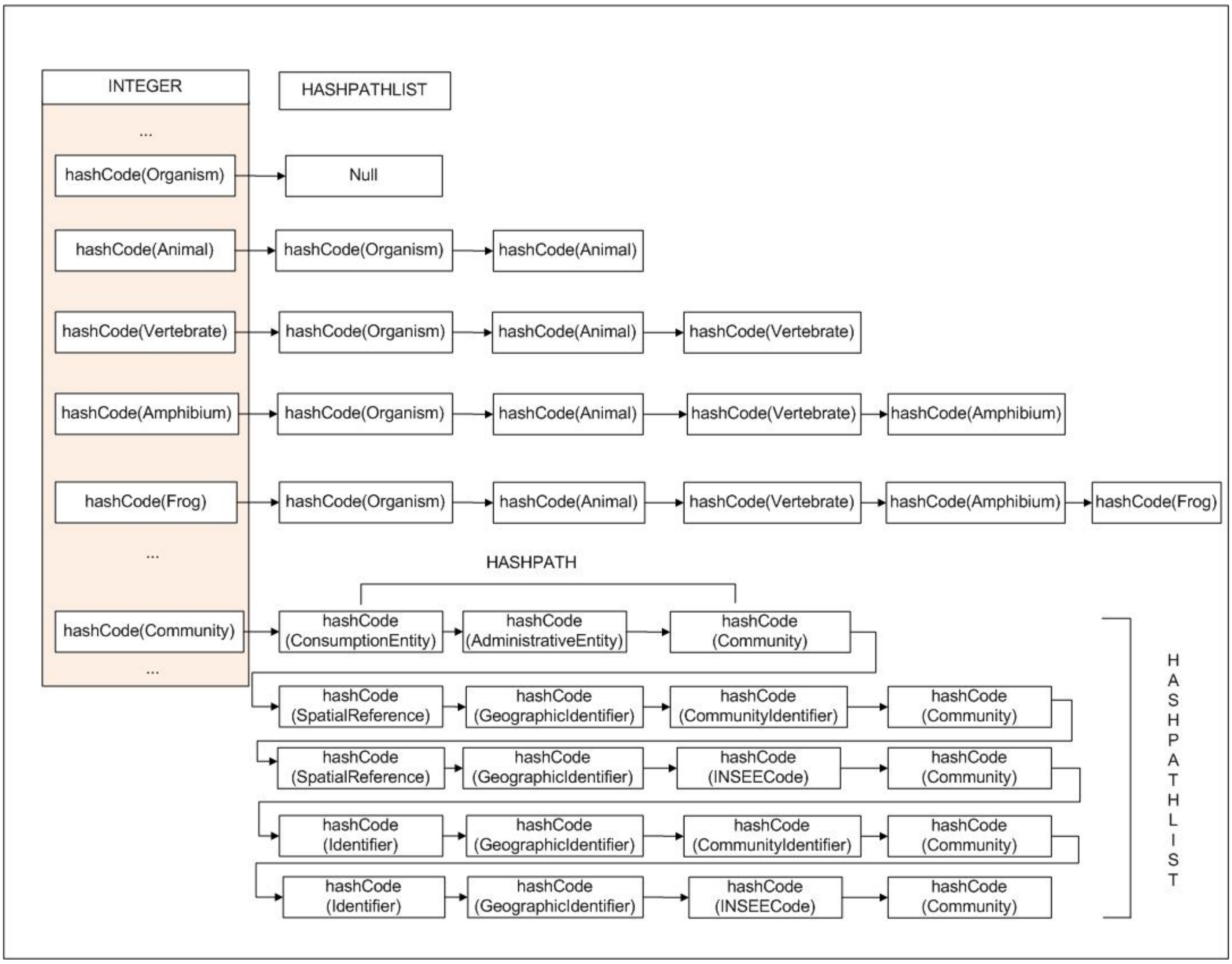

Figure 16: Part of the Map<concept, HashPathList>

Additionally, to limit the complexity of the above structure, for each *HashPath* we create a *BigInteger*[14] which comprises all the *hashCode* values of the respective *HashPath*. The set of all *BigIntegers of a concept* constitutes the *BigIntegerList* structure. So, for each concept we form a pair of type *<concept, BigIntegerList>*. The set of all these pairs is a map named after *MapConceptBigIntegerList* and this is the final structure stored in the memory.

The described algorithm for the formation of the *MapConceptBigIntegerList* runs for all the concepts of the given ontology is:

**Algorithm: Fill in the structure MapConceptBigIntegerList**

```
concept
ConceptPathList
MapConceptBigIntegerList <concept, BigIntegerList>
function fillInMapConceptBigIntegerList
         create an empty HashPathList
         take the absolute value of the concept.hashCode
         for each ConceptPath of the ConceptPathList do
              create an empty HashPath
              for each concept of the ConceptPath do
                   take the absolute value of the concept.hashCode
                   add this value to the HashPath
              end for
              add the HashPath to the HashPathList
         end for
         create the BigIntegerList corresponding to the created HashPathList
         MapConceptBigIntegerList.add(abs(concept.hashCode), BigIntegerList )
end function
```

The three described algorithms are executed sequencially.

A part of the created Map<concept,BigIntegerList> after the parsing of all concepts given in the ontology is illustrated in Figure 17.

[14] http://download.oracle.com/javase/1.4.2/docs/api/java/math/BigInteger.html

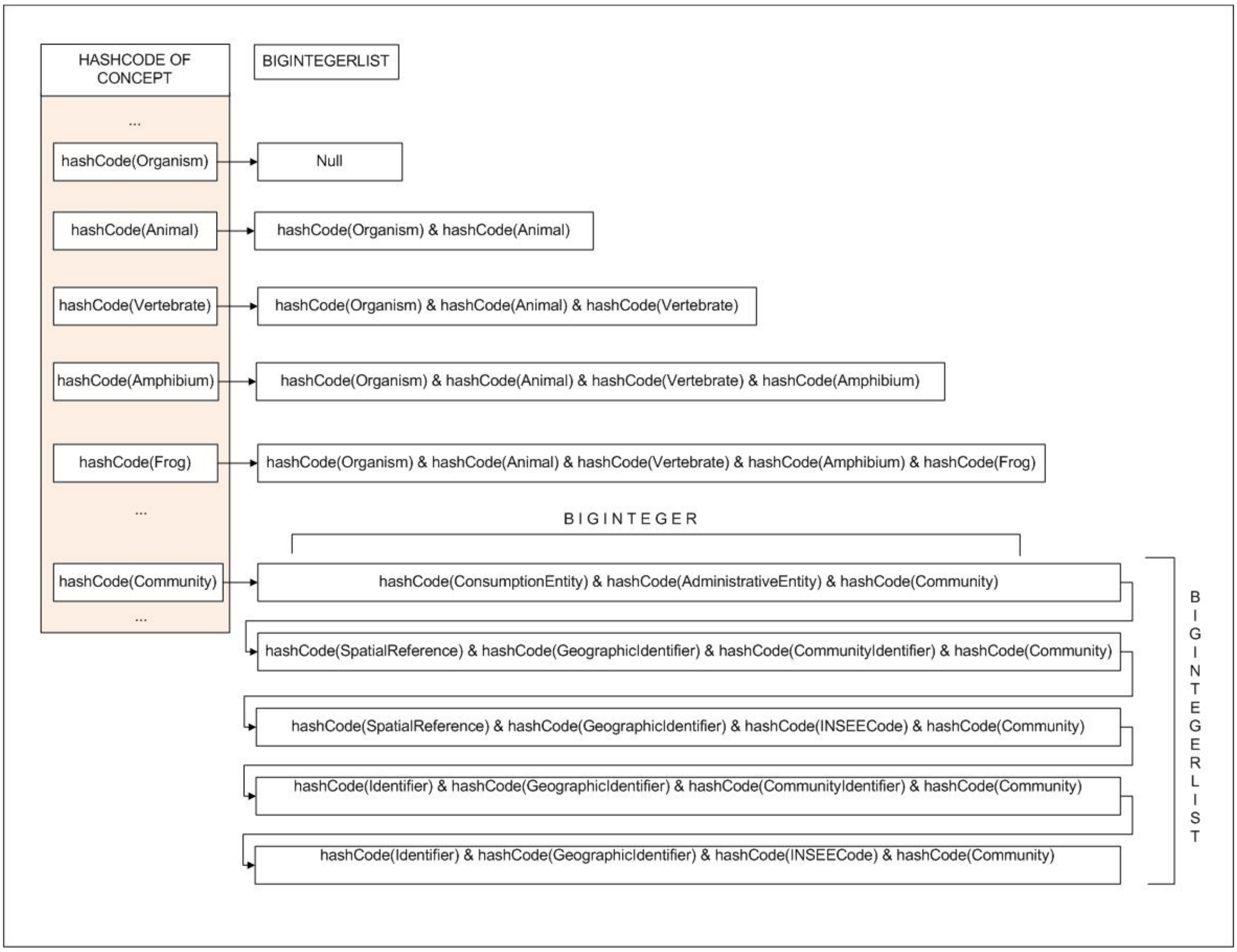

Figure 17: Part of the Map<concept,BigIntegerList>

A class diagram formed by the union of all the aforementioned concepts is located in Appendix 3.

### 3.3.2 Write Algorithm

To implement the write method of the SCS Engine, we extend the typical form of the write method in JavaSpaces as follows:

```
Lease write (Entry entry, Transaction txn, long lease,
      MetaInformation sinfo)
      throws TransactionException, RemoteException
```

As one can easily assume, the MetaInformation is the added attribute which comprises all the semantic annotations that will be associated with the inserted element. MetaInformation's basic characteristics have been described in the Information Model, according to which, the supported MetaInformation types in terms of this thesis are the WSML and RDFS. The Feature attribute in terms of WSML and RDFS MetaInformation contains the concept's URI, the respective BigIntegerList, and an identifier which uniquely characterizes each element.

From a conceptual point of view, when performing the write operation in the SCS Engine, we first write the MetaInformation into the Meta-Info based Index Tree of the Data Manager according to the implemented mechanism which has been described in paragraph 3.3.1, and then we write the actual information, namely the Entry, into the Type-based Index Tree of the Data Manager, by invoking the standard write method provided by the JavaSpaces. The write method of the SCS Engine is also responsible for the connection between the MetaInformation stored into the Meta-Info based Index Tree and the Entry stored into the Type-based Index Tree.

### 3.3.3 Read Algorithm

To implement the read method of the SCS Engine, we use as a basis the typical form of the read method in JavaSpaces. The result is as follows:

```
ResultsList read (Entry tmpl, Transaction tnx, SpaceQuery query)
            throws TransactionException, RemoteException
```

The read operation of the SCS Engine differs from the standard read operation of the JavaSpaces in three ways. First, there is no parameter timeout to specify how long the client is willing to wait until he/she receives a result entry. Now, the client does not wait for a specific period: if there is a result at the time he/she makes the read operation, then it is returned back to him/her, otherwise a null result is returned. Second, the new read operation has an additional parameter, the SpaceQuery query. This attribute allows the execution of both, syntactic and semantic, read operations. This thesis emphasizes on the semantic read operations. The SpaceQuery includes among others the syntactic match degree floor, the semantic match degree floor (hereafter we will refer to it as semantic_match_degree_floor), the type of MetaInformation, and an identifier. In case of a syntactic read, the SyntacticQuery implements the SpaceQuery, while in case of a semantic read, the SemanticQuery implements the SpaceQuery. Additionally, when performing a semantic read, according to the type of MetaInformation we are searching for, there is the need for a class to extend the SemanticQuery. For example, when searching for WSML or RDFS MetaInformation then we have to form the respective WSMLSpaceQuery or RDFSSpaceQuery to extend the SemanticQuery with the concept's URI we are searching for. The last difference is in the returned value of the read operation. The typical read of the JavaSpaces returns one and only result, which is the first entry into the space matching the given criteria, while the read operation of the SCS Engine returns a list of all the results matching the given criteria. Figure 18 graphically illustrates the class diagram for the concepts participating in the read operation.

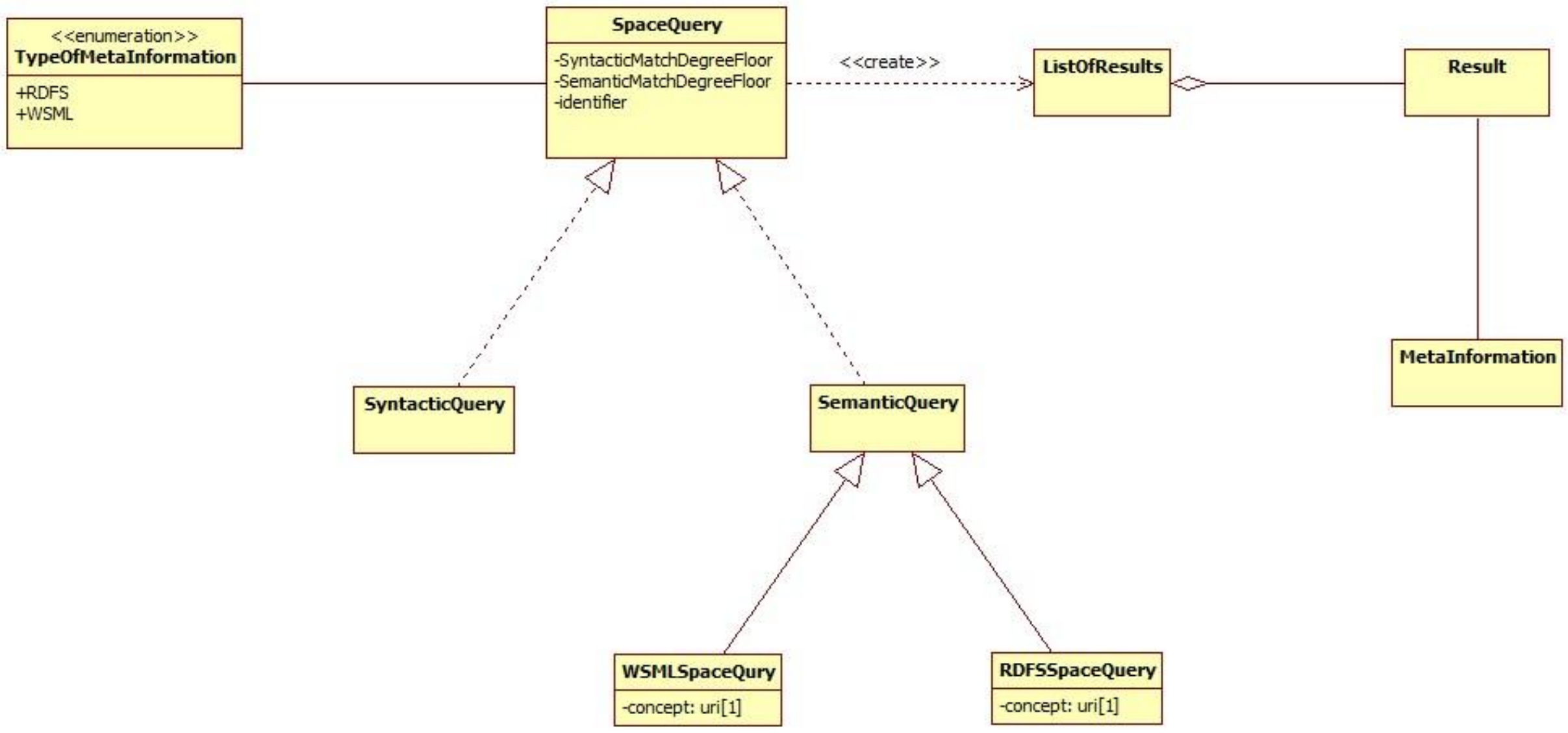

Figure 18: Class Diagram for the concepts participating to the read operation

From a conceptual point of view, when performing a semantic read operation in the SCS Engine, the client gives the MetaInformation along with a semantic match degree floor. The former indicates what kind of information the client is searching for, while the latter determines how close to that kind of information the returned results will be. The read operation searches in the Meta-Info based Index Tree of the Data Manger for stored MetaInformation elements which match the given and for stored MetaInformation elements. When referring to MetaInformation, we are particularly interested to its type (RDFS or WSML) and to its concept's URI. The type of MetaInformation defines to which indexing scheme of the Meta-Info based Index Tree we will perform the search, while the concept's URI along with the semantic_match_degree_floor determine which concepts of the Meta-Info based Index Tree will be returned. Afterwards, the engine follows the links from the selected data stored in the Meta-Info based Index Tree to the respective real Java object stored in the Typed-based Index Tree, and put that object along with its MetaInformation in the ResultList.

To find relevant concepts of a concept we use the $S_{dice}$ metric which determines the semantic similarity between concepts within an ontology. The $S_{dice}$ is determined by Valerie Cross [54] as follows: Let X represent the set of all parents from the root to concept c1 and Y represent the set of all parents from the root to concept c2. Let also f(X) represent a function which simply computes the cardinality of the set X and f(X∩Y) represent the cardinality of the intersection of the parents on the path from the root to c1 and the parents on the path from the root to c2. Assuming that the weights between all child-parent links are 1, then the $S_{dice}$ is computed by the formula:

$$S_{dice}(X,Y) = 2 * f(X \cap Y) / [f(X) + f(Y)]$$

The range of the $S_{dice}$ is [0, 1]. The $S_{dice}$ for two concepts having no common ancestor is 0, while the $S_{dice}$ for two identical concepts is 1.

## ***Example***

$S_{dice}$ plays the role of the syntactic match degree floor. To better understand how does that metric work, we give a set of examples, based on the SwinDomainOntology, a part of which is depicted in Figure 19. We want to execute a read operation given that the MetaInformation in the SpaceQuery is RDFS and the concept's URI of that MetaInformation is "Frog" (the real URI is http://swing.uni-muenster.de/core/Swing/Frog, but for simplicity reasons we refer to it as "Frog"). We also give the semantic_match_degree_floor = 0.5. We will concentrate on how $S_{dice}$ works, so we will compute every $S_{dice}$ between the concept URI "*Frog*" and each concept of the ontology.

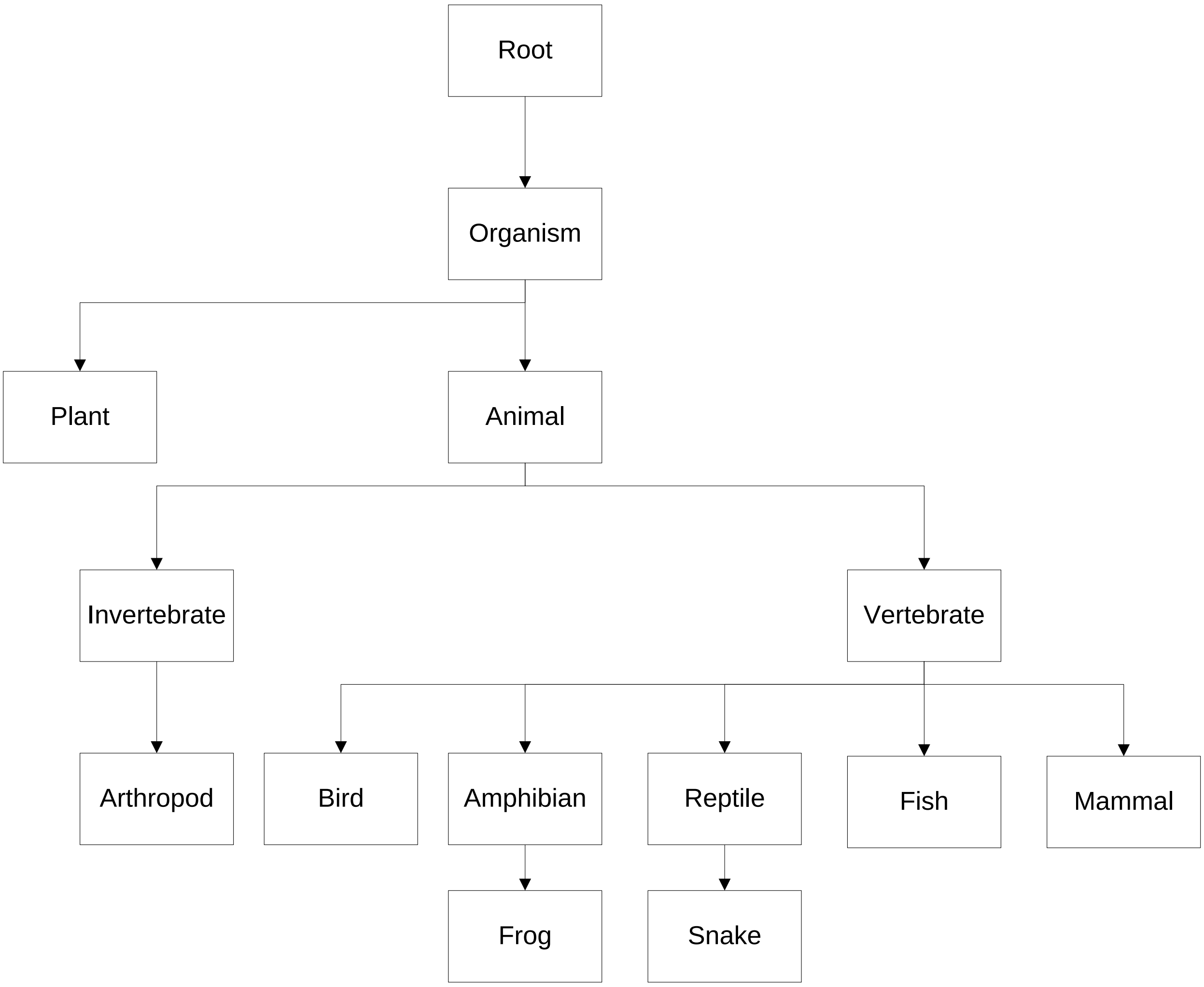

Figure 19: Visualized part of the SwingDomainOntology

For all the computations of the $S_{dice}$ it is F(Frog) = #(Organism, Animal, Vertebrate, Amphibian, Frog) = 5

- $S_{dice}$ between Frog and Organism

  F(Organism) = #(Organism) = 1

  F(Frog∩Organism) = #(Organism) = 1

  $S_{dice}$ (Frog, Organism) = 2 * 1 / (5+1) = 0.333

- $S_{dice}$ between Frog and Plant

F(Plant) = #(Organism, Plant) = 2

F(Frog∩Plant) = #(Organism) = 1

$S_{dice}$ (Frog, Plant) = 2 * 1 / (5+2) = 0.286

- $S_{dice}$ between Frog and Animal

  F(Animal) = #(Organism, Animal) = 2

  F(Frog∩ Animal) = #(Organism, Animal) = 2

  $S_{dice}$ (Frog, Animal) = 2 * 2 / (5+2) = 0.571

- $S_{dice}$ between Frog and Invertebrate

  F(Invertebrate) = #(Organism, Animal, Invertebrate) = 3

  F(Frog∩ Invertebrate) = #(Organism, Animal) = 2

  $S_{dice}$ (Frog, Invertebrate) = 2 * 2 / (5+3) = 0.5

- $S_{dice}$ between Frog and Vertebrate

  F(Vertebrate) = #(Organism, Animal, Vertebrate) = 3

  F(Frog∩ Vertebrate) = #(Organism, Animal, Vertebrate) = 3

  $S_{dice}$ (Frog, Vertebrate) = 2 * 3 / (5+3) = 0.75

- $S_{dice}$ between Frog and Arthropod

  F(Arthropod) = #(Organism, Animal, Invertebrate, Arthropod) = 4

  F(Frog∩ Arthropod) = #(Organism, Animal) = 2

  $S_{dice}$ (Frog, Arthropod) = 2 * 2 / (5+4) = 0.444

- $S_{dice}$ between Frog and Bird

  F(Bird) = #(Organism, Animal, Vertebrate, Bird) = 4

  F(Frog∩ Bird) = #(Organism, Animal, Vertebrate) = 3

  $S_{dice}$ (Frog, Bird) = 2 * 3 / (5+4) = 0.667

- $S_{dice}$ between Frog and Amphibian

  F(Amphibian) = #(Organism, Animal, Vertebrate, Amphibian) = 4

  F(Frog∩ Amphibian) = #(Organism, Animal, Vertebrate, Amphibian) = 4

  $S_{dice}$ (Frog, Amphibian) = 2 * 4 / (5+4) = 0.889

- $S_{dice}$ between Frog and Reptile

F(Reptile) = #(Organism, Animal, Vertebrate, Reptile) = 4

F(Frog∩ Reptile) = #(Organism, Animal, Vertebrate) = 3

$S_{dice}$ (Frog, Reptile) = 2 * 3 / (5+4) = 0.667

- $S_{dice}$ between Frog and Fish

  F(Fish) = #(Organism, Animal, Vertebrate, Fish) = 4

  F(Frog∩ Fish) = #(Organism, Animal, Vertebrate) = 3

  $S_{dice}$ (Frog, Fish) = 2 * 3 / (5+4) = 0.667

- $S_{dice}$ between Frog and Mammal

  F(Mammal) = #(Organism, Animal, Vertebrate, Mammal) = 4

  F(Frog∩ Mammal) = #(Organism, Animal, Vertebrate) = 3

  $S_{dice}$ (Frog, Mammal) = 2 * 3 / (5+4) = 0.667

- $S_{dice}$ between Frog and Snake

  F(Snake) = #(Organism, Animal, Vertebrate, Reptile, Snake) = 5

  F(Frog∩ Snake) = #(Organism, Animal, Vertebrate) = 3

  $S_{dice}$ (Frog, Snake) = 2 * 3 / (5+5) = 0.6

- $S_{dice}$ between Frog and Frog

  F(Frog∩ Frog) = #(Organism, Animal, Vertebrate, Amphibian, Frog) = 5

  $S_{dice}$ (Frog, Snake) = 2 * 5 / (5+5) = 1

- $S_{dice}$ (Frog, conceptX) = 0 , where conceptX does not belong to the set {Organism, Plant, Animal, Invertebrate, Vertebrate, Arthropod, Bird, Amphibian, Reptile, Fish, Mammal, Frog, Snake}

Since we gave the semantic_match_degree_floor = 0.5, the ResultList will contain concepts stored in the Meta-Info based Index Tree which have $S_{dice}$ (Frog, Concept) > 0.5. So, the concepts contained in the ResultList are {Animal, Vertebrate, Bird, Amphibian, Reptile, Fish, Mammal, Snake, Frog}.

It is easily understandable that a high price to the semantic_match_degree_floor parameter returns results close related to the concept we are searching for, while a low price to the semantic_match_degree_floor parameter returns also results not close related to the concept. When semantic_match_degree_floor=1, then the returned concepts are identical with the concept to read. Accordingly, when semantic_match_degree_floor=0 then all the concepts are returned. In both cases, the returned results contain both the actual information and the associated meta-information elements.

### 3.3.4 Take Algorithm

The take operation has many similarities with the read operation of the SCS engine. In fact, they differ in two aspects, which will be later explained. To implement the take method of the SCS Engine, we extend the typical form of the take method in JavaSpaces, as follows:

```
ResultsList take (Entry tmpl, Transaction tnx, long timeout,
     MetaInformation sinfo)
     throws RemoteException, InterruptedException,
     TransactionException
```

The added attribute is the MetaInformation which was described in detail in the write operation. The returned value is now the ResultsList which returns a list of all the results matching the given criteria. Another difference, indistinct from the declaration of the method is the parameter of type Entry.

From a conceptual point of view, when performing a semantic take operation in the SCS Engine, the client gives the MetaInformation which indicates what kind of information the client is searching for. This information will be retrieved to the client and simultaneously, it will be deleted from the SCS. The take operation searches in the Meta-Info based Index Tree of the Data Manger for stored MetaInformation elements which match the given. When referring to MetaInformation, we are particularly interested to its type (RDFS or WSML) and to its concept's URI. The type of MetaInformation defines to which indexing of the Meta-Info based Index Tree we will perform the search-and-delete, while the concept's URI determine which concepts of the Meta-Info based Index Tree will be returned and deleted from SCS. Afterwards, the engine follows the links from the selected data stored in the Meta-Info based Index Tree to the respective real Java object stored in the Typed-based Index Tree, and put that object along with its MetaInformation in the ResultList. The MetaInformation along with the connected java object are deleted from the SCS.

To conclude, we refer to the two differences between take and read operation: First, in the read method the client gives along with the MetaInformation, a semantic match degree floor, which has as an effect to return not only results identical to the given MetaInformation, but also results related to that MetaInformation. Second, in the read operation we return the matched results leaving the SCS intact, while in the take operation the returned results are also deleted from the SCS.

# CHAPTER 4
# EVALUATION

In this chapter we evaluate the SCS Engine. We present and discuss the results of measurements taken in the context of a series of experiments concerning the performance of the write, read, and take operations.

The test system was a 64-bit Microsoft Windows 7 PC with Intel® Core™ i7 with a CPU at 1.60GHz and 4.00 GB RAM. The Java version was JDK™ 6 and the VM used an initial heap size of 256 MB and a maximum heap size of 256 MB.

## 4.1 SCS Engine Case Study

The first step for the evaluation of the SCS Engine is to specify the ontology which will be used at initialization time from the SCS Engine to extract relevant information. We decided to use the SwingDomainOntology[15] which was first introduced in CHAPTER 3 and described in Appendix 2. During the initialization, the ontology is first parsed from Sesame Repository and then SCS Engine creates the structure Map<concept,BigIntegerList> described in detail in Paragraph 3.3.1. After this initialization, SCS Engine is ready to be used from external sources and tested from us. For the evaluation of the basic operations, we introduced a prototype implementation in Java[16] that plays the role of external sources, which write, read, and take Objects along with their MetaInformation based on the concepts described in SwingDomainOntology. An analytical evaluation for each basic operation according to predefined criteria is presented below.

## 4.2 Performance of the write operation

The evaluation of performance of the write operation was held by taking measurements of the execution time of the writes based on the following metrics:

1) The size of the Objects to be stored in the SCS.
2) The single-threaded or multi-threaded environment where the writes are executed.

As far as the first metric is concerned, we introduce a client who performs write operations for Objects with size ranging from 1KB to 51MB in a single-threaded environment. For small sizes, up to 50KB, the Object written in the SCS are random Strings. For bigger sizes, ranging from 89KB to 45MB, the Objects are parsed Document Object Model (DOM) Objects. To produce DOM Objects, we parse the content of XML documents, which are responses to Web Sensor[17] and Web Feature Service[18] calls defined in pilots of Envision project. We examine how the size of the Object to be inserted in the SCS affects the performance of the write operation.

As far as the second metric is concerned, we introduce a client who performs write operations for Objects with fixed size of 32 KB in a multi-threaded environment. We examine how the number of the threads executing concurrently affects the performance of the write operation.

[15] http://kenai.com/projects/envision/sources/runtime/show/SimpleUseCase/LinkedData%20Examples?rev=164
[16] http://java.com/en/
[17] http://www.opengeospatial.org/projects/groups/sensorweb
[18] http://www.opengeospatial.org/standards/wfs

### 4.2.1 Expected Results

We aim at succeeding:

(i) small write time independently of the size of the Object stored in the SCS,

(ii) small write time for a big number of processes executing concurrently

The explanation of each goal follows:

(i) the size of the Objects to be stored in the SCS ranges from 1KB to 51MB, as these objects derive from responses to (Open Geospatial Consortium) OGC[19] service calls

(ii) the number of external sources that would want to write semantically annotated Objects in the SCS at the same time is not predefined. As a result, we have to deal with the case where many write operations are executed concurrently.

### 4.2.2 Measurement Results

**Case 1 : f(SizeOfObject) = average-write-time**

Table 8 presents the experimental results referring to the average time of the write operation depending on the size of the inserted Objects. More particularly, each pair of values represents the average time computed from the insertion of 3430 Objects in relation to the size, e.g. from the insertion of 3430 Objects of size 50 KB along with their MetaInformation we calculated that the average execution time for the write operation is 0,121609799 ms.

| Average Time (ms) | Size of Object (KB) |
|---|---|
| 0,146398367 | 1 |
| 0,107319918 | 2 |
| 0,120443278 | 3 |
| 0,138524351 | 4 |
| 0,144356955 | 5 |
| 0,106153397 | 10 |
| 0,121609799 | 15 |
| 0,108486439 | 20 |
| 0,158977034 | 50 |
| 0,101487314 | 64 |
| 0,134149898 | 89 |
| 0,106445028 | 1165 |
| 0,131233596 | 3414 |
| 0,128025663 | 5890 |
| 0,127734033 | 13214 |
| 0,169145523 | 26044 |
| 0,135899679 | 45838 |
| 0,132400117 | 52245 |

Table 8: Average write-time in relation to the size of the Object written in the SCS in a single-threaded environment

[19] http://www.opengeospatial.org/

Table 8 leads us to the following observations:

(i) The execution time of the write operation is low, belonging to the range (0.1, 0.2) ms. As we mentioned in Paragraph 3.3.2, the building of the RDFS indexing takes place during the execution of the write operation. Also, the RDFS indexing is based on the given ontology. Taking into consideration that the involvement of ontologies in most of the times means the use of reasoners which increase dramatically the time in every application, one would expect high execution time for the write operation. In our case this is not applicable: the parsing of the given ontology is performed once: at the initialization of the SCS Engine. There we extract the Map<concept,BigIntegerList> structure which is stored in memory and keeps all the information necessary for each concept. So, when performing the write operation given the Object, along with the respective MetaInformation, the time needed to build the appropriate path of the RDFS indexing is reduced, as we extract the information relevant to the given concept from the Map<concept,BigIntegerList> and not from the ontology.

(ii) The execution time of the write operation is independent of the size of the Object to be written in the SCS. In Paragraph 3.3.2 we explain that during the write operation, we first build the appropriate path in the RDFS Indexing and then we invoke the standard write method provided by the JavaSpaces in order to write the real Object in the space. Therefore, the fact that the size of the Object to be written in the SCS Engine does not affect the execution time of the write operation is solely due to the efficient implementation of the JavaSpaces. This behaviour first strengthens our choice to use the JavaSpaces as our basis and more particularly the in-memory implementation. Afterwards, this fact adds another advantage to the SCS Engine, as in the environmental domain we have to deal with large data volumes so we want our mechanisms to be as independent as possible from the size of the Objects.

Figure 20 graphically illustrates the data collected in Table 8, evaluating thus the write operation in a single-threaded environment.

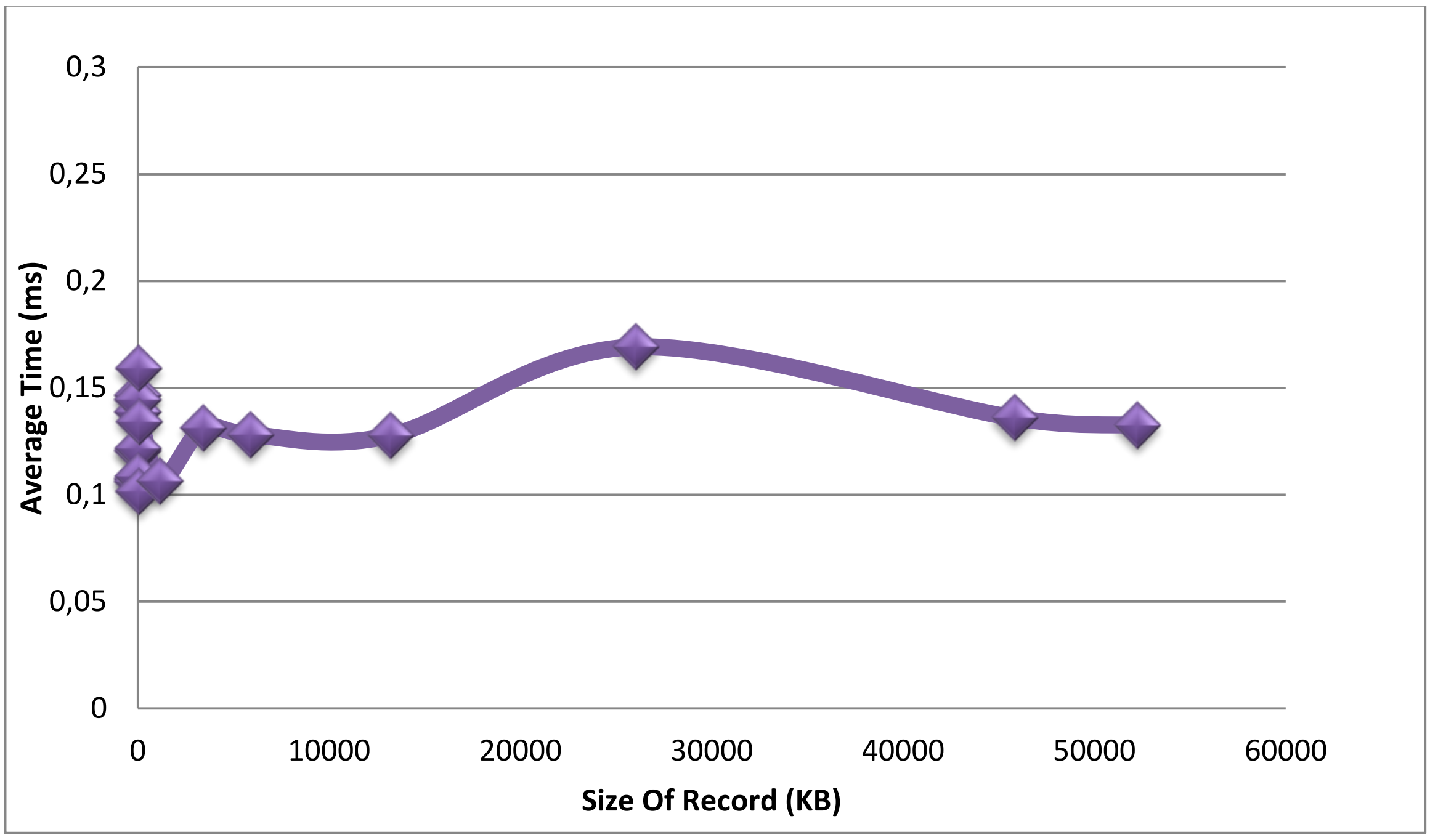

Figure 20: Average write-time in relation to the size of the written Objects in a single-threaded environment

**Case 2 : f(NumberOfThreads) = average-write-time**

Table 8 presents the experimental results referring to the average time of the write operation depending on the number of threads running simultaneously. More particularly, each pair of values represents the average time computed from the insertion of 3000 Objects each one of size 32 KB, in relation to the number of threads running concurrently e.g. from the insertion of 3000 Objects of size 32 KB along with their MetaInformation we calculated that the average execution time for the write operation when having 300 threads executing concurrently is 0,966 ms.

| Average Time (ms) | Number Of Threads |
|---|---|
| 0,386666667 | 10 |
| 0,748 | 50 |
| 0,814 | 100 |
| 0,966 | 300 |
| 1,43 | 500 |
| 1,426 | 1000 |
| 1,400333333 | 1500 |
| 1,5223615 | 3000 |

Table 9: Average write-time in relation to the number of threads executing concurrently

As one may easily infer, the average time is generally increasing when there is an increase in the number of threads running simultaneously. This is expected as the computing resources (i.e. memory and CPU time) are shared among the concurrent writes. In addition, we observe that the smallest average time in Table 9 (0,386666667 ms) outreaches the biggest average time in Table 8 (0,169145523 ms). This behaviour is normal, as in the measurements in Table 8, the CPU is entirely given to each write operation from its begging to its end, while in the measurements in Table 9 the CPU is shared between many (in our case 10) concurrent write operations. What we should underline here is that SCS Engine allows multiple clients to write in the space, adhering to the ACID (atomicity, consistency, isolation, durability) properties. This feature is inherited again from the JavaSpaces.

Figure 21 graphically illustrates the data collected in Table 9, evaluating thus the write operation in a multi-threaded environment.

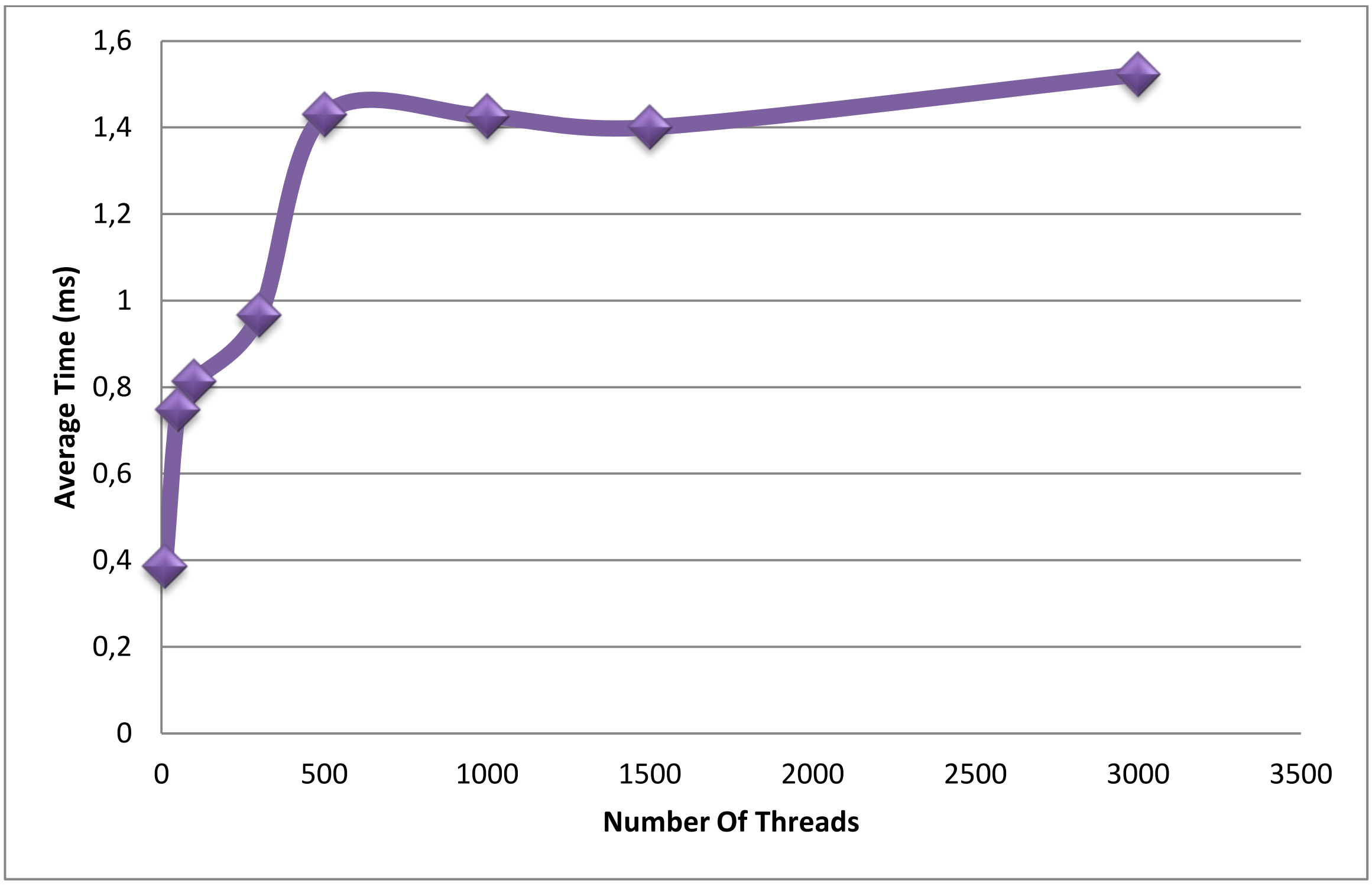

Figure 21: Average write-time in a multi-threaded environment

### 4.2.3 Comparative study between SCSEngine and JavaSpaces

In this chapter we present a brief evaluation of the write operation in JavaSpaces. The ultimate goal is to compute the complexity that the Meta-Info based Index tree adds to the JavaSpaces concerning the write operation. In other words, we examine the added complexity to the JavaSpaces by the semantically annotation of information.

Table 10 presents the experimental results referring to the average time of the write operation in JavaSpaces depending on the size of the inserted Objects. More particularly, each pair of values represents the average time computed from the insertion of 3430 Objects in relation to the size of Object, e.g. from the insertion of 3430 Objects of size 64 KB we calculated that the average execution time for the write operation is 0,040333333 ms.

| Average Time (ms) | Size Of Record (KB) |
|---|---|
| 0,047 | 32 |
| 0,040333333 | 64 |
| 0,035333333 | 89 |
| 0,048666667 | 1165 |
| 0,049 | 3414 |
| 0,056 | 5890 |
| 0,059333333 | 13214 |
| 0,062666667 | 26044 |
| 0,055333333 | 45838 |
| 0,066666667 | 52245 |

Table 10: Average write-time in relation to the size of the Object written in the JavaSpaces in a single-threaded environment

Table 11 presents the comparison results between the average write time in JavaSpaces and the average write time in SCS Engine, as well as the growth rate when passing from JavaSpaces to SCS Engine. The increase ranges from 49,6% to 73,7%.

| Size of Object (KB) | Average Write Time in JavaSpaces (ms) | Average Write Time in SCS Engine (ms) | Growth Rate |
|---|---|---|---|
| 64 | 0,040333333 | 0,101487314 | 0,602577589 |
| 89 | 0,035333333 | 0,134149898 | 0,736613046 |
| 1165 | 0,048666667 | 0,106445028 | 0,542799998 |
| 3414 | 0,049 | 0,131233596 | 0,626620001 |
| 5890 | 0,056 | 0,128025663 | 0,562587698 |
| 13214 | 0,059333333 | 0,127734033 | 0,535493152 |
| 26044 | 0,062666667 | 0,169145523 | 0,629510342 |
| 45838 | 0,055333333 | 0,135899679 | 0,592836912 |
| 52245 | 0,066666667 | 0,132400117 | 0,49647577 |

Table 11: Comparison between the average write time in SCS Engine and the average write time in JavaSpaces

### 4.3 Performance of the read operation

In this paragraph we evaluate the performance of the semantic read operation. The evaluation was held by taking measurements of the execution time of the read operations based on the following metrics:

1) The value of the $S_{dice}$ parameter, with a range [0,1]. The smaller the value of the $S_{dice}$ is, the bigger the number of the returned results we take. $S_{dice}$=0 means that we search for all the Objects stored in the SCS Engine, independent of their MetaInformation, while $S_{dice}$=1 means that we search for Objects with MetaInformation same with the given MetaInformation in the read operation. A value of the $S_{dice}$ between (0,1) means that we search for Objects with MetaInformation more or less close to the given MetaInformation in the read operation.

2) The size of the retrieved from the SCS Engine Objects.

3) The single-threaded or multi-threaded environment where the reads are executed.

As far as the first metric is concerned, we introduce a client who performs read operations in a single-threaded environment where the size of each Object stored in the SCS is 1 KB. Totally, SCS Engine stores 3430 Objects along with their MetaInformation. We examine how the value of the $S_{dice}$ parameter affects the performance of the read operation.

As far as the second metric is concerned, we introduce a client who performs semantic read operations in a single-threaded environment where the size of the Object stored in the SCS varies from 1 KB to 51 MB. Again, the total number of Objects stored in the SCS Engine along with their MetaInformation is 3430. In this case, we examine how the size of the retrieved Objects affects the performance of the read operation.

As far as the third metric is concerned, we introduce a client who performs read operations for Objects with fixed size of 32 KB in a multi-threaded environment. We examine how the number of the threads executing concurrently affects the performance of the read operation.

### 4.3.1 Expected Results

Generally, we want to guarantee in each case the elimination of execution time of read operations. The rapid information retrieval is essential to the adaptation of a process. To clarify this need we give an example: suppose that a business process contains an invocation to a service which computes the temperature of a room. Suppose also that an external source has written the temperature of the required room in the SCS. It is easy to understand that the execution of a read operation with the proper arguments can give to the process the required data and as a result the invocation of the service which gives the temperature can be omitted. So, the existence of relevant data in the SCS is the first important thing. Nevertheless, it is not the only one. The relevant temperature should be fed to the process before the invocation of the service. If the temperature is fed to the process after the invocation, then it is not taken into consideration as it is already available from the service. So, in such a case although we have the information in the SCS we cannot exploit it and the process is executed without any adaptation. To be able to exploit the data stored in the SCS Engine, we should provide efficient read operations.

More particularly we aim at succeeding:

(i) small retrieval time independently of the size of the Object stored in the SCS,

(ii) small retrieval time for large values of the $S_{dice}$ parameter (ranging from 0,8 to 1), and

(iii) small retrieval time for small number of processes executing concurrently

The explanation of each goal follows:

(i) the size of the Objects stored in the SCS ranges from 1KB to 51MB,

(ii) we are interested in MetaInformation elements very close to the given MetaInformation, and

(iii) the number of processes that need to adapt their execution at the same time is small, and the number of reads that are executed concurrently by a process is also small

### 4.3.2 Measurement Results

**Case 1 : f($S_{dice}$) = average-read-time**

Table 12 presents the experimental results referring to the average time of the read operation depending on the value of the $S_{dice}$ parameter. More particularly, each pair of values represents the average time computed from the read of 686 Objects of size 32 KB in relation to the value of the Sdice parameter, e.g. from the read of 686 Objects of size 32 KB we calculated that the average execution time for the read operation when Sdice=0,5 is 45,20262391 ms.

| Average Time (ms) | Sdice |
|---|---|
| 2021,549563 | 0 |
| 69,52915452 | 0,1 |
| 69,2244898 | 0,2 |
| 56,55976676 | 0,3 |
| 46,27696793 | 0,4 |
| 45,20262391 | 0,5 |
| 34,39212828 | 0,6 |
| 28,11078717 | 0,7 |
| 21,70991254 | 0,8 |
| 19,20845481 | 0,9 |
| 19,1574344 | 1 |

Table 12: Average read-time in relation to the value of the $S_{dice}$ parameter in a single-threaded environment

As expected, small prices in $S_{dice}$ parameter result in slow read operations, while high prices in $S_{dice}$ parameter result in quick read operations. The explanation for this behavior is very simple: A high price in the $S_{dice}$ parameter means that the SCS Engine with the Matchmaker will look for Objects with MetaInformation very close to the given. A low price in the $S_{dice}$ parameter means that the SCS Engine with the Matchmaker will look for Objects with MetaInformation close or not so close to the given. Consequently, the number of the returned results given a high $S_{dice}$ will be smaller than the number of the returned results given a low $S_{dice}$. Considering that the read operation lasts longer when it has to return more results, we come up to the desired conclusion.

Figure 22 graphically illustrates the data collected in

Table 12, evaluating thus the performance of the read operation in terms of the $S_{dice}$ parameter. We note that the figure does not illustrate the case where $S_{dice}$=0. This is an extreme case where the returned results for every read operation independent of the given MetaInformation are all the Objects stored in the SCS. This is also why the price of the average time is so high.

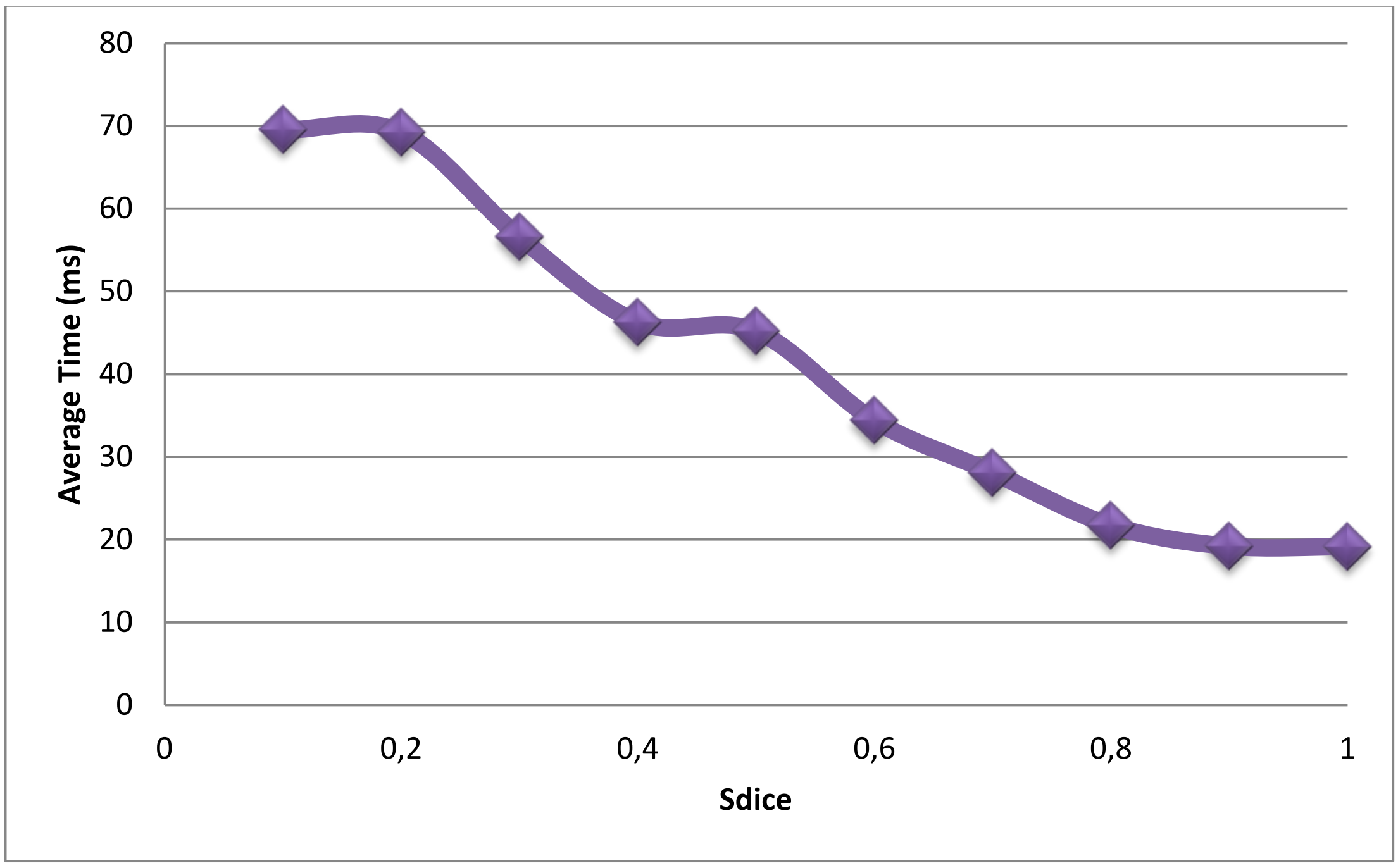

Figure 22: Average read-time in relation to the values of the $S_{dice}$ in a single-threaded environment

**Case 2 : f(SizeOfObject) = average-read-time**

Table 13 presents the experimental results referring to the average time of the read operation depending on the size of the retrieved Objects. More particularly, each pair of values represents the average time computed from the read of 686 Objects of size ranging from 1 KB to 51MB, e.g. from the read of 686 Objects of size 3414 KB we calculated that the average execution time for the read operation is 17,91107872 ms.

| Average Time (ms) | Size of Object (KB) |
|---|---|
| 17,93002915 | 1 |
| 18,54081633 | 32 |
| 18,15014577 | 64 |
| 18,00728863 | 89 |
| 18,4154519 | 1165 |
| 17,91107872 | 3414 |
| 17,66909621 | 5511 |
| 17,9212828 | 13214 |
| 18,57142857 | 26044 |
| 18,12682216 | 45838 |
| 17,78279883 | 52245 |

Table 13: Average read-time in relation to the size of the retrieved Objects in a single-threaded environment

Just like in the case of the write operation, the read operation is also independent of the Size of the Object which is stored in the SCS and retrieved to the client. The explanation of this behavior is similar to the explanation we gave in the write operation: The retrieval of the real

Object is based totally on the in-memory implementation of JavaSpaces which has the same performance for large and small Objects. Once again and for the same reasons, the fact that the read operation doe not increase when the size of the Objects increases adds another advantage to the SCS Engine, as in the environmental application domain we have to deal with large data volumes.

Figure 23 graphically illustrates the data collected in Table 13, evaluating thus the read operation in single-threaded environment.

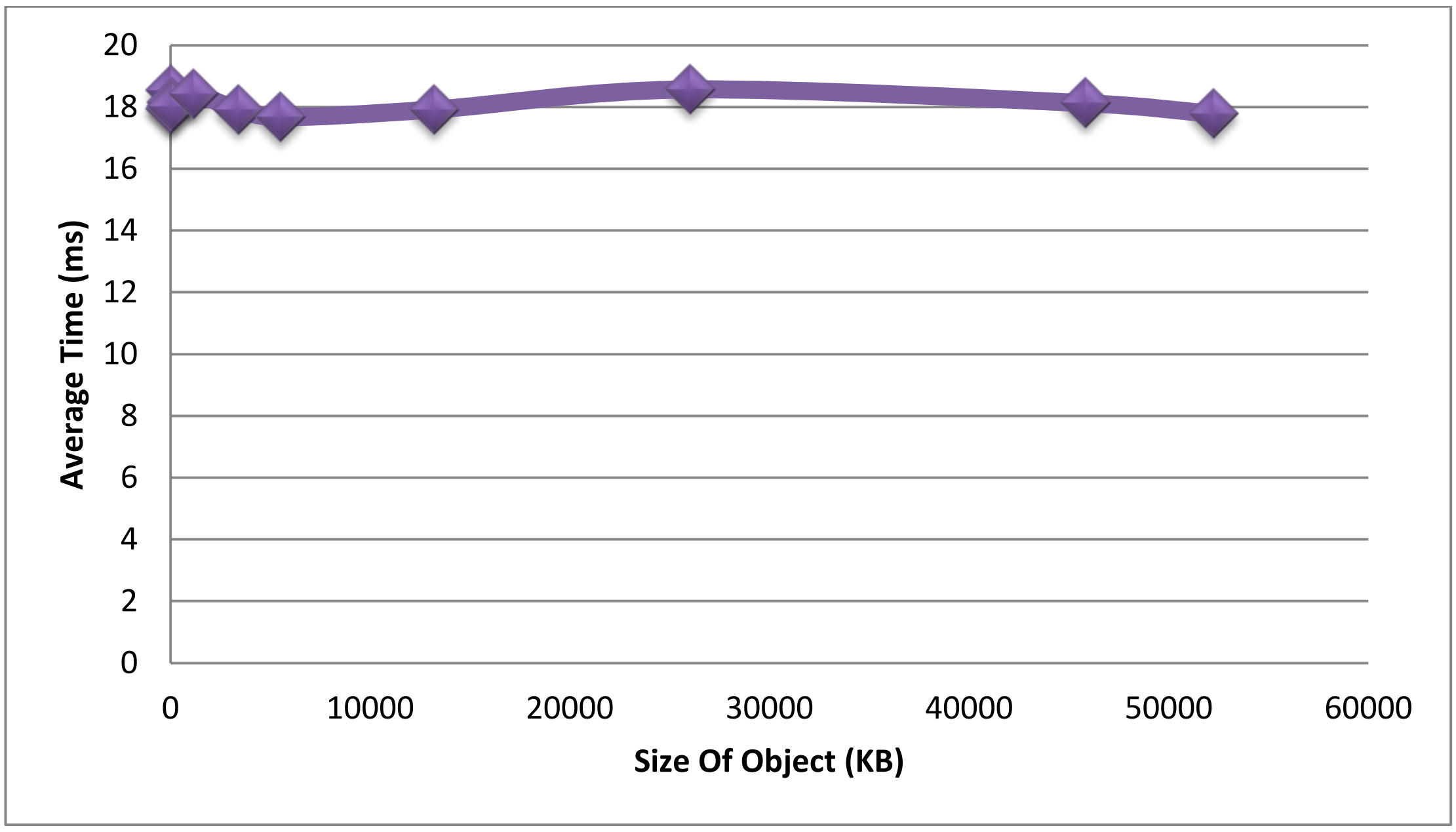

Figure 23: Average read-time in relation to the size of retrieved Objects in a single-threaded environment

What we should underline here is the efficiency of the read operation. The ultimate goal of the SCS Engine is to exploit available information within its environment to succeed the adaptive execution of business processes. More particularly, it is of high importance that the relevant data are gathered as quickly as possible.

As a result, the rapid execution of retrieval operations is a first-priority need. From the experimental results, we can see that the performance is quite satisfactory: From 0,01767 seconds to 0,01857 seconds we can retrieve relevant information of any size.

**Case 3 : f(NumberOfThreads) = average-read-time**

Table 14 presents the experimental results referring to the average time of the read operation depending on the number of threads running simultaneously. More particularly, each pair of values represents the average time computed from the retrieval of 600 Objects each one of size 32 KB, in relation to the number of threads running concurrently e.g. for the retrieval of 600 Objects of size 32 KB along with their MetaInformation we calculated that the average execution time for the read operation when having 100 threads executing concurrently is 3656,676667 ms.

| Average Time (ms) | Number Of Threads |
|---|---|
| 79,29833333 | 5 |
| 122,23 | 10 |
| 200,565 | 20 |
| 504,0683333 | 50 |
| 630,8616667 | 60 |
| 3656,676667 | 100 |
| 7246,006667 | 150 |
| 9799,753333 | 200 |
| 10525,55 | 300 |
| 13998,81333 | 600 |

Table 14: Average read-time in relation to the number of threads executing concurrently

Just like in the case or the write operation, the average time is generally increasing when increasing the number of threads running simultaneously. This is entirely expected as the CPU is shared among the concurrent reads. What we should underline here is that SCS Engine allows multiple clients to read in the space, adhering to the ACID (atomicity, consistency, isolation, durability) properties. This feature is inherited again from the JavaSpaces. Figure 24 graphically illustrates the data collected inTable 14, evaluating thus the read operation in a multi-threaded environment.

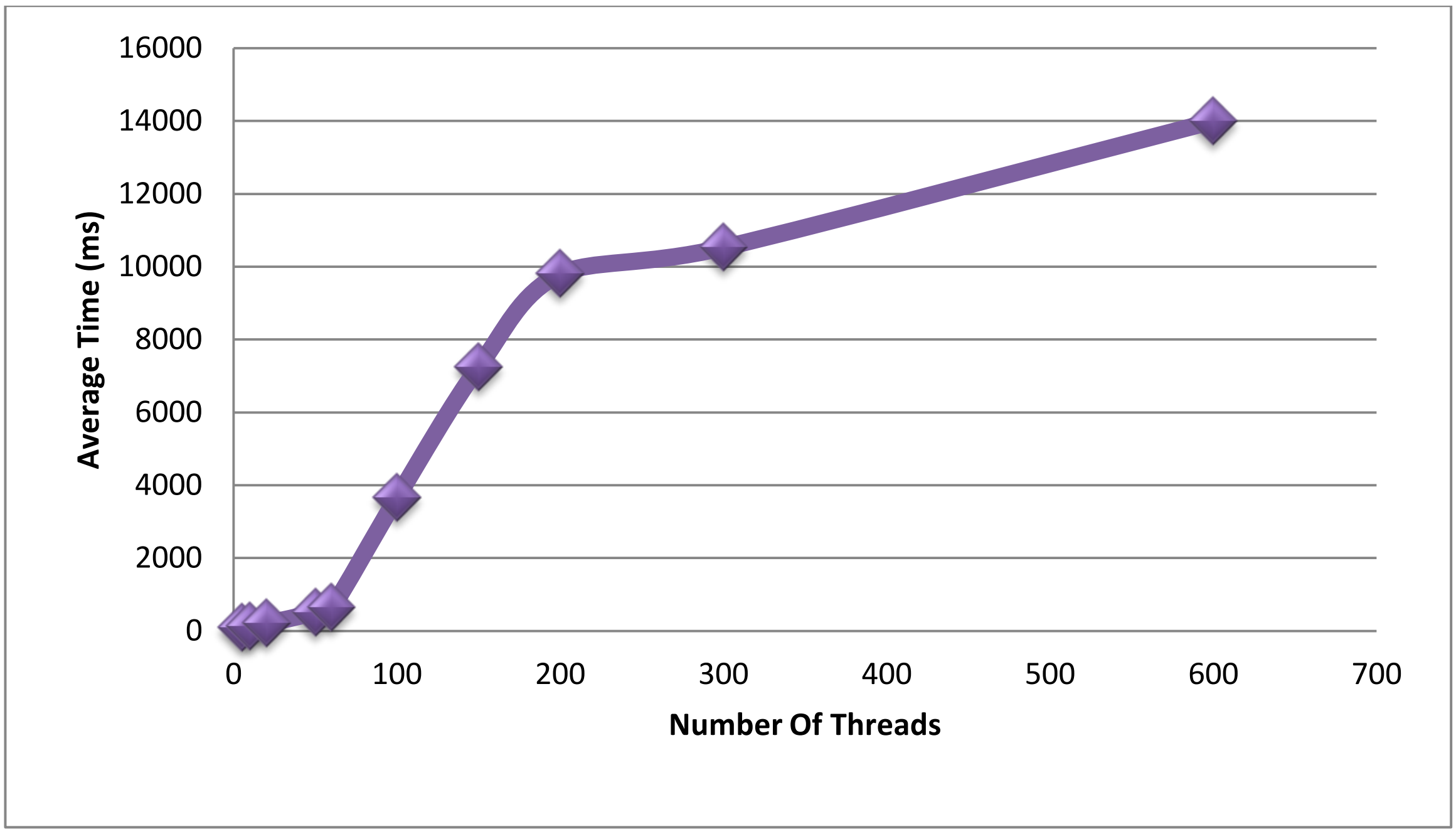

Figure 24: Average read-time in a multi-threaded environment

### 4.3.3 Comparative study between SCSEngine and JavaSpaces

In this chapter we compute the complexity that the Meta-Info based Index tree adds to the JavaSpaces concerning the read operation.

Table 15 presents the experimental results referring to the average time of the read operation in JavaSpaces and the average time of the read operation in SCS Engine when the size of the Object to be retrieved is 1KB. The average time for both cases has been computed from the retrieval of 343 Objects.

Table 15 also shows the growth rate when passing from JavaSpaces to SCS Engine. The increase is 87,2%.

| Size of Object (KB) | Average Read Time in JavaSpaces (ms) | Average Read Time in SCS Engine (ms) | Growth Rate |
|---|---|---|---|
| 1 | 1,743440233 | 13,6568915 | 0,872339893 |

Table 15: Comparison between the average read time in SCS Engine and the average read time in JavaSpaces

## 4.4 Performance of the take operation

The evaluation of the take operation was held by measuring the execution time based on the following metrics:

1) The size of the retrieved and deleted from the SCS Objects.
2) The single-threaded or multi-threaded environment where the take operations are executed.

As far as the first metric is concerned, we introduce a client who performs take operations in a single-threaded environment where the size of the Object stored in the SCS varies from 1 KB to 51 MB. Totally, SCS Engine stores 3430 Objects along with their MetaInformation. In this case, we examine how the size of the retrieved and deleted from the SCS Object affects the performance of the take operation.

As far as the second metric is concerned, we introduce a client who performs take operations for Objects with fixed size of 32 KB in a multi-threaded environment. We examine how the number of the threads executing concurrently affects the performance of the take operation.

### 4.4.1 Expected Results

The expected results in the take operation are similar to the expected results in the read operation. More particularly we aim at succeeding:

(i) small retrieval -with concurrent deletion- time independently of the size of the Object stored in the SCS and

(ii) small retrieval -with concurrent deletion- time for small number of processes executing concurrently

The explanation of each goal follows:

(i) the size of the Objects stored in the SCS ranges from 1KB to 51MB and

(ii) the number of processes that need to adapt their execution at the same time is small, and the number of takes that are executed concurrently by a process is also small

### 4.4.2 Measurement Results

**Case 1 : f(SizeOfObject) = average-take-time**

Table 16 presents the experimental results referring to the average time of the take operation depending on the size of the retrieved and deleted Objects. More particularly, each pair of values represents the average time computed from the take of 686 Objects of size ranging from 1 KB to 51MB, e.g. from the take of 686 Objects of size 3414 KB we calculated that the average execution time for the take operation is 6,771137026 ms.

| Average Time (ms) | Size Of Object (KB) |
|---|---|
| 6,983965015 | 1 |
| 7,660349854 | 32 |
| 6,645772595 | 64 |
| 6,268221574 | 89 |
| 6,720116618 | 105 |
| 6,82361516 | 1165 |
| 6,771137026 | 3414 |
| 6,842565598 | 5511 |
| 7,868804665 | 13214 |
| 7,182215743 | 26044 |
| 7,29154519 | 45838 |
| 7,017492711 | 52245 |

Table 16: Average take-time in relation to the size of the Object taken from the SCS

Just like in the cases of the write and read operations, the take operation is also independent of the size of the Object which is stored in the SCS and retrieved back to the client with its simultaneous deletion from the space. The explanation of this behavior is similar to the explanation we gave in the write and read operations: The retrieval with the simultaneous deletion of the real Object is based totally on the in-memory implementation of JavaSpaces which has the same performance for large and small Objects. Once again and for the same reasons, the fact that the take operation does not increase when the size of the Objects increases adds another advantage to the SCS Engine.

Figure 25 graphically illustrates the data collected in Table 16, evaluating thus the take operation in a single-threaded environment.

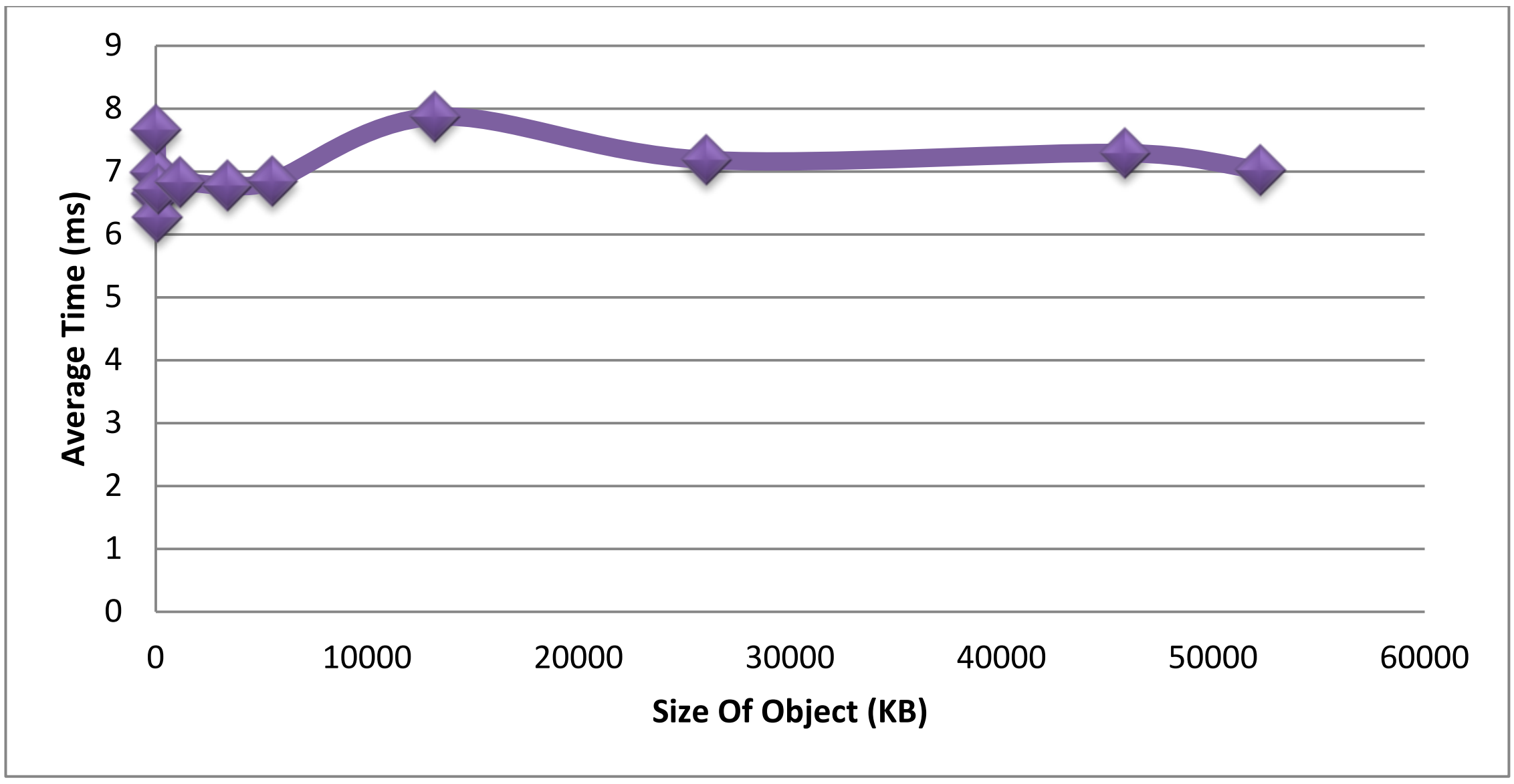

Figure 25: Average take-time in relation to the size of the retrieved Objects in a single-threaded environment

**Case 1 : f(NumberOfThreads) = average-take-time**

Table 17 presents the experimental results referring to the average time of the take operation depending on the number of threads running simultaneously. More particularly, each pair of values represents the average time computed from the retrieval with concurrent deletion of 300 Objects each one of size 32 KB, in relation to the number of threads running concurrently e.g. for the retrieval with concurrent deletion of 300 Objects of size 32 KB along with their MetaInformation we calculated that the average execution time for the take operation when having 50 threads executing concurrently is 299,6166667 ms.

| Average Time (ms) | Number Of Threads |
|---|---|
| 37,37666667 | 5 |
| 63,13333333 | 10 |
| 119,1433333 | 20 |
| 194,5233333 | 30 |
| 299,6166667 | 50 |
| 478,134375 | 80 |
| 975,45 | 100 |
| 1225,336667 | 150 |
| 2203,856667 | 300 |

Table 17: Average take-time in relation to the number of threads executing concurrently

Just like in the cases of the write and read operations, the average time is generally increasing when increasing the number of threads running simultaneously. This is entirely expected as the CPU is shared among the concurrent takes. What we should underline here is that SCS Engine allows multiple clients to take Objects from the space, adhering to the ACID (atomicity, consistency, isolation, durability) properties. This feature is inherited again from the JavaSpaces.

Figure 26 graphically illustrates the data collected in Table 17, evaluating thus the take operation in a multi-threaded environment.

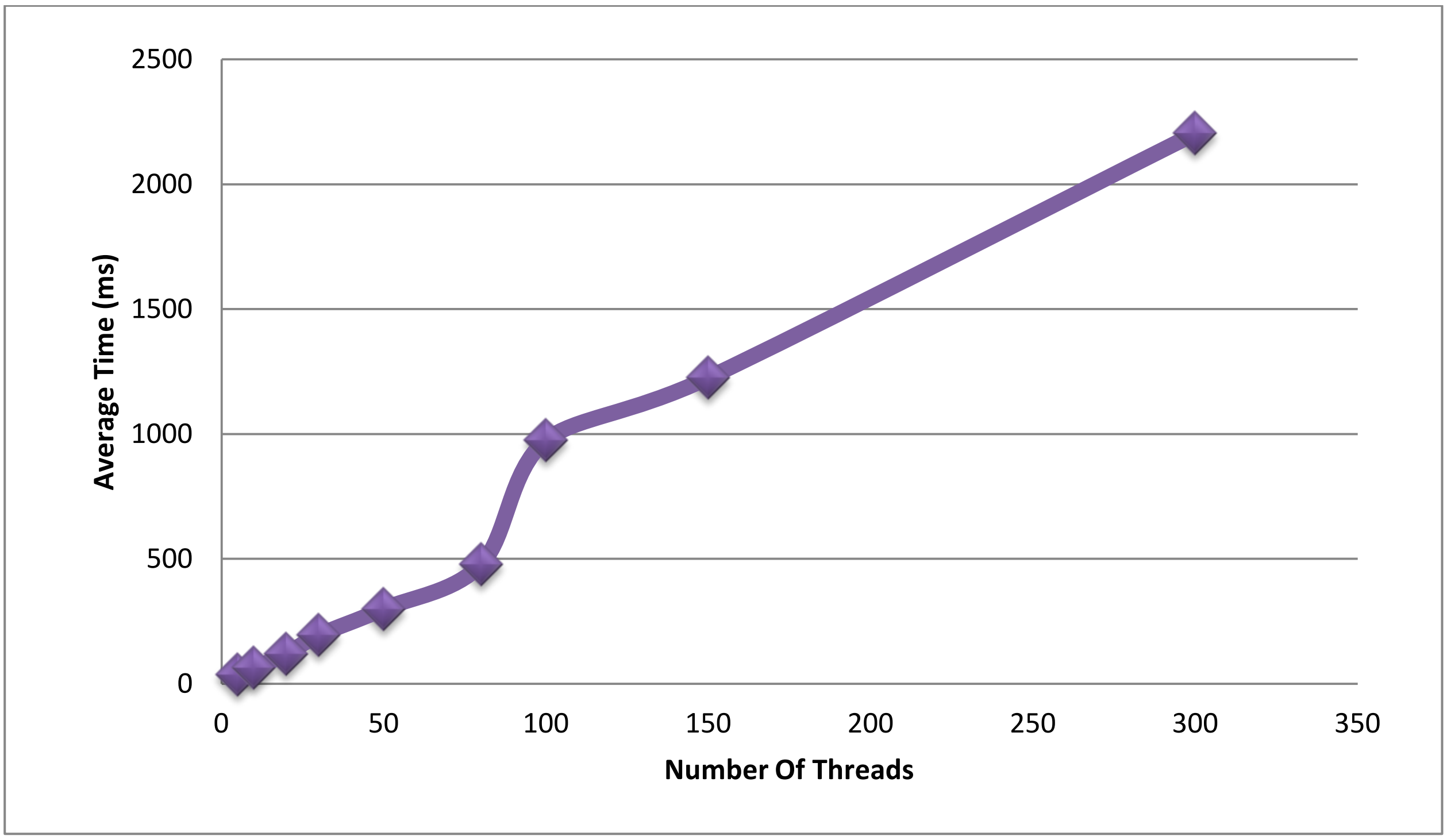

Figure 26: Average take-time in a multi-threaded environment

### 4.4.3 Comparative study between SCSEngine and JavaSpaces

In this chapter we compute the complexity that the Meta-Info based Index tree adds to the JavaSpaces concerning the take operation.

Table 18 presents the experimental results referring to the average time of the take operation in JavaSpaces and the average time of the take operation in SCS Engine when the size of the Object to be retrieved and deleted from the space is 1KB. The average time for both cases has been computed from the retrieval and deletion of 343 Objects.

Table 18 also shows the growth rate when passing from JavaSpaces to SCS Engine. The increase is 51%.

| Size Of Object (KB) | Average Take Time in JavaSpaces (ms) | Average Take Time in SCS Engine (ms) | Growth Rate |
|---|---|---|---|
| 1 | 1,251461988 | 2,555555556 | 0,510297483 |

Table 18: Comparison between the average take time in SCS Engine and the average take time in JavaSpaces

# CHAPTER 5
# CONCLUSIONS AND OUTLOOK

## 5.1 Summary

The provision of adaptable service oriented processes is a vision pursued by several research communities. Adaptation is a high-level goal required in numerous application domains such as crisis management, e-Commerce, and environmental applications. Athanasopoulos and Tsalgatidou [1] specified an approach able to facilitate the provision of such adaptable processes envisioned from the principles of data-driven adaptation. The so-called Adaptive Execution Infrastructure comprises an execution engine along with a space containing information and appropriate adaptation algorithms. This thesis specified in detail, implemented, and evaluated a basic component of the proposed architecture: the SCS Engine constituting a mechanism responsible for the collection and handling of information.

A study of the state-of-the-art in the semantic TupleSpace domain and more particularly, an analysis of the four most prominent approaches (sTuples, Semantic Web Spaces, Triple Space Computing, Conceptual Spaces), highlighted the need for the proposal of a new TupleSpace implementation, i.e. the SCS Engine, which can address unresolved matters. The specified architecture of the SCS Engine (i) laid the foundation for a space able to manipulate and exchange semantically annotated information, (ii) is easily extendable, and (iii) can effortlessly be incorporated to the Adaptive Execution Infrastructure as the provided interface allows the communication with the other components. The process of initialization of the SCS Engine, namely the parsing and analysis of given ontologies, ensured the ability to manipulate information enhanced with semantics in an efficient, simple, and easy way. The implementation of the supported operations (write, read, take) made possible the storage, search, and deletion of contextual information stemming from external sources.

The evaluation of the SCS Engine drew the attention to the performance of these three basic operations, making thus our engine ready to be used for the purpose it was created for. It became clear that all the three operations are independent of the size of Object. The values of the average time for the write operation vary from 0,1 to 0,2 ms, for the read operation from 16 to 18 ms, and for the take operation from 6 to 8 ms[20]. This satisfactory performance for the write, read, and take operation is first attributed to the use of the in-memory implementation of the JavaSpaces. In addition, the performance of each operation can be attributed to: (i) write: the mechanisms for the indexing of the meta-information are based totally on the use of the in-memory structure created at the initialization of the SCS. As a result, we do not burden the write operation during the creation of the Meta-Info based Index Tree. (ii) read: the existence of an indexing mechanism made the retrieval more efficient. Also, the structure created during the initialization keeps for each concept all paths which end up to it, beginning from the root, so the $S_{dice}$ between two concepts can be computed very quickly. Consequently, finding which meta-information elements satisfy a given query is an easy step. (iii) take: retrieving from an index (in our case this index is the Meta-Info based Index Tree) meta-information elements matching exactly a given meta-information element is an easy task. After the aforementioned explanation, it became clear that the decisions made through the implementation of the SCS Engine were correct.

The overall assessment of the SCS Engine showed that the use of the TupleSpace proved to be a good choice that can be applied to a wider mechanism catering for the adaptation of business processes. The fact that more and more efforts propose the use of the TupleSpace by business processes ([55], [56], [57], [58]) enforces our choice. Another observation for the SCS Engine is the following: since SCS Engine is open and extendable, the interested programmer can take it and use it in another system of any type for the exchange of semantically annotated information. Last but not least, the idea that a process can also communicate with external sources through the use of a TupleSpace, may inspire the scientists to invent more types of communication in a service chain.

[20] All computations refer to single-threaded environments

## 5.2 SCS role in terms of Envision Project

As we mentioned in CHAPTER 1, the Envision Execution Infrastructure is a core part of the Envision platform and aims to facilitate the execution and adaption of service chains. The principle adopted by the design, implementation, and operation of the Infrastructure is that "a service chain, comprising heterogeneous services, should be able to use the information available within its environment and adapt its execution accordingly" [1].

The main goals of the Envision Execution Infrastructure are: (i) the accommodation of the adaptive service chaining based on collected information, (ii) the enhancement of the distributed service chain execution, and (iii) the collection of information from several sources. To achieve these goals, three main components have been introduced, with the SCS Engine playing the role of the first one. In the sequel, we give a very short description of the functionality of the other two components. We underline that the proposed approach is rather generic and applicable to several application domains.

- The **Service Orchestration Engine** enables the execution of service chains as well as the monitoring and reconfiguration of their respective instances. Service chains are BPEL processes and include the interaction of multiple heterogeneous services, such as Web Services and OGC services. A key feature of this component is its distributed architecture.
- The **Process Optimizer** enables the provision of adaptable service processes, which are executed by the Service Orchestration Engine, and the extraction of information query templates that are executed by the SCS Engine. To do so, Process Optimizer implements an AI planner in order to facilitate the discovery of process plans which control the execution and adaptation of service processes.

The interested reader can refer to the respective ENVISION deliverables ([10], [59]) to get a more detailed analysis about the architecture and the implementation of these two components.

### 5.2.1 Supported Interfaces in terms of Envision Project

In terms of interoperability with the other two components, specific interfaces should be realized. We have to indicate that these interfaces have not yet been implemented in terms of this thesis, but they will be, in terms of ENVISION project. Following, we give some details about these interfaces:

- The Process Optimizer will extract appropriate Information Discovery Query Templates which are used to constantly query for information related to a process instance. SCS engine uses as input the Query Templates, and in combination with the appropriate input from the Orchestration engine, it will execute the extracted queries. The execution of these queries will possibly lead to discovery of required information which would be used for the adaptation service chains at runtime.

- The Service Orchestration Engine will accommodate the execution of service chains given by the Process Optimizer. The Orchestration Engine may provide the SCS Engine with instance specific information which will trigger the instantiation of appropriate Query Templates, produced by the Process Optimizer. In other words, a Query Template in SCS Engine waits for the appropriate instance specific information from the Orchestration Engine so as to be transformed to a query ready for execution. Depending on the query results, if related information is collected, the SCS Engine will inform the Orchestration Engine. The Orchestration Engine will decide if this information will be used during the execution of the process chain, namely, if the service chain will be executed according to the adaptation caused from the existence of this information, or if the service chain will be executed as is.

Figure 27 graphically illustrates the interaction among SCS Engine, Service Orchestration Engine, and Process Optimizer.

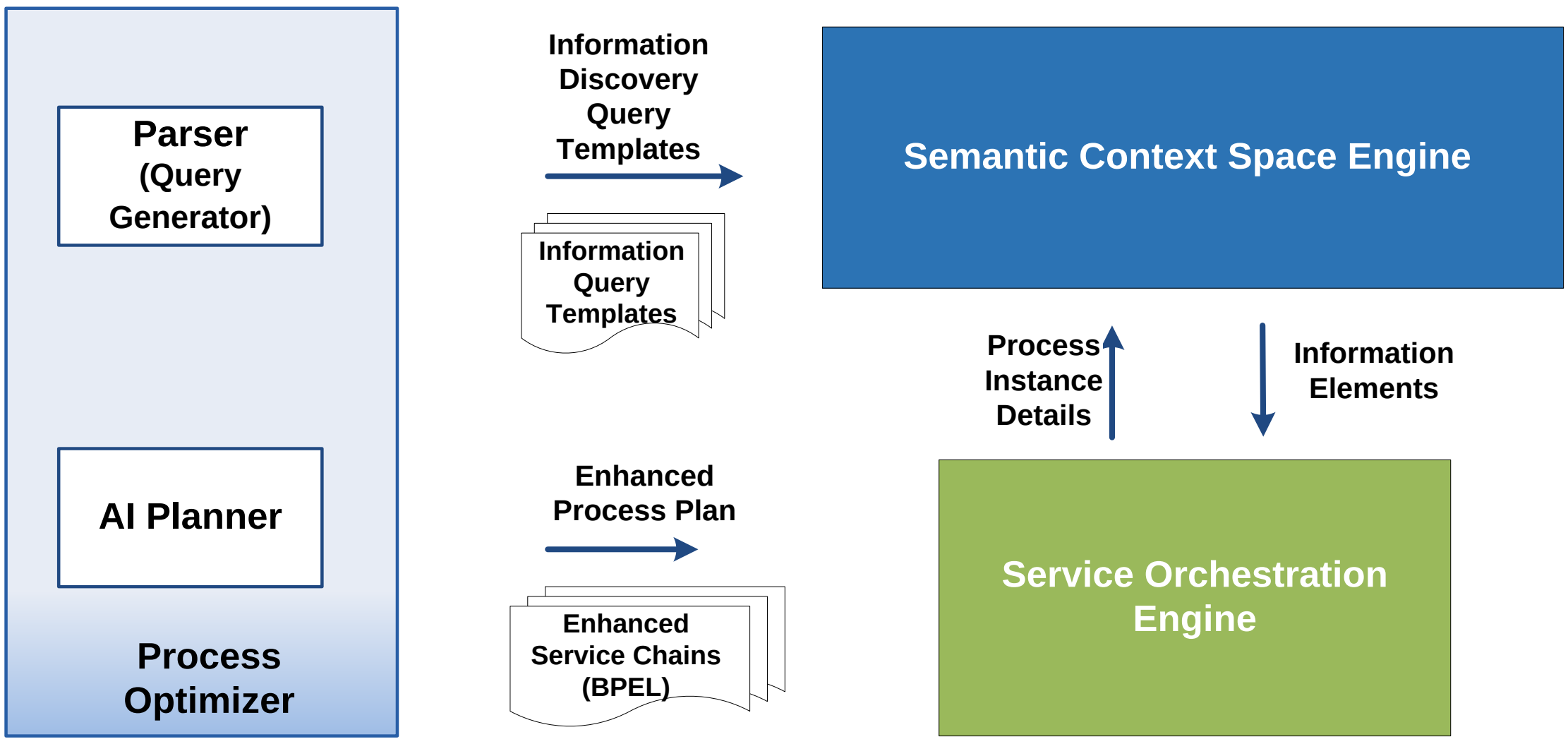

Figure 27: Interaction among the main components of the Envision Execution Infrastructure

## 5.3 Open Issues and Research Challenges

The SCS Engine proposed in this thesis incorporates ideas and techniques derived from a wide research and technological scope: environmental domain, TupleSpace, and data-driven adaptation. Taking into account the current needs in these fields, as well as the requirements in terms of Envision project, we quote our future work which will be integrated in the current version of the SCS Engine.

- Support for scope management operations. The logical grouping of information in information scopes, as well as the specification of associations among those scopes are two basic characteristics of the SCS Engine. To add this functionality to our engine, the implementation of the following scope management operations is essential: (i) createScope, (ii) removeScope, (iii) findScope, (iv) getScopes, (v) addAffiliation, and (vi) removeAffiliation. These operations are already provided from previous work done in the SCS Engine. What is left now is the provision of additional supplementary write, read, and take operations which will take into account (apart from the Object and its MetaInformation) the scope in which we want to execute the respective operation. Finally, an evaluation of these operations should also take place.

- Support for the specification of spatial and temporal features. This need arises from the environmental domain where it is not enough to enhance the real information with its MetaInformation. For example, suppose that we write in the space the information “25” along with the MetaInformation “temperature in Celsius degrees”. This information may be helpful in some occasions but misleading in some others: assume that the temperature written in the space has been measured in Athens and we are looking for the temperature of Paris or assume that the temperature has been measured in Athens in month May and we are looking for the temperature of Athens in month February. In cases like this, the spatial and temporal characteristics of the information are essential.
- Provision of a subscribe-notify communication pattern for better integration with other components, e.g. between the SCS Engine and the Service Orchestration Engine. As we illustrated in Paragraph 2.2.2, JavaSpaces provide (apart from the write, read, and take operations) the notify operation which notifies a requestor when entries that match a given template are written into the space.
- Loading of additional ontologies for the initialization and the use of the SCS Engine. In the current implementation, the SCS Engine uses a single ontology described in RDFS or WSML. Our goal is provide the client with the ability to add ontologies after the initialization of which he/she will use for the execution of the write, read, and take operations, something that until now is left to the server of our engine.

# ABBREVIATIONS

| | |
|---|---|
| Context Aware Computing | CAC |
| Artificial Intelligence | AI |
| Information and Communication Technologies | ICT |
| Model as a service | Maas |
| Service Oriented Computing – Software Engineering – Systems Development Laboratory | S3Lab |
| Semantic Context Space | SCS |
| Uniform Resource Identifier | URI |
| Triple Space Computing | TSC |
| Triple Space Transfer Protocol | TSTP |
| Application programming interface | API |
| Web Service Execution Environment | WSMX |
| Semantic Web Services described using Web Service Modeling Ontology | WSMO |
| Conceptual Spaces | CSpaces |
| Web Service Execution Environment | WSMX |
| Directed Acyclic Graph | DAG |
| FOL | First Order Logic |
| Remote Method Invocation | RMI |
| Universal Description Discovery and Integration | UDDI |
| Java Virtual Machine | JVM |
| Java Naming and Directory Interface | JNDI |
| Topology Suite | JTS |
| Web Service Modeling Language | WSML |
| Resource Description Framework | RDFS |
| Ontology Representation and Data Integration | ORDI |
| SHA | Secure Hash Algorithm |
| Document Object Model | DOM |
| Open Geospatial Consortium | OGC |

# APPENDIX 1

## Jini Framework – Apache River

A Jini system [48] is a distributed system based on the idea of federating groups of users and the resources required by those users, making use of the Java technology. The overall goal is *"to turn the network into a flexible, easily administered tool with which resources can be found by human and computational clients"* [60]. When talking about resources we mean hardware devices, software programs, or a combination of them. More specifically, the system is designed to support a dynamic network which enables the addition and deletion of services in a flexible way. Additionally, Jini provides a framework for knowledgeable services, namely services that are aware of the environment in which they operate, thus requiring less configuration and administration. Finally, Jini resolves the problem of network and system instability, by giving importance to the process and system location. Responsibility for Jini has been recently transferred to Apache[21] under the project name "River".

According to the Jini Architecture Specification [48], a Jini system comprises the following three parts:

1. **The infrastructure**: a set of components that enables building a federated Jini system. The components are: (i) *A distributed security system*, responsible for extending the Java platform's security model to the world of distributed systems. It defines how entities are identified and how they get the rights to perform actions. This system is integrated into Remote Method Invocation (RMI)[22] which in turn defines the base language within which the Jini technology-enabled services communicate. (ii) *Discovery and join protocols*, specifying the way a service of any kind becomes part of a Jini system. More specifically, it defines how to discover, become part of, and advertise supplied services to the other members of the federation, (iii) *The lookup service*, just another service of the network which plays the role of a repository for services. It can be compared to the Universal Description Discovery and Integration (UDDI)[23] in Web Services, as it enables services to register and applications to search and find the needed services. The lookup service is considered as the central point of contacts between the users and the system.
2. **A programming model**: a set of interfaces that enables the construction of reliable distributed services. Each service has an interface that specifies a communication protocol through which other services or applications can communicate it. The set of all those interfaces make up the distributed extension of the standard Java programming language model that constitutes the Jini programming model. The most significant interfaces are: (i) *The leasing interface*, responsible for allocating and freeing resources. To succeed that, the concept of "resource leasing" was introduced defining a time-based resource allocation. A lease is a grant of guaranteed access over a time period, indicating that nothing lasts forever. (ii) *The event and notification interfaces* which enable the event-based communication between Jini technology-enabled services, by extending the event model used by JavaBeans TM[24], one example of the Jini technology-enabled service, to support the distributed environment. (iii) *The transaction interfaces* support a two-phase commit protocol, enabling thus applications using the Jini framework to coordinate state changes: all of the changes made to the group occur atomically or none of them occur.

[21] http://www.apache.org/
[22] http://www.oracle.com/technetwork/java/javase/tech/index-jsp-136424.html
[23] http://en.wikipedia.org/wiki/Universal_Description_Discovery_and_Integration
[24] http://download.oracle.com/javase/tutorial/javabeans/index.html

3. **Services**: the entities within the federation which offer functionality to other members of the federation. Usually, Jini services are java objects and they define the interface through which a person, a program, or another service, can invoke them, as well as the offered operations. Jini services advertise themselves by registering to the Jini Lookup Service. Afterwards, clients can discover and use them as we have previously explained. Examples of Jini services are: a printing service, a JavaSpaces service (JavaSpaces are analyzed in Paragraph 2.2), and a transactional manager.

The aforementioned three parts can be combined in such an extent that there is no possibility to distinct them concretely. The Jini Architecture combining the infrastructure, the programming model, and the services along with the corresponding components in the familiar Java application environment is shown in Table 19.

| | Infrastructure | Programming Model | Services |
|---|---|---|---|
| Java Based | Java VM | Java API's | Java Naming and Directory Interface (JNDI)[25] |
| | RMI | JavaBeans | Enterprise beans |
| | Java Security | … | Topology Suite (JTS)[26] |
| Java + Jini | Discovery/Join | Leasing | Printing |
| | Distributed Security | Transactions | Transaction Manager |
| | Lookup | Events | JavaSpaces Servers |

Table 19: Jini Architecture in segments

Following, we graphically illustrate the communication between the Service, the Lookup Service, and the Client. We underline that action *discovery* takes place when a service is looking for a lookup service with which to register, action *join* occurs when a service has found a lookup service and wants to join it, and action lookup takes place when a client needs to locate and invoke a service.

[25] http://www.oracle.com/technetwork/java/jndi/index.html
[26] http://www.vividsolutions.com/jts/JTSHome.htm

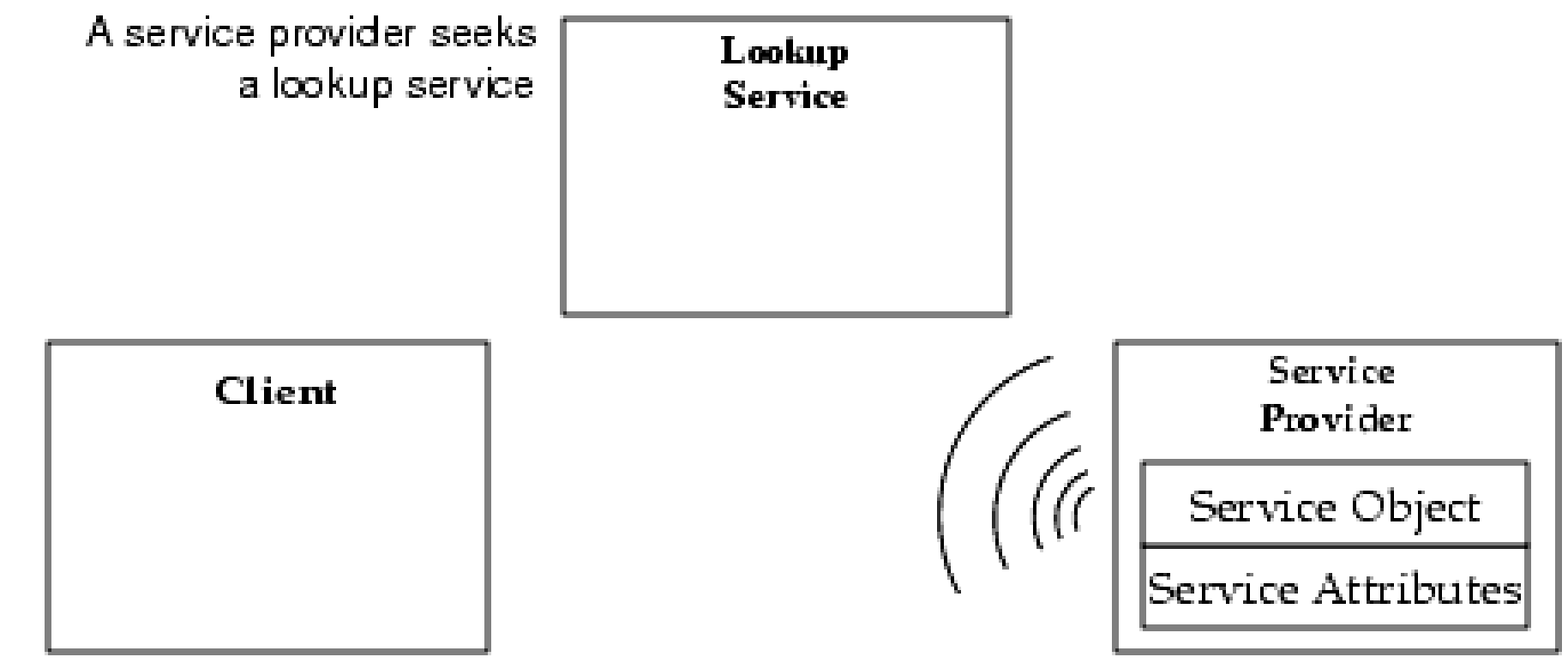

Figure 28: Action Discovery[48]

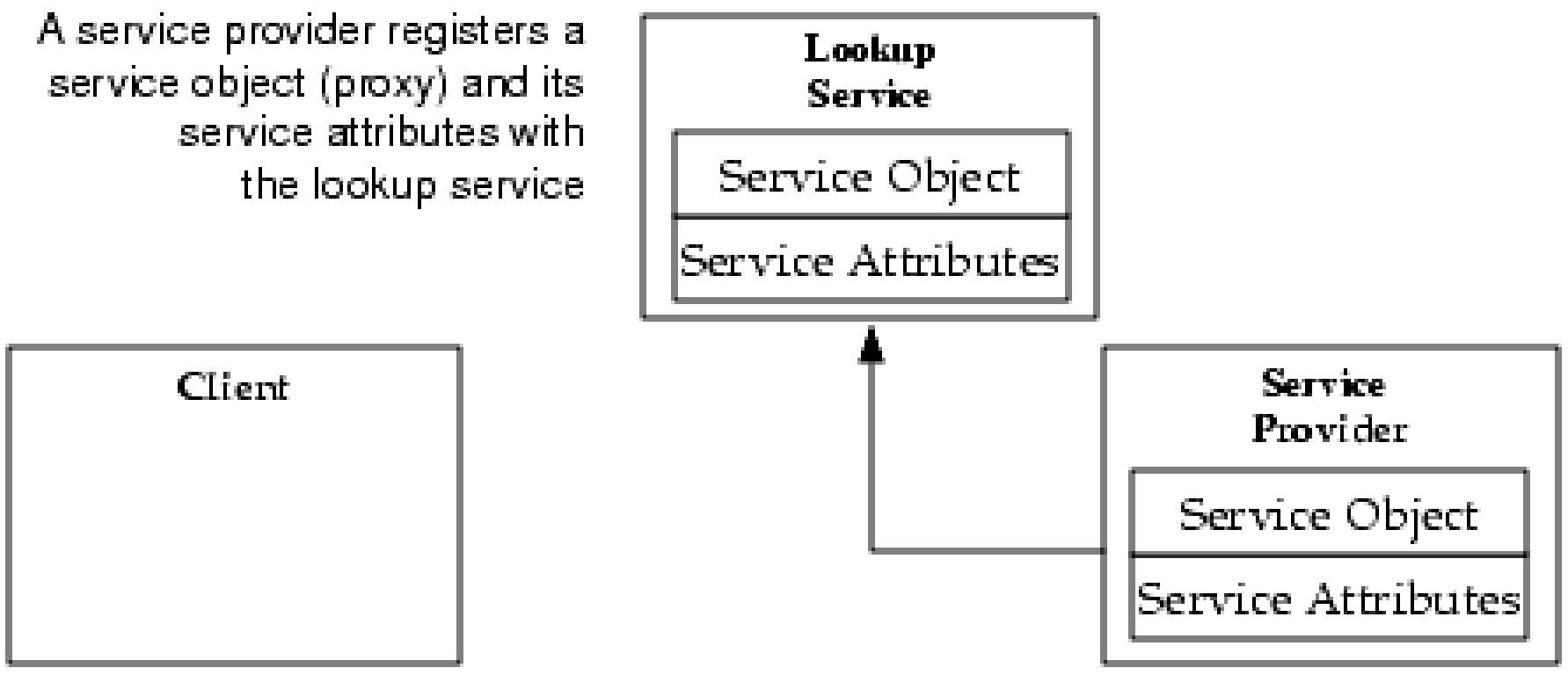

Figure 29: Action Join [48]

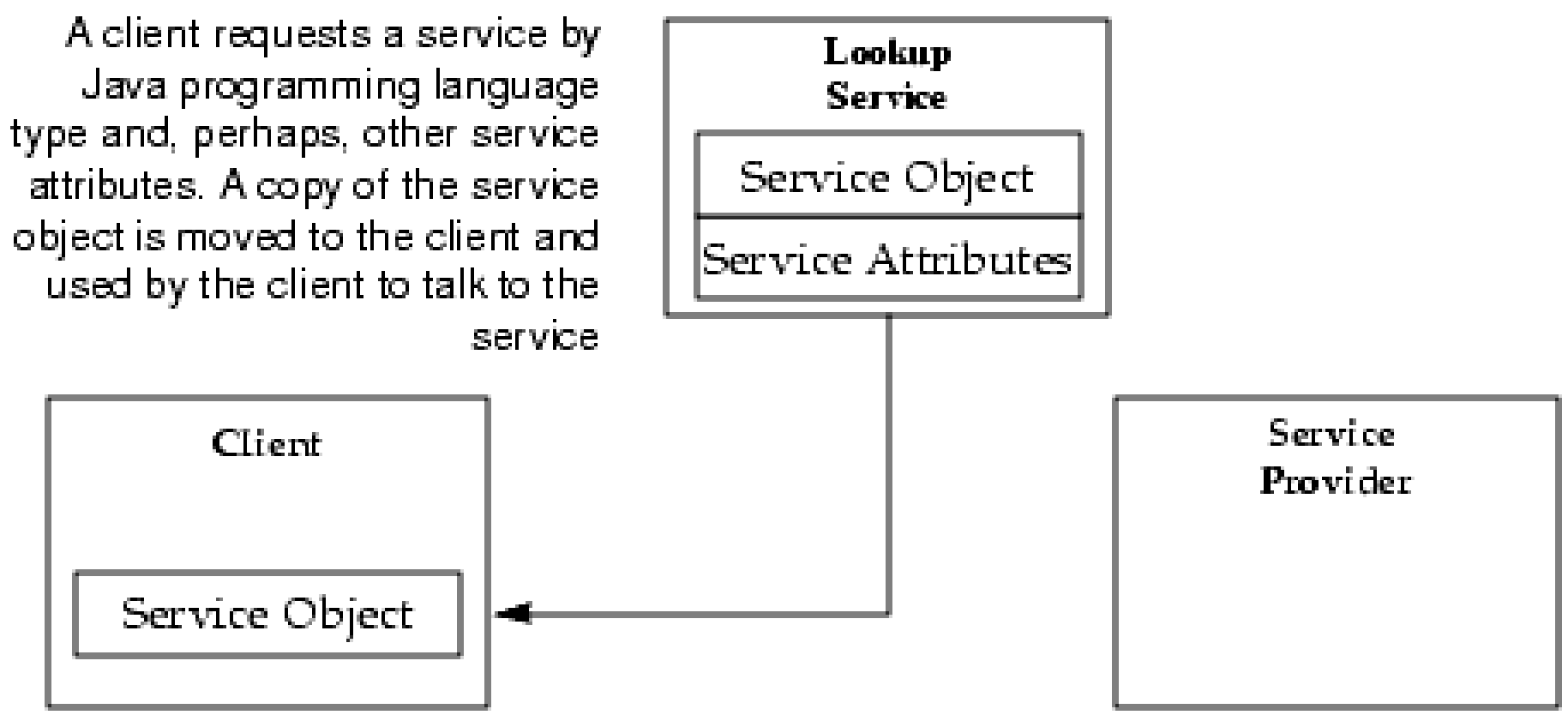

Figure 30: Action Lookup [48]

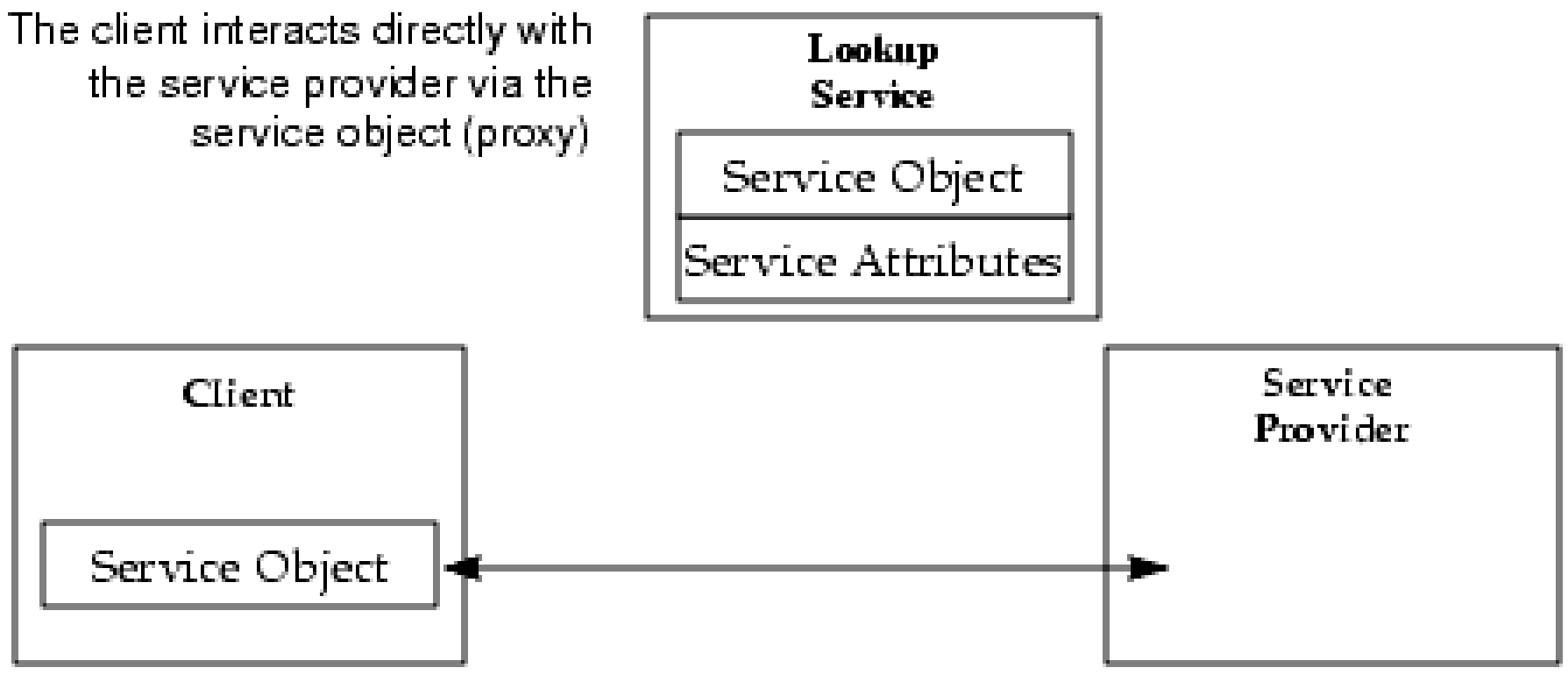

Figure 31: Service Invocation [48]

# APPENDIX 2

SwingDomainOntology was first defined in terms of SWING project[27] and is still in use in Envision. It belongs to the category of Domain Ontologies. Domain ontologies were developed to support several tasks in the SWING application like semantically annotating service capabilities and service contents, formulating goals for discovery, and specifying workflows for execution. They also capture the specific view of an information community independent of how the information is encoded. More particularly, SwingDomainOntology defines 343 concepts – the most significant of those are illustrated in Table 20.

[27] http://138.232.65.156/swing/tools.html

| Concept Name | French Name | Synonyms | further Relations (to Data Types) | Description | Comments |
|---|---|---|---|---|---|
| Activity | activité | | | Manipulation and transformation of material from ressource to product | |
| Administrative Entity | | | | | |
| Aggregate | granulat | | hasName, hasFunctionalProperties | Aggregate products | functional properties - determine the use of the aggregate like chemical properties, color, hardness, …; Aggregates are also described as "raw materials" (evidently not all raw materials are aggregates) Targeted by modelling Aggregates as subconcept of construction material |
| AggregateRessource | ressource en granulat | | hasAmount (in tons) | Quantity economically available at a given time | |
| BindingMaterial (cement) | liant, agglomérant | cement / (F) ciment | | | Describes large family of binding materials to consolidate aggregates |
| Building | construction, bâtiment | | | | BTP = Batiment et Travaux Publics (Building and Public Works) |
| Community | | | | | |
| Company | compagnie | (F) Société | Adress, name => hasLocation Location with a range specifiedBy Address | | |
| Concrete | béton | | hasAmount (in tons or cubic meters) This is not a characteristic property, hasAmount are consumption rates... | | Production often expressed in tons, delivery often expressed in cubic meters |
| Consortium | | | | | |
| ConstructionApplication | application a la construction | | Amount of consumption | Expresses the general classes of aggregate use (in construction) | |
| ConstructionApplicationSite | site d'application à la construction | (F) lieu du chantier | Location (end point of aggregate transportation) | | |
| ConstructionMaterial | matériaux de contruction | | hasFunctionalProperties2 | | same as the ones that are functional properties for the Aggregates? |
| ConstructionSiteManagement | | | | | |
| ConstructionSiteOwner | | | | | |
| ConsumptionEntity | | | | | |
| ConsumptionRate | taux de consommation | | hasValue | | |
| Country | | | | | |
| Department | | | | | |
| Extraction | extraction | | ofType (Drilling, Blasting, Shoveling, Sawing) | | |
| Geology | géologie | | hasLithology | | |
| IndustrialSite | site industriel | | Location, Type (Consumers of Quarry Product e.g. Cement Industry, Paper Industry, ...) | | |
| Infrastructure | infrastructure | | | | |
| Legislation | législation | (F) réglement, décret, circulaire | has reference for quarrying | | |
| Lithology | lithologie | rock type / (F) nom roche | | | determined by geology and quantity |
| Location | | | | | |
| MineralResource | ressource minérale | | hasValue, hasAmount | | |
| NaturalResource | ressource naturelle | | hasValue | | has natural origin (minerals, forestry, fish, water, landscape) |
| Non-ExpandableResource | ressource non consommable | | hasValue, hasContext | | |
| Non-RenewableResource | ressource non renouvelable | | hasValue, hasContext | | |
| Owner | | | | | |
| Population | | | | | |
| Processing | traitement | | type (washing, sieving, re-assembling / mixing) | | |
| Product | produit | | hasPrice, has Functional Properties | | Defined by Functional Properties / Physico-chemical characteristics |
| Production | production | | | | |
| ProductionCapacity | capacité de production | | | Capacity often expressed in Tons per Year (tpa) / Tons per Day (tpd) | |
| ProductionRate | taux de production | | | | |
| Quantity | | | | | |
| Quarry | carrière | | HasLocation | | |
| QuarryAdministration | administration de carrière | | | | |
| QuarryLocation | localisation de carrière | | HasCoordinates | | |
| QuarryManagement | gestion de carrière | | | | |
| QuarryOwner | | | | | |
| Railway | chemin de fer | | HasAmount | | Expresses the amount of aggregates needed for railroad construction |
| Regional | | | | | |
| RenewableRessource | ressource renouvelable | | HasAmount | | |
| Roads | routes | | HasAmount | | Expresses the amount of aggregates needed for road construction |
| SiteManagement | gestion de site | | | | |
| Topography | topographie | | | | |
| Transportation | transport | | HasCapacity / HasCapability | | |

Table 20: Basic Concepts defined in SwingDomainOntology[61]

# APPENDIX 3

Figure 32 illustrates the class diagram for the concepts participating to the creation of the structure during the initialization of the SCS Engine described in Paragraph 3.3.1.1.

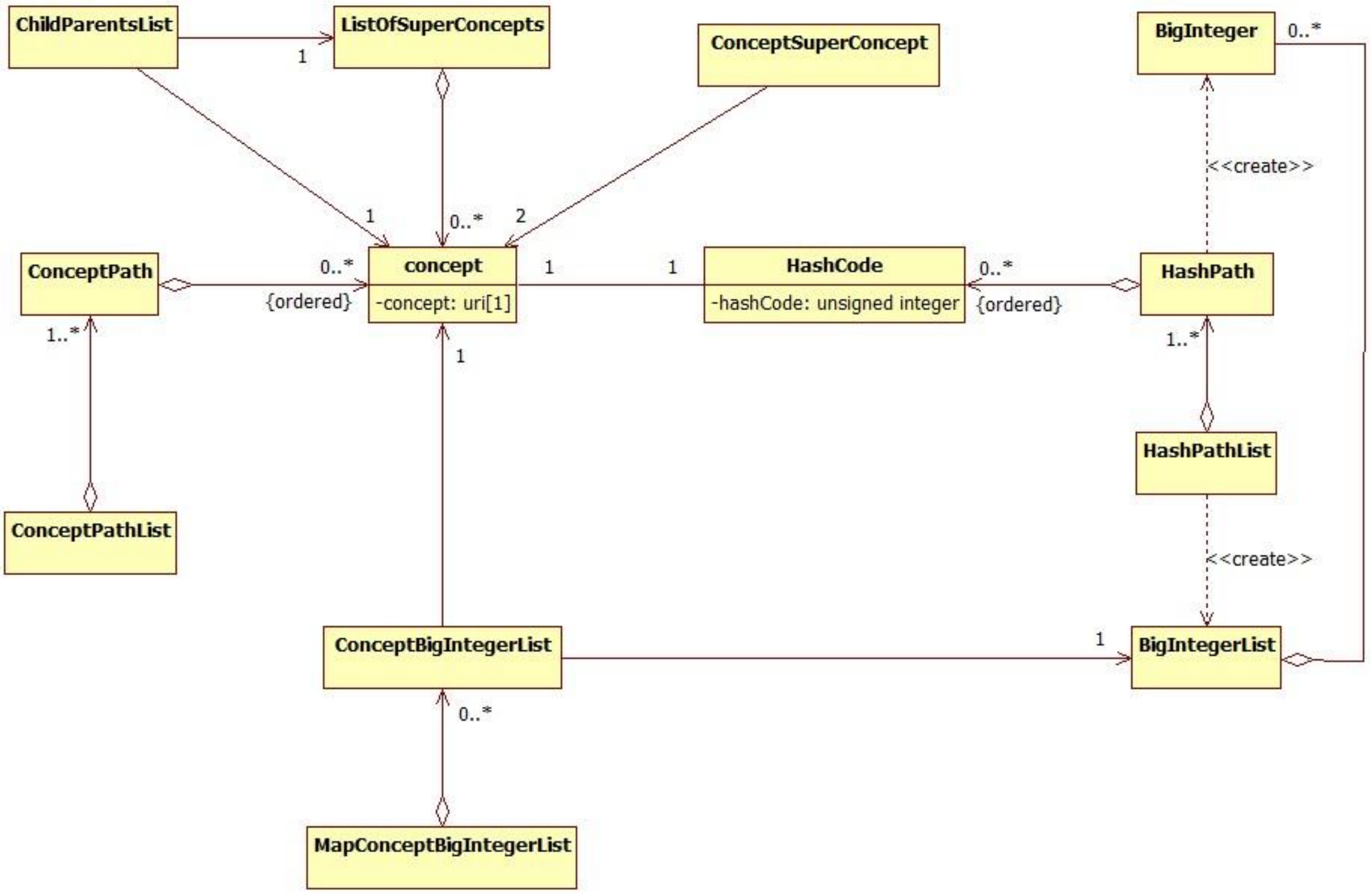

Figure 32: Class Diagram for the concepts participating to the creation of the assistant structure

The meaning of each class is as follows:

Concept: the uri referring to a concept of the given ontology

HashCode: the output of the hashCode function referring to a concept

ConceptPath: the path of concepts from the root to the respective concept

ConceptPathList: the list of all ConceptPaths of a concept

ConceptSuperConcept: a list of pairs <concept, concept> where the second concept is the parent of the first concept

ListOfSuperConcepts: a list of all the superconcepts of a concept

ChildParentsList: a list of pairs of <concept, ListOfSuperConcepts>

HashPath: the path of hashcodes from the root to the respective concept

HashPathList: the list of all ConceptPathLists of a concept

BigInteger: the union of a HashPath in a single number

BigIntegerList: the list of all BigIntegers concerning a concept

ConceptBigIntegerList: a list of pairs of <concept, BigIntegerList>

MapConceptBigIntegerList: a map of ConceptBigIntegerLists